%% file: omnitune.tex
\def\buildtechreport{}

\ifdefined\buildtechreport
\documentclass[sigconf,nonacm]{acmart}
\else
\documentclass[sigconf]{acmart}

\AtBeginMaketitle{%
  \catcode`\,=12
  \input{edbt-macros}%
  \catcode`\,=13
}%
\fi

\AtBeginDocument{%
  }

\input{preamble}
\newif\iffullversion
\ifdefined\buildtechreport \fullversiontrue \else \fullversionfalse \fi
\newcommand{\trref}[1]{\iffullversion Appendix~\ref{#1}\else our technical report~\cite{omnitune_techreport}\fi}
\newcommand{\trcite}[1]{\iffullversion Appendix~\ref{#1}\else\cite{omnitune_techreport}\fi}

\begin{document}

\title{A General LLM Framework for Query Refinement}

\author{Eldar Hacohen}
\affiliation{%
  \institution{Bar-Ilan University}
  \city{Ramat Gan}
  \country{Israel}
}
\email{eldar.hacohen@live.biu.ac.il}

\author{Yuval Moskovitch}
\affiliation{%
  \institution{Ben-Gurion University of the Negev}
  \city{Beer-Sheva}
  \country{Israel}
}
\email{yuvalmos@bgu.ac.il}

\author{Amit Somech}
\affiliation{%
  \institution{Bar-Ilan University}
  \city{Ramat Gan}
  \country{Israel}
}
\email{somecha@cs.biu.ac.il}

\renewcommand{\shortauthors}{Hacohen et al.}

\begin{abstract}
Numerous studies have explored the SQL query refinement problem, where the objective is to minimally modify an input query so that its output satisfies a specified set of constraints. However, these works typically target restricted classes of queries or constraints.
We present \sysName, a general framework for refining arbitrary SQL queries using LLM-based optimization by prompting (OPRO).
\sysName employs a two-step OPRO scheme that explores promising refinement subspaces and samples candidates within them, supported by concise history and skyline summaries for effective feedback.
Experiments on a comprehensive benchmark demonstrate that \sysName handles both previously studied refinement tasks and more complex scenarios beyond the scope of existing solutions.
\end{abstract}


\maketitle

\graphicspath{ {./images/} }

    \input{introduction}

\input{problem-definition}

\input{methods}

    \input{system}

    \input{evaluation}

    \input{related}

    \input{conclusion}

\section*{Artifacts}
\new{The source code of \system{}, the benchmark (datasets, queries, constraint and distance
functions), and scripts reproducing all experiments are publicly available at
\url{https://github.com/analysis-bots/OmniTune}, with instructions for running them via
Docker or locally in Python. The repository also includes the Refinement Problem Wizard
user interface~\cite{HacohenMS25} (directory \texttt{demo/})\iffullversion.\else{} and the full version of this
paper, including all appendices~\cite{omnitune_techreport}.\fi}
\newpage
\balance
\bibliographystyle{ACM-Reference-Format}
\bibliography{references}

\iffullversion
\clearpage
\appendix

\input{case_studies_appendix}

\input{mean-optimality-appendix}

\input{parameters_appendix}

\fi

\end{document}

%% file: preamble.tex
\usepackage{alltt}
\usepackage[utf8]{inputenc}
\usepackage{hyperref}
\usepackage{graphicx}
\usepackage{caption}
\usepackage{fancyvrb}
\usepackage{array}
\usepackage{cprotect}
\usepackage{makecell}
\usepackage{multirow}
\usepackage{adjustbox}
\usepackage{enumitem}
\usepackage{xcolor,colortbl}
\usepackage[normalem]{ulem}
\usepackage[noend, linesnumbered, ruled, vlined]{algorithm2e}
\usepackage{tabularx}
\usepackage{makecell}
\usepackage{subcaption}
\usepackage{amsmath}
\usepackage{amsfonts}
\usepackage{bm}
\usepackage{verbatimbox}
\usepackage{multirow}
\usepackage{makecell}
\usepackage{booktabs} 
\usepackage{listings}
\usepackage{xcolor}
\usepackage{dsfont}
\usepackage[most]{tcolorbox}
\usepackage{stfloats}
\usepackage{balance}
\usepackage{float}

\newtcolorbox{graybox}[2][]{
  colback=gray!25,      %
  colframe=gray!90,     %
  coltitle=white,
  fonttitle=\bfseries,
  title=#2,
  boxrule=0.9pt,
  arc=2pt,
  left=2pt, right=2pt, top=1pt, bottom=1pt,
  #1
}

\DeclareMathOperator*{\argmin}{arg\,min}

\lstdefinestyle{mypython}{
    language=Python,
    basicstyle=\ttfamily\small,
    keywordstyle=\bfseries\color{blue},
    stringstyle=\color{orange},
    commentstyle=\color{teal},
    showstringspaces=false
}

\SetKwComment{Comment}{// }{}

\makeatletter
\def\old@comma{,}
\catcode`\,=13
\def,{%
  \ifmmode%
    \old@comma\discretionary{}{}{}%
  \else%
    \old@comma%
  \fi%
}
\makeatother
\theoremstyle{definition}

\newtheorem{problem}{Problem}

\newcommand{\pred}{\ensuremath{Preds(Q)}\xspace}
\newcommand{\preds}{\pred}
\newcommand{\predVal}{\ensuremath{\theta}\xspace}
\newcommand{\allAss}{\ensuremath{\mathcal{S}(Q)}\xspace}
\newcommand{\allSS}{\ensuremath{\mathcal{S^+}(Q)}\xspace}
\newcommand{\dom}[1]{\ensuremath{\text{Dom}(#1)}\xspace}
\newcommand{\dist}{\ensuremath{\Delta}\xspace}
\newcommand{\const}{\ensuremath{\psi}\xspace}

\newcommand{\prob}{\textsc{General Query Refinement}}
\newcommand{\sysName}{\textsc{OmniTune}\xspace}
\newcommand{\system}{\sysName}

\newcommand{\probins}{\ensuremath{\mathcal{QRP}}\xspace}
\newcommand{\probfull}{\ensuremath{\mathcal{QRP}=\langle Q, D, \const, \varepsilon, \Delta \rangle}\xspace}
\newcommand{\probtuple}{\ensuremath{\langle Q, D, \const, \varepsilon, \Delta \rangle}\xspace}

\newcommand{\Hist}{\mathcal{H}}
\newcommand{\skyline}{\mathsf{Sky}(\Hist)}
\newcommand{\sky}{\mathsf{Sky}}
\newcommand{\hist}{\mathcal{H}[\Theta_t]}

\newcommand{\origv}[1]{\textcolor{red}{\sout{#1}}}
\newcommand{\refv}[1]{\textcolor{green!50!black}{\textbf{#1}}}
\newcommand{\good}[1]{\textcolor{green!45!black}{#1}}

\long\def\new#1{{\color{black}#1}}

%% file: introduction.tex
\section{Introduction} 
The problem of SQL query refinement has been studied in previous work~\cite{mishra2009refinement,KoudasLTV06,MusleaL05,ChuC94,moskovitch2023diversity,campbell2024topk,suraj2022range,MoskovitchLJ22}, with the goal of minimally modifying a given query so that its results satisfy predefined constraints. 
Early work~\cite{mishra2009refinement}, for example, focuses on refining \texttt{WHERE}-clause predicates in conjunctive Select--Project--Join (SPJ) queries to meet \emph{cardinality} constraints, adjusting queries that return too few or too many tuples. 
More recent studies~\cite{moskovitch2023diversity,campbell2024topk,suraj2022range} consider constraints defined over specific subgroups in the query output.  

While effective for their specific target settings, existing approaches typically assume fixed notions of structured constraints and refinement, and are often restricted to relatively simple query classes. For example, Li et. al. \cite{moskovitch2023diversity} considers only SPJ queries with cardinality constraints, the work of ~\cite{campbell2024topk} extends this setting to SPJ with an \verb|ORDER BY| clause and considers cardinality constraints over subgroups within the top-$k$ results, and \cite{suraj2022range} consider range queries and are limited to the refinement of numerical predicates with constraint over a single binary attribute. 
In practice, however, users may wish to refine a much broader range of queries—including those with \texttt{GROUP BY} clauses or subqueries—and may impose application-specific definitions of both constraints and refinement criteria. 
To illustrate this gap, we next consider an example of a group-by query refinement problem.

\begin{example}\label{ex:running}

In Middle-earth, the Knightly Order plans to launch a pilot continental scholarship program aimed at identifying and supporting young, promising warriors. To this end, the Order tasks its chief analyst, Galadriel, with reviewing the pool of applicants and proposing initial selection criteria for the program’s first cycle. These criteria include:
(i) a minimum GPA threshold,
(ii) an approved set of majors, and
(iii) a minimum representation quota per participating region.

Senior leadership further specifies two program-level constraints that must be satisfied by any proposed parameterization:
(1) geographic coverage---at least eight distinct regions must be represented in the pilot cohort; and (2) gender balance---the overall ratio of female to male recipients should be approximately 1:1.

Galadriel examines the applications database (see Table~\ref{tab:db} for a sample), consisting of five attributes: ID, gender, region, major, and grade point average (GPA). She issues an initial SQL query\footnote{Note that the function \textit{COUNT\_IF} is not supported in all databases, yet can be easily implemented in native SQL: \texttt{COUNT(CASE gender=F THEN 1 ELSE 0 END;)}.}, as depicted in Figure~\ref{fig:scholarship-query}, representing an intuitively suitable criteria: GPA threshold of 3.5, majors are \textit{Tactics} and \textit{Archery}, and minimum region-quota of 100 students.

\begin{table}[t]
    \centering
    \small
    \begin{tabular}{|c|c|c|c|c|c|}\hline
id & gender & region & major & GPA \\\hline
1 & F & Shire & Archery & 3.7 \\
2 & M & Rivendell & Swordcraft & 3.8 \\
3 & M & Mordor & Tactics & 3.7 \\
4 & F & Rohan & Melee & 3.7 \\
5 & M & Erebor & Scouting & 3.8  \\
\(\cdots\) & \(\cdots\) & \(\cdots\) & \(\cdots\) & \(\cdots\) \\
9996 & F & Rohan & Swordcraft & 3.8  \\
9997 & F & Khazad-dûm & Stealth & 3.6  \\
9998 & M & Lothlórien & Tactics & 3.9  \\
9999 & F & Shire & Melee & 3.6  \\
10000 & M & Rohan & Archery & 3.5  \\\hline
\end{tabular}
    \caption{The Middle-earth Students dataset $D$.}
    \label{tab:db}
\end{table}

\begin{figure}[t]
\small
\centering
{
\begin{tabular}{l}
\verb"SELECT region, avg(GPA), COUNT_IF(gender = F), COUNT(*)"\\
\verb"FROM MiddleEarth_Applicants"\\
\verb"WHERE GPA > 3.5 AND major IN ('Tactics', 'Archery')"\\
\verb"GROUP BY region"\\
\verb"HAVING COUNT(*) > 100"\\
\end{tabular}
}
\caption{Initial Scholarship Query $Q$} 
\label{fig:scholarship-query}
\end{figure}

Galadriel now examines the results (see Table~\ref{tab:result}); unfortunately, none of the constraints are satisfied:
\begin{enumerate}[leftmargin=*]
    \item The initial scholarship design includes only 5 regions, fewer than the required 8.
    \item The total number of female students is \(170\), out of \(632\) which is significantly below the required ratio of \(0.5\).
\end{enumerate}

Consequently, Galadriel needs to \emph{refine} the selection criteria in order to meet the stated constraints. \qed

\end{example}

Unfortunately, even the simple query of the example above is not supported by existing query refinement frameworks~\cite{mishra2009refinement,campbell2024topk,moskovitch2023diversity,suraj2022range}, as none support the refinement of group-by queries with predicates in the HAVING clause, as well as free-form constraints.  
Addressing this would require formally defining the refinement problem, design a constraint satisfaction framework, and explore adaptations of existing solutions to support these query settings. Alternatively, refining the query manually can be a tedious and repetitive task.
To this end, we present a \new{general} framework for SQL query refinement problems. Given a database $D$ and starter query $Q$, as well as notions for refinement distance and constraints (represented as Python functions), our framework iteratively seeks a refined query that is as close as possible to $Q$ while (approximately) satisfying the constraints. 

Our solution builds on the optimization-by-prompting (OPRO) paradigm~\cite{yang2024opro,chen-etal-2024-prompt},
where an LLM is iteratively prompted to propose candidate refinements based on a textual
specification of the task, a description of the search domain, and the history of past
candidates. However, as we show empirically, a direct application of this paradigm performs
poorly, due to the vast combinatorial search space and the long,
low-information interaction histories that fail to effectively guide subsequent proposals.
We therefore design a dedicated two-step OPRO scheme for the query refinement problem.
First, the LLM selects a promising refinement \emph{subspace} that aggregates many potential refinements. Then, a subsequent LLM samples individual refinements that aim to satisfy the constraints while \new{minimizing the distance from~$Q$}.

We then design \system{}, an efficient prototype system that instantiates our
query-refinement OPRO scheme in practice. \system{} addresses several core challenges,
including representing an exponentially large space of refinement subspaces without
materialization, constructing concise, query-dependent textual descriptions of the
database and search domain, and providing informative yet compact guidance to LLM-based
optimizers. 

At a high level, \system{} operates in two phases: (1) a lightweight pre-processing
phase that derives tailored representations of the database and refinement domain, and
(2) an optimization loop that alternates between selecting promising refinement subspaces
and sampling concrete refinement queries 
via two specialized LLM modules. Throughout
optimization, \system{} maintains a skyline of candidate refinement queries
that captures the
best observed trade-offs between query distance and constraint satisfaction; this skyline
serves primarily as a multi-objective signal that guides both subspace selection and refinement query
generation, and additionally enables an effective early-stopping mechanism
when the skyline stabilizes across iterations.

Our empirical evaluation demonstrates the effectiveness of our framework across a wide range of query refinement tasks: (i) Top-$k$ queries with group-representation constraints~\cite{campbell2024topk},
(ii) range queries with group-ratio constraints~\cite{suraj2022range},
(iii) diversity-aware SPJ queries~\cite{moskovitch2023diversity},
and (iv) complex queries beyond the scope of prior dedicated approaches.
We compare \system{} against strong state-of-the-art LLMs, including
OpenAI GPT-5.6-Luna~\cite{openai2026gpt56systemcard} and Google Gemini 3.0
Flash-Thinking~\cite{google2025gemini3flash}.
Our results show that \system{} achieves performance close to that of specialized refinement solutions on their native problem classes, while substantially outperforming LLM-based baselines and extending effective query refinement to previously unsupported settings. 

\begin{table}[t]
    \centering
    \small
    \begin{tabular}{|c|c|c|c|}\hline
\textbf{region} & \textbf{Avg.~GPA} & \textbf{Count\_if(gender=F)} & \textbf{Count(*)} \\\hline
Rivendell & 3.9 & 50  & 150 \\
Gondor & 3.7 & 33  & 125 \\
Lothlórien & 3.8 & 32  & 120 \\
Mordor & 3.8 & 34  & 135 \\
Erebor & 3.7 & 21 & 105 \\\hline
\end{tabular}
    \caption{Query output \(Q(D)\)}
    \label{tab:result}
\end{table}

\new{We presented an early prototype of \system{} in~\cite{HacohenMS25}, comprising a Refinement Problem Wizard that translates natural-language specifications into formal query refinement problems, and an early prototype of the LLM-based solver engine. This work focuses on the latter, which we substantially redesign, extend, and evaluate.}

\paragraph*{Contributions \& Paper Outline} We next outline our contribution and the organization of the paper.
\begin{itemize}[leftmargin=*]
    \item We formulate the \prob\ problem, \new{which} provides a
    unified abstraction encompassing prior refinement settings and a wide range of new,
    previously unsupported scenarios (Section~\ref{sec:problem}).

    \item We devise a dedicated LLM-based optimization-by-prompting (OPRO) scheme for query
    refinement, addressing the challenges of large, combinatorial refinement spaces via
    structured subspace selection and query refinement sampling (Section~\ref{sec:OPRO}).

    \item We design \system{}, an end-to-end prototype instantiating our
    OPRO scheme, including query-conditioned database representations and skyline-based
    multi-objective optimization (Section~\ref{sec:history}).

    \item We empirically evaluate \system{} on a diverse collection of query refinement
    tasks adopted from prior work~\cite{campbell2024topk,suraj2022range,moskovitch2023diversity},
    as well as on more complex query patterns beyond the scope of existing solutions.
    The results show that \system{} achieves performance close to specialized approaches
    while substantially outperforming strong LLM baselines (Section~\ref{sec:eval}).
\end{itemize}
We survey related work in Section~\ref{sec:related} and conclude in
Section~\ref{sec:conclusion}.

%% file: problem-definition.tex
\section{Preliminaries and Problem Definition}\label{sec:problem}
In this section, we provide an overview of the \prob\ problem. Notations are summarized in Table~\ref{tab:notation_summary}.

\paragraph{Refinable predicates, supported queries.} We build on the \new{widely-accepted} notion of query refinement \new{first} introduced in~\cite{mishra2009refinement}. 
Given a query~$Q$ over a dataset~$D$, a refinement of~$Q$ modifies one or more of its \emph{refinable} predicates. We call a predicate in an SQL query \emph{refinable} if it is of the form $\langle A~\mathsf{op}~C \rangle$, where $A$ is an attribute in~$D$, $\mathsf{op}$ is a comparison operator, and $C$ is a literal constant. The attribute~$A$ may be either a \emph{base} (schema-native) attribute or a \emph{derived} attribute, such as one obtained via aggregation or column-wise computation.
If~$A$ is \emph{numeric}, then $\mathsf{op} \in \{<,\leq,=,\geq,>\}$ and $C \in \mathbb{R}$. 
If $A$ is \emph{categorical}, then $\mathsf{op} = \mathsf{IN}$ and $C$ denotes a set of categorical values, such that $C \subseteq \dom{A}$, where $\dom{A}$ is the set of distinct values appearing in column $D(A)$. We consider general SQL queries containing \verb|WHERE| and/or \verb|HAVING| clauses, with at least one refinable predicate. 
Let $\pred$ denote the set of all \textit{refinable} predicates in a query~$Q$. 
For each predicate $p \in \pred$, we write $p.A$, $p.\mathsf{op}$, and $p.C$ to denote its attribute, operator, and literal(s), respectively.

\begin{example}\label{ex:running-cont}
Consider again the dataset~$D$ shown in Table~\ref{tab:db} and the query~$Q$ in Figure~\ref{fig:scholarship-query}. 
The set of refinable predicates in~$Q$ consists of three predicates. 
The first two appear in the \verb|WHERE| clause: 
$p_1 = \langle \texttt{GPA}~> 3.5 \rangle$ and 
$p_2 = \langle \texttt{major}~\mathsf{IN}~ \{\texttt{Tactics}, \texttt{Archery}\} \rangle$.
The third predicate appears in the \verb|HAVING| clause and is given by 
$p_3 = \langle \texttt{COUNT(*)}~ > 100 \rangle$.
\qed

\end{example}

\paragraph* {Query refinement assignment} 
Following~\cite{mishra2009refinement}, a valid refinement consists of modifying the literal(s) in $p.C$ for any predicate $p \in \preds$. 
For predicates over numerical attributes, refinement is performed by directly altering the value in $p.C$, whereas for categorical predicates, refinement is achieved by adding to or removing items from the set specified by $p.C$.

A \emph{refinement assignment} is denoted by $\theta = \{\theta.p \mid p \in \preds\}$, where each component $\theta.p$ specifies a (possibly unchanged) refined literal $p.C'$ for predicate $p$.
We denote by $Q_{\theta}$ the refined query obtained by applying assignment $\theta$ to $Q$, and by $\theta_0$ the \textit{baseline} assignment, corresponding to the original query, that is, $\theta_0.p = p.C$ for all $p \in \preds$.

\begin{example}\label{ex:refinement} 
    Consider again the database $D$ and query $Q$, as depicted in Table~\ref{tab:db} and Figure~\ref{fig:scholarship-query} (resp.).  The baseline assignment $\predVal_0=\{3.5,(\texttt{Tactics}, \texttt{Archery}),100\}$, corresponding to the original query $Q$. A possible refinement assignment is  $\predVal'=\{3.6,(\texttt{Tactics}),80\}$, which represents the following refined query $Q_{\theta'}$ (modifications in \textbf{bold}):
 {\small
\begin{alltt}
    SELECT region, AVG(GPA), COUNT_IF(gender = F), count(*)
    FROM MiddleEarth_Applicants
    WHERE GPA > \textbf{3.6} AND major IN \textbf{('Tactics')}
    GROUP BY region
    HAVING COUNT(*) > \textbf{80} \qed \end{alltt} 
}   
\end{example}

In the remainder of the paper, we use 'refinement' to denote both the assignment $\predVal$ and the resulting query $Q_\predVal$ interchangeably. We further denote the set of \textit{all} possible assignments for $Q$ by $\allAss$.

\paragraph*{Query output constraints} Expanding the line of work on query refinement for output constraint satisfaction~\cite{campbell2024topk,suraj2022range,rodeo,moskovitch2023diversity,erica}, we consider general constraints over the output. Since achieving perfect satisfaction may be infeasible~\cite{campbell2024topk}, we adopt the notion of satisfaction deviation, relax the goal of exact satisfaction, and allow for approximate satisfaction. To this end, we define the output constraints as a function that measures the degree of violation of a query’s result with respect to the constraints for a given database. 

\begin{definition}[Constraint deviation function]
    Given a refinement assignment $\predVal$, a constraints deviation function $\const:\allAss \to [0,1]$ depicts the degree of violation of the output of the refined query $Q_\theta(D)$. Given a tolerance $\varepsilon\in[0,1)$, we say that an assignment $\predVal$ is $\varepsilon$-satisfying, if $\const(\predVal) \leq\varepsilon$.
\end{definition}

By formulating constraints as functions, we enable the expression of any general constraint on the query output $Q(D)$.

\begin{example}
Recall the constraints specified in Example~\ref{ex:running}:
(i)~\textit{A $1$:$1$ female-to-male ratio}; and
(ii)~\textit{At least $8$ distinct regions are represented}. We first define a deviation function for the geographic coverage constraint. This function counts the number of regions returned by $Q_\theta(D)$ and returns the normalized slack when fewer than $8$ regions are selected, and $0$ otherwise:
\[
f^{1}(\theta) =
\begin{cases}
\dfrac{8 - |Q_\theta(D)|}{8}, & \text{if } |Q_\theta(D)| < 8, \\
0, & \text{otherwise.}
\end{cases}
\]

The gender-balance deviation is measured by the (normalized) distance of the fraction of female students from the desired value $1/2$, aggregated across all regions:
\[
f^{2}(\theta)
=
2 \cdot
\left|
\frac{\text{SUM(\#Female) in } Q_\theta(D)}
     {\text{SUM(Count(*)) in } Q_\theta(D)}
- \frac{1}{2}
\right|.
\]

Using $f^{1}$ and $f^{2}$, we define the overall constraint deviation function $\const$ as their average:
\[
\const(\theta) = \tfrac{1}{2}\bigl(f^{1}(\theta) + f^{2}(\theta)\bigr).
\]

Consider again the baseline assignment $\theta_0$ corresponding to the original query $Q$ (Figure~\ref{fig:scholarship-query}) and its output $Q(D)$, shown in Table~\ref{tab:result}.
The query result contains $5$ distinct regions, and the ratio of female students among all selected students is $0.267$.
Therefore,
\[
\const(\theta_0)
=
\tfrac{1}{2}
\left(
  \tfrac{8 - 5}{8}
  + 2 \cdot |0.267 - 0.5|
\right)
=
\tfrac{1}{2} (0.375 + 0.466)
=
0.421.
\]
If $\varepsilon$ is set to $0.1$, then $\theta_0$ is not $\varepsilon$-satisfying.  \qed 
\end{example}

In our solution, as described in Section~\ref{sec:OPRO}, the constraint deviation function is provided to \system{} as a Python code (see examples in our Github repository~\cite{our_github_repository}). \new{To facilitate the specification of such functions, in~\cite{HacohenMS25}, we presented a user-friendly Wizard that translates natural-language specifications into formal query refinement problems.}

\paragraph*{Refinement distance} 
The query refinement problem aims to make \emph{minimal} modifications to the query to satisfy the constraints.
Minimality may be defined in various ways, such as based on the set of output tuples of the queries~\cite{suraj2022range,campbell2024topk}, known as outcome-based distance~\cite{campbell2024topk}, or by the semantic distance in query predicates~\cite{moskovitch2023diversity,campbell2024topk}, referred to as predicate-based distance~\cite{campbell2024topk}. It may also involve a combination of both approaches. 
To keep our framework general, we use a refinement distance function $\dist_{\predVal_0,D}:\allAss\to\mathbb{R}_{\geq 0}$  that, given a refinement $\predVal\in\allAss$, returns a value in $\mathbb{R}_{\geq 0}$ that reflects the distance of $\predVal$ from $\predVal_0$ (possibly with respect to $D$, as in~\cite{suraj2022range}). For brevity, when unambiguous, we write $\dist(\theta)$ in place of $\dist_{\theta_0,D}(\theta)$.

\begin{example}\label{ex:distance}
 \new{An example refinement distance function supported in \system{} is the predicate-based distance defined in~\cite{campbell2024topk}}, %
 where the distance is measured with respect to each refinable predicate in the query. For a numerical predicate, the distance is defined as $\frac{|p.C - p.C'|}{p.C}$, where $p.C'$ is the constant value of the predicate $p$ in the refined query. The difference is normalized to show deviation from the original value of $p$. For categorical attributes, the distance is measured using Jaccard distance over the sets $p.C$ and $p.C'$. The overall distance is then the sum of the per-predicate distance over all the refinable attributes.
Consider again the refinement assignment $\predVal'=\{3.6,[\texttt{Tactics}],80\}$ and the baseline assignment $\predVal_0=\{3.5,[\texttt{Tactics}, \texttt{Archery}],100\}$, described  in Example~\ref{ex:refinement}. Using a predicate-based $\dist(\theta')$ we obtain
$$\dist(\theta') = \frac{|3.6-3.5|}{3.5} + \big(1-\frac{1}{2}\big) +  \frac{|100-80|}{100} = 0.029 + 0.5 + 0.2 =0.729$$
\end{example}

We are now ready to formally define the \prob\ problem.

\begin{problem}[\prob]\label{prob}

Given a database $D$, a query $Q$ with baseline assignment $\theta_0$,
a constraint-deviation function $\psi$ with tolerance threshold $\varepsilon$,
and a query distance measure $\dist_{\theta_0,D}$,
we define an optimal refinement, denoted $\theta^*$, as:
\[
\theta^*
\;=\;
\argmin_{\theta \in \allAss \,:\, \psi(\theta) \le \varepsilon}
\dist_{\predVal_0,D}(\theta).
\]

\end{problem}
As discussed above, our formulation subsumes a wide range of prior work
(e.g.,~\cite{campbell2024topk,suraj2022range,moskovitch2023diversity}),
each of which addresses a restricted variant of the problem
(see Section~\ref{sec:related}),
while also enabling many previously unsupported cases involving arbitrary
constraints, refinement distance measures, and query types.
We refer to a specific instantiation of the problem as a tuple
\[
\mathcal{QRP} = \langle Q, D, \const, \varepsilon, \Delta \rangle
\]

We next present an OPRO scheme for solving any instance of the \prob\ problem, followed by the \system{} architecture that instantiates this approach.
As shown in our experimental evaluation, our approach achieves performance close to
problem-specific solutions, while naturally extending to a much broader class of
instances covered by the \new{general} formulation.

%% file: methods.tex
\section{OPRO Scheme for Query Refinement}\label{sec:OPRO}

Our solution builds on the optimization-by-prompting (OPRO) paradigm~\cite{yang2024opro,chen-etal-2024-prompt},
a recent and promising approach for general black-box optimization.
In OPRO, an LLM-based agent is given a natural-language specification of the objective
and search domain, and iteratively proposes candidate solutions.
Each selection is conditioned on the full interaction history, enabling the agent to
adapt its decisions based on previously explored candidates and their evaluated outcomes.

We next describe a naïve, straightforward instantiation of the OPRO paradigm,
followed by our dedicated, two-step OPRO scheme for the \prob\ problem.

\paragraph{Naïve OPRO for Query Refinement}
Given an instance of the query refinement problem $\probfull$,
the naïve approach repeatedly prompts the LLM to modify the query $Q$
so as to produce an optimal refinement assignment $\theta^*$.
For a fixed horizon of $T$ iterations, at each iteration
$1 \le t \le T$, the LLM is asked to generate a new candidate refinement assignment $\theta_t$,
while being provided with the complete interaction history of prior attempts,
including both distance and constraint-deviation scores, namely,
\[
\left\{
\langle \theta_{t'}, \Delta(\theta_{t'}), \psi(\theta_{t'}) \rangle
\right\}_{t' = 1}^{t' < t}.
\]

However, as demonstrated in our experimental evaluation
(Section~\ref{sec:eval}),
this naïve OPRO approach is largely ineffective.
The underlying reasons are twofold:
(i) the combinatorial size of the refinement search space, and
(ii) the growing history of previous attempts, which, when presented in raw form,
provides limited informative signal to guide subsequent refinements.

\vspace{2mm}
We therefore propose a two-step OPRO scheme.
In the first step, the agent selects a promising refinement \emph{subspace}
that contains multiple candidate assignments.
In the second step, the agent samples individual assignments from the selected subspace
with the goal of identifying $\varepsilon$-satisfying refinements that minimize
the query distance from the original query $Q$.
We next formally define refinement subspaces and then present our OPRO scheme.%

\setlength{\textfloatsep}{1pt}
\begin{table}[t]
\centering
\small
\setlength{\tabcolsep}{1pt}
\begin{tabular}{ll}
\toprule
\textbf{Notation} & \textbf{Description} \\
\midrule
$D; A; D(A)$ & Database; Attribute; Column (att. $A$ projected on $D$). \\
$Q; Q(D) $ & Query; Query results over $D$. \\
$\pred$ & The set of refinable predicates in $Q$. \\
$p=\langle p.A, p.op, p.C\rangle$ & Attribute, operator and constant of predicate $p$. \\
$\dom{A}$ & Set of all distinct values in column $D(A)$. \\
$\theta;~\theta_0$ & A refinement assignment for $Q$; Baseline assignment. \\
$\theta^*$ & A minimal-distance $\varepsilon$-satisfying refinement assignment. \\
$\mathcal{S(Q)}$ & The space of all individual assignments for query $Q$. \\
$\Theta$; $\allSS$ & Refinement subspace; Domain of all subspaces for $Q$. \\
$\psi, \varepsilon$ & Output constraints deviation function, tolerance thresh. \\
$\dist_{\theta_0}$ (or simply $\dist$) & Refinement distance function. \\
$T$; $K$ & Horizon (\# of iterations); \# of samples per iteration. \\
$\Hist$; $\Hist_t$ & Refinement history; History at iteration $t$. \\
$\hist$ & Local refinements history within subspace \(\Theta_t\). \\
\bottomrule
\end{tabular}
\vspace{0.4em}
\caption{Notations Summary }
\label{tab:notation_summary}
\end{table}

\paragraph*{Refinement Subspace}
A refinement \textit{subspace}, denoted $\Theta$, defines a \textit{range} of assignments, by restricting the admissible values for each predicate in $\pred$.
For predicates with numerical attributes, this restriction takes the form of an interval over the value domain, while for categorical attributes it is defined using set inclusion, following the partial-order formulation of~\cite{alex2025partial}.

\begin{definition}[Refinement Subspace]
Given a query $Q$ and database $D$, a refinement subspace is a collection
$\Theta \;=\; \{\Theta.p \mid p \in \texttt{Preds}(Q)\},
$
where each component $\Theta.p$ constrains the admissible values for $p.C$.
For numerical attributes, $\Theta.p$ is an interval contained in
$\bigl[\min(\dom{p.A}), \max(\dom{p.A})\bigr]$.
For categorical attributes, $\Theta.p$ is a set-inclusion range
$[C_{\min}, C_{\max}]$ with
$C_{\min} \subseteq C_{\max} \subseteq \dom{p.A}$.
\end{definition}

An assignment $\theta$ is \emph{compatible} with a subspace $\Theta$, denoted by $\theta \in \Theta$, 
if it assigns to each predicate a value that lies within the range permitted by that subspace. Namely, 
$\theta \in \Theta \iff \theta.p \in \Theta.p \quad \forall p \in \pred $.

\begin{example}\label{ex:3.3}
    Recall again from our running example, $\pred$ contains three predicates over the \texttt{GPA}, \texttt{major}, and \texttt{COUNT(*)} attributes of query $Q$ (see Figure~\ref{fig:scholarship-query}). In Example~\ref{ex:refinement} we examine two refinement assignments: $\predVal_0=\{3.5,(\texttt{Tactics}, \texttt{Archery}),100\}$, and $\predVal'=\{3.6,(\texttt{Tactics}),80\}$.

    Now, let us consider the following refinement \textit{subspace}: $$\Theta = \{[3.4,3.7],[(\texttt{Tactics}),(\texttt{Tactics},\texttt{Melee},\texttt{\texttt{Stealth}})],[70, 120]\}$$

    See that $\theta' \in \Theta$, as $3.5 \in [3.4,3.7]$, $(\texttt{Tactics}) \in$ $[(\texttt{Tactics})$, $(\texttt{Tactics},\texttt{Melee},\texttt{\texttt{Stealth}})]$, and $80 \in [70, 120]$; whereas $\theta_0 \notin \Theta$ because $(\texttt{Tactics}, \texttt{Archery}) \nsubseteq (\texttt{Tactics},\texttt{Melee},\texttt{\texttt{Stealth}})$. \qed      
    
\end{example}

We denote the set of all possible refinements \textit{subspaces} for query $Q$ by $\allSS$.

\begin{algorithm}[t]
\small
\caption{Two-Step OPRO Scheme for Query Refinement}
\label{alg:global-local-search}
\SetAlgoLined
\DontPrintSemicolon

\SetCommentSty{textnormal}
\SetKwComment{tcp}{// }{}

\KwIn{Query refinement instance $\probfull$; Refinement subspaces domain $\allSS$; horizon $T$; samples per iteration $K$}
\KwOut{$\varepsilon$-satisfying assignment $\theta$ with min. refinement distance}

$\Hist \gets \emptyset$ \tcp*{Initialize refinement history} \label{alg:init-hist}
$SAT \gets \emptyset$ \tcp*{Initialize constraint-satisfying set} \label{alg:init-sat}

\For{$t = 1$ \textbf{to} $T$}{ \label{alg:for-t}
    $\Theta_t \gets \mathcal{M}^s(\allAss,\Hist)$ \tcp*{Suggest subspace} \label{alg:suggest-subspace}
    \If{$\Theta_t \notin \Hist$}{%
         $\Hist[\Theta_t] \gets \emptyset$ \tcp*{Initialize subspace-specific history} \label{alg:init-subspace}
    }%

    \For{$k = 1$ \textbf{to} $K$}{ \label{alg:for-k}
        $\theta_k \gets \mathcal{M}^a(\Theta_t,\Hist[\Theta_t])$ 
           \tcp*{Suggest refinement} \label{alg:suggest-theta}

        \For(\tcp*[f]{Update history}){$\Theta \in \Hist$}{ \label{alg:for-history}
            \If{$\theta_k \in \Theta$}{%
                $\Hist[\Theta] \gets \Hist[\Theta] 
                    \cup \{\theta_k\!:\!(\Delta(\theta_k)),\psi(\theta_k)\}$ 
                    \label{alg:update-hist}
            }%
        }%

        \If(\tcp*[f]{If constraints are met}){$\psi(\theta_k) \le \varepsilon$}{\label{alg:if-sat}
            $SAT \gets SAT \cup \{\theta_k\}$ \label{alg:update-sat}
        }%
    }%
}%

\Return $\argmin_{\theta \in SAT} \Delta(\theta)$ 
\label{alg:return}

\end{algorithm}

Building on the definition of refinement subspaces, we devise an OPRO scheme for the query refinement problem.
Our two-step OPRO scheme is depicted in Algorithm~\ref{alg:global-local-search}. We next describe the inputs and algorithm steps: 

\begin{figure*}[t]
    \centering
    \includegraphics[width=0.9\linewidth]{images/FinalFigure2.pdf}
    \caption{\system{} Workflow \& Architecture}
    \label{fig:system-flow}
\end{figure*}

\noindent\textbf{Inputs.}
The system takes as input:
(1) an instance of the \prob\ problem $\probfull$;
(2) the set $\allSS$, representing the domain of all refinement subspaces
(we describe below how a concise textual representation of $\allSS$ is constructed
without explicitly materializing it);
and (3) the horizon $T$, specifying the maximum number of iterations, and the number of assignments sampled from each subspace $K$, as detailed
below.

\vspace{1mm}
\noindent\textbf{Initialization \& History Management.} As mentioned above, in our iterative OPRO scheme, we keep the history of previously chosen assignments. We use a key-value structure, denoted $\Hist$ (Line~\ref{alg:init-hist}), in which the keys are unique subspaces $\Theta_0,\Theta_1,\dots$; and the values are lists of all individual assignments examined thus far, that are compatible with each $\Theta \in \Hist$, along with their distance and satisfaction values. Namely, $\Hist[\Theta]=\{\langle \theta, \Delta(\theta),\psi(\theta) \rangle \mid \theta \in \Theta  \}$. We also initialize the set $SAT$ (Line~\ref{alg:init-sat}), which will contain all $\varepsilon$-satisfying assignments, throughout the optimization process.

\vspace{1mm}
\noindent\textbf{Subspace-Selection Step (LLM Call \#1).}
At each iteration $t$, we abstractly model the first step as an LLM call
that selects the next \emph{refinement subspace} $\Theta_t \in \allSS$.
This call, denoted $\mathcal{M}^s$, takes as input the optimization history $\Hist$
(Line~\ref{alg:suggest-subspace}) and outputs a candidate subspace $\Theta_t$.
In Section~\ref{sec:history}, we detail how this abstract call is instantiated in
\system{} by the \textit{SubspaceLM} module, which provides the LLM with a concise,
textual representation for $\allSS$ and $\Hist$.

\vspace{1mm}
\noindent\textbf{Assignment-Sampling Step (LLM Call \#2).}
Given the selected subspace $\Theta_t$, we issue $K$ consecutive LLM calls that sample
\emph{individual} refinement assignments
$\{\theta_k \in \Theta_t\}_{k=1}^{K}$.
We denote this abstract LLM call by $\mathcal{M}^a$
(Line~\ref{alg:suggest-theta}), where each call is conditioned on the
subspace-specific history $\Hist[\Theta_t]$, which contains assignments previously
sampled from $\Theta_t$.
In \system{}, this call is instantiated by the \textit{AssignmentLM} module
(Section~\ref{sec:history}).
After each call, the global history $\Hist$ is updated, such that each previously
generated subspace $\Theta \in \Hist$ is appended with the currently sampled assignment
$\theta_k$ whenever $\theta_k \in \Theta$
(Lines~\ref{alg:for-history}--\ref{alg:update-hist}).
Any $\theta_k$ that is $\varepsilon$-satisfying is appended to $SAT$
(Lines~\ref{alg:if-sat}--\ref{alg:update-sat}).

\vspace{1mm}
\noindent\textbf{Termination \& Return value.}
Finally, after $T$ iterations, we return the set of $\varepsilon$-satisfying assignments
stored in $SAT$ that attain the minimum query distance from the baseline assignment $\theta_0$ (Line~\ref{alg:return}).

\vspace{2mm}
We next describe the architecture of \system{},
designed in accordance with Algorithm~\ref{alg:global-local-search}.
The architecture incorporates several key components, including concise yet effective
textual representations of $D$ and $\allSS$, mechanisms for summarizing subspace-history
information $\Hist[\Theta_t]$, skyline-based objective definitions for realistic optimization,
and an early-stopping mechanism.

%% file: system.tex
\section{\system{} Architecture}
\label{sec:history}

We designed \system{}, an efficient system that implements the two-step OPRO scheme for query refinement, as described above. 
\system{} overcomes numerous challenges, such as producing concise textual representations for the database $D$, subspace domain $\allSS$, and history $\Hist$, as well as introducing realistic objectives for the LLM calls, and early stopping mechanisms. 

\paragraph*{\new{Formulating the refinement problem}}
\new{\system{} receives the constraint-deviation function $\const$ and the refinement
distance $\dist$ as executable Python code (Section~\ref{sec:problem}). Importantly, users need not
write this code themselves: the Refinement Problem Wizard of our prototype~\cite{HacohenMS25} lets them load a database, enter the query $Q$, and describe the constraints in natural
language. An LLM then generates $\const$ via in-context learning, given the schema of $D$,
the query $Q$, and implementation guidelines and examples (e.g., an output range of $[0,1]$, normalized
across constraints); the generated code can be further edited by the user. For $\dist$,
users select a predicate-based or an outcome-based distance, or describe a custom one in
natural language. The Wizard code is available in our repository~\cite{our_github_repository}.}

\paragraph*{Workflow Overview}
Figure~\ref{fig:system-flow} illustrates the overall architecture and execution
flow of \system{}.
Given a query refinement problem instance $\probfull$, \system{} operates in two
main phases.
In the \emph{pre-processing} phase, \system{} constructs concise yet informative
textual representations of the database $D$, tailored to the input query $Q$,
as well as a representation of the refinement subspace domain $\allSS$, which
serves as the search space for subspace generation.
In the \emph{optimization loop} phase, \system{} iteratively issues LLM calls via two
specialized modules, in accordance with Algorithm~\ref{alg:global-local-search}:
The \textit{SubspaceLM} module generates a candidate refinement subspace
$\Theta_t$, while the \textit{AssignmentLM} module samples individual refinement
assignments from $\Theta_t$.
To guide this process, \system{} maintains a compact summary of the interaction
history $\Hist$, which provides the \textit{SubspaceLM} with high-level statistics
over previously explored subspaces $\{\Theta_{t'} \mid t' < t\}$.
In addition, both modules are guided by skyline-based objectives derived from the
set of previously observed assignments.
Intuitively, the skyline maintains a set of high-quality assignments that represent
the best trade-offs observed so far between query distance and constraint
satisfaction.

Finally, \system{} employs a skyline-based early-stopping mechanism that directly
builds on this trade-off view:
when the skyline remains unchanged for several consecutive iterations,
indicating that no new assignment improves the current trade-offs between distance
and satisfaction,
the system halts and returns the best assignment found so far.
We next elaborate on each part of the system. We start with the preprocessing phase and then describe in detail the optimization loop modules. \new{For output examples, see Figure~\ref{fig:ex_texts}; the full prompt scheme is available in our repository~\cite{our_github_repository}.}

\begin{figure}[t]
    \centering
    \includegraphics[width=0.95\linewidth]{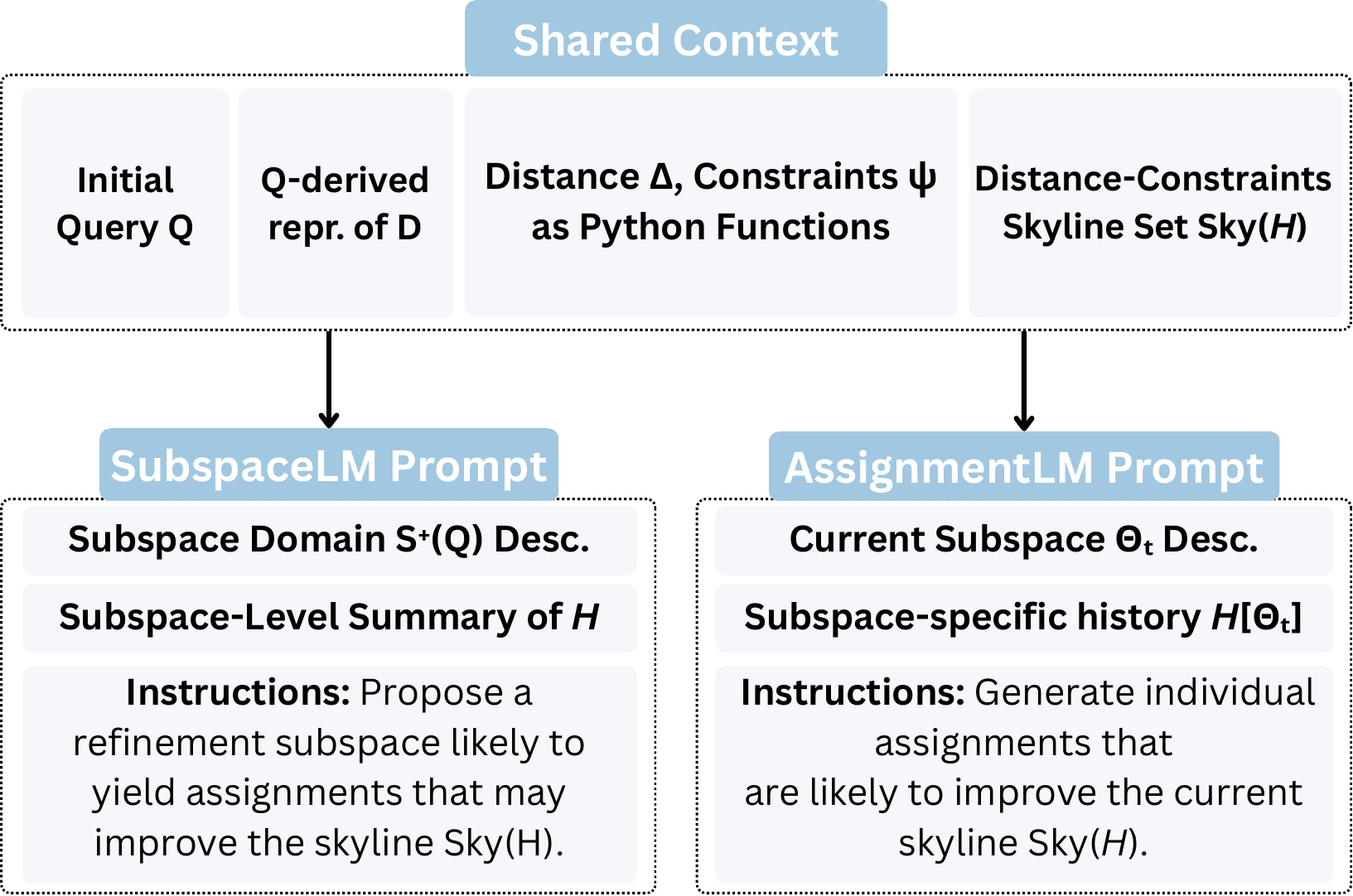}
    \caption{Prompts Formulation}
    \label{fig:promts}
\end{figure}

\subsection{Pre-processing phase}
As mentioned above, in this phase, given the problem instance $\probfull$,
we first generate a concise description of the database $D$ and refinement subspace domain $\allSS$:

\vspace{1mm}
\noindent\textbf{1. $Q$-driven Database description. }
Producing an effective and concise database schema description is a well-recognized challenge in LLM-based data management tasks, including text-to-SQL~\cite{trummer2024generating,wherearewetodatnl2sql}, LLM-based schema matching~\cite{liu2024magneto}, and LLM-driven BI and analytics~\cite{siriusbi}. Overly long or exhaustive descriptions often degrade LLM performance (see, e.g., the context rot problem~\cite{levy2024same}), motivating the need for concise and focused task descriptions.

To that end, rather than providing an exhaustive description of the entire database $D$, we derive a concise, query-conditioned representation tailored to the given query refinement instance. %
 This representation includes only the tables and attributes that appear in the query $Q$. Specifically, for each table referenced in $Q$, we describe its primary and foreign keys, as well as the names and data types of the attributes used in $Q$. In addition, we provide summary statistics for each such attribute: for numeric columns, the mean, standard deviation, and quartile values; and for categorical columns, the prevalence of each category. \new{Figure~\ref{fig:ex_schema} shows examples of the textual representations used by \system{}.}

\vspace{1mm}
\noindent\textbf{2. Description of the Subspace Domain $\allSS$.}
The goal in this component is to provide the \textit{SubspaceLM} module with a
compact yet expressive description of the refinement subspace domain $\allSS$,
from which refinement subspaces $\Theta_t$ can be generated.
Explicitly materializing $\allSS$ is infeasible, as it contains all possible
combinations of per-predicate value ranges and would far exceed the context
limitations of current LLMs.
Instead, we construct a concise characterization of $\allSS$ by specifying, for
each refinable predicate $p \in \pred$, the admissible range of values for
$p.C$.
For \emph{base attributes} appearing directly in the database $D$, this process
is straightforward: for numerical attributes, we compute the minimum and maximum
values observed in the data, and for categorical attributes, we provide the set
of distinct values.
However, handling \emph{derived} or \emph{computed} attributes—such as those
appearing in a \texttt{HAVING} clause—poses a greater challenge.

To address this case, we employ an LLM-assisted approach in which the model is
prompted to generate auxiliary SQL queries that compute the minimum and maximum
values of each derived field.
For example, in our running scenario (Example~\ref{ex:refinement}), where the
\texttt{HAVING} clause includes the refinable predicate \texttt{COUNT(*) > 100},
the LLM produces the following query:%
{\small
\begin{alltt} 
    SELECT MIN(size), MAX(size)
    FROM (SELECT COUNT(*) AS size
          FROM MiddleEarth_Applicants
          GROUP BY region)
    \end{alltt} } 
\vspace{-13pt}
\new{Figure~\ref{fig:ex_ss} shows an example representation of $\allSS$; the exact prompt is available in our repository~\cite{our_github_repository}.}

\input{fig_ex_texts}

\subsection{Optimization loop, prompt formulations}
In accordance with Algorithm~\ref{alg:global-local-search}, the main optimization
loop runs for a horizon of $T$ iterations.
At each iteration $t$, \system{} invokes the \textit{SubspaceLM} module to generate
a refinement subspace $\Theta_t$, and then issues $K$ calls to the
\textit{AssignmentLM} module to sample individual assignments from $\Theta_t$,
while maintaining the interaction history $\Hist$.
We next describe the \textit{SubspaceLM} and \textit{AssignmentLM} modules, with a
focus on the formulation of their respective prompts (see Figure~\ref{fig:promts} for illustration).
In particular, we introduce two techniques that substantially enhance optimization
performance, as demonstrated in our experimental evaluation
(Section~\ref{sec:eval}):
(i) a skyline-based objective, used in place of naïve instructions to generate the
“best” subspace or assignment; and
(ii) a concise, subspace-level summary of the history $\Hist$, employed by the \textit{SubspaceLM} module. In~\cite{our_github_repository}, we provide the exact prompt templates, as well as numerous example workflows of \system{}.

\paragraph*{Shared information}

The following items are included in both the SubspaceLM and AssignmentLM prompts:

\vspace{1mm}
\noindent\textbf{1. Description of the instance.}
We begin with a brief, task-agnostic overview of the query refinement problem, then describe the input instance $\probtuple$ as follows: The $Q$-derived representation of the database $D$, as
described above, together with the query $Q$ itself.
In addition, the distance function $\Delta$ and the constraint-deviation function
$\psi$ are provided as executable Python code.
Code-based representations are widely regarded as particularly effective for
LLM-based reasoning and instruction following~\cite{coderepr}.

\vspace{1mm}
\noindent\textbf{2. Skyline-based Objectives.} 
Prior work has shown that clear, context-rich instructions can substantially
improve zero-shot LLM performance across domains
(e.g.,~\cite{Sivarajkumar2024,sumanathilaka-etal-2024-llms}).
However, in our setting, naively prompting an LLM to generate an ``optimal''
refinement subspace $\Theta_t$ or assignment $\theta_k$ proves ineffective
(see Section~\ref{sec:eval}).

Existing OPRO solutions~\cite{yang2024opro} typically instruct the LLM, at each iteration, to
surpass the ``best'' solution obtained so far with respect to a single objective.
In contrast, our setting is inherently multi-objective, as 
assignments must be $\varepsilon$-satisfying w.r.t. $\psi$ while
simultaneously minimizing the distance $\dist$.
To this end, we devise a skyline-based objective.

Intuitively, a \textit{skyline}~\cite{skyline2001} captures the set of ``best'' assignments by retaining
only those for which no other assignment achieves both a smaller query distance
$\Delta(\theta)$ and a strictly better constraint deviation $\psi(\theta)$.
In our setting, all $\varepsilon$-satisfying assignments are treated as equivalent with
respect to constraint satisfaction, which we capture via the constraints-\textit{threshold} deviation function
$$\psi_\varepsilon(\theta) \triangleq \min\{\psi(\theta),~\varepsilon\}$$
An assignment $\theta \in \Hist$ is said to belong to the skyline set, denoted
$\theta \in \skyline$, if and only if it is not dominated by any other assignment
$\theta' \in \Hist$, that is,
$$\begin{aligned}
    \theta \in \skyline
\Leftrightarrow
\nexists\, \theta' \in \Hist :&\Big(\big(\Delta(\theta') < \Delta(\theta)
 \land
 \psi_\varepsilon(\theta') \le \psi_\varepsilon(\theta)\big)\\&~~\lor
\big(\psi_\varepsilon(\theta') < \psi_\varepsilon(\theta)
 \land
 \Delta(\theta') \le \Delta(\theta)\big)\Big)
\end{aligned}$$
\new{Figure~\ref{fig:ex_sky} present an example of the skyline set $\skyline$ representation in \system{}}

\paragraph{The SubspaceLM prompt (LLM Call \#1).}
Given the problem instance description and the current skyline, as introduced above,
the LLM is instructed to propose a refinement subspace likely to yield assignments
that may improve the skyline.
Rather than providing the full interaction history $\Hist$, we supply a concise yet
informative \emph{subspace-level} summary. %
Specifically, for each previously explored subspace $\Theta_{t'}$ with $t' < t$,
we present aggregate statistics over the assignments sampled from that subspace,
as recorded in $\Hist[\Theta_{t'}]$, including the total number of sampled assignments, as well as the median and standard
deviation of their distance and constraint-deviation scores.
\new{An example of the representation is given in Figure~\ref{fig:ex_historysum}.}

\paragraph{AssignmentLM Prompt (LLM Call \#2).}
Given the refinement subspace $\Theta_t$ selected by the \textit{SubspaceLM} module,
the \textit{AssignmentLM} is invoked $K$ times to generate individual assignments
$\theta_k$ that are likely to improve the current skyline $\sky(\Hist_t)$ where $\Hist_t$ is the history $\Hist$ at iteration $t$.
As in the subspace-generation step, providing the full history $\Hist$ is
unnecessary; instead, the prompt includes only the subspace-specific history
$\Hist[\Theta_t]$, containing assignments previously sampled from the current
subspace.

\paragraph{Early stopping criteria}
Finally, we introduce an early-stopping mechanism~\cite{prechelt2002early} based on the skyline
set $\skyline$ defined above, which allows \system{} to terminate the optimization
process before reaching the full horizon $T$, thereby reducing LLM
token consumption.
At iteration $t$, \system{} halts if:
$$
\sky(\Hist_t) = \sky(\Hist_{t-u})
$$
That is, if the skyline set remains unchanged for $u$ consecutive iterations.
Intuitively, this condition indicates that the system has failed to identify a new
assignment that dominates previously observed ones with respect to both refinement
distance $\dist$ and constraint satisfaction $\psi$.
In such cases, \system{} terminates and returns the $\varepsilon$-satisfying
assignment in $\mathit{SAT}$ with minimum distance from the baseline assignment (see Line~\ref{alg:return} in Algorithm~\ref{alg:global-local-search}).
In our implementation, we set $u = 2$.

%% file: fig_ex_texts.tex
\begin{figure}[]
\centering
\small

\begin{subfigure}{\linewidth}
\centering
\begin{verbatim}
{ "table":"MiddleEarth_Applicants","primary_key":"id",
  "attributes":{
    "GPA":{"type":"float","domain":[2.0,4.0],
        "stats":{"mean":3.12,"std":0.41,
        "q1":2.85,"median":3.10,"q3":3.40}},
    ...
    "gender":{"type":"category","domain":["F","M"],
      "stats":{"distribution":[{"value":"F","p":0.46},
                               {"value":"M","p":0.54}]}}}
\end{verbatim}
\caption{$Q$-driven Database Schema Description.}
\label{fig:ex_schema}
\end{subfigure}

\vspace{0.5em}

\begin{subfigure}{\linewidth}
\centering
\small
\begin{verbatim}
[{"attribute": "gpa", 
    "value_range": {"min_val":2, "max_val":4}},
 { "attribute": "major", 
   "value_range": {"min_val":[], "max_val":      
        [Archery, Melee, Spearmanship, Tactics, 
        Camouflage, Scouting, Swordship, Horsemanship]}},
 { "attribute": "COUNT(*)", 
   "value_range": {"min_val":34, "max_val":452}}]

\end{verbatim}
\caption{Subspaces Domain Definition  \(\allSS\)}
\label{fig:ex_ss}
\end{subfigure}

\vspace{0.5em}

\begin{subfigure}{\linewidth}
\centering
\small
\begin{verbatim}

[{ "assignment_id:1, "vals":{3.7,"(Tactics,Archery)",90},
      "dev":0.49, "dist":0.20 },
 { "assignment_id:7, "vals":{3.6,"(Tactics,Archery)",80},
      "dev":0.24, "dist":0.23 },
  ...
 { "assignment_id:13, "vals":{3.9,"(Tactics,Melee)",100},
      "dev":0.20, "dist":0.57 }]
\end{verbatim}
\caption{Skyline Set $\skyline$.}
\label{fig:ex_sky}
\end{subfigure}

\vspace{0.5em}

\begin{subfigure}{\linewidth}
\centering
\begin{verbatim}
[{ "subspace_id":1,
    "subspace":{"gpa":[3.4,3.8],
                "major":["Tactics","Archery"],
                "Count(*)":[60,90]},
    "history_size":5,
    "stats":{"deviation":{"med":.49,"std":.25},
            "distance":{"med":.20,"std":.03}} }, 
 { "subspace_id":2, ...]
\end{verbatim}
\caption{Subspace-level History Summary.}
\label{fig:ex_historysum}
\end{subfigure}

\caption{Example textual representations of \system{} objects}
\label{fig:ex_texts}
\end{figure}

%% file: evaluation.tex
\section{Experimental Setup and Evaluation }
\label{sec:eval}

\begin{figure}[t]
\small
\setlength{\tabcolsep}{3pt}
\centering
\begin{tabular}{p{1.8cm} p{6.2cm}}
\toprule
\textbf{Class} & \textbf{Databases Used} \\
\midrule
Top-$k$~\cite{campbell2024topk} & Astronauts~\cite{astrodb}, Law Students~\cite{lawstudentsdb}, MEPS~\cite{mepsdb}, TPC-H~\cite{tpch} \\
Range~\cite{suraj2022range} & Texas Tribune~\cite{texastribune}, COMPAS~\cite{compasdb}, Housing Prices~\cite{housingdb}, Fraud Detection~\cite{frauddb} \\
Diversity~\cite{moskovitch2023diversity} & Healthcare~\cite{healthcaredb}, ACSIncome~\cite{acisincome}, COMPAS, Students~\cite{studentsperfdb} \\
Complex (new) & Law Students, Texas Tribune, TPC-H \\
\bottomrule
\end{tabular}
\addtocounter{figure}{-1}\stepcounter{table}%
\renewcommand{\thefigure}{\arabic{table}}\renewcommand{\theHfigure}{table.\arabic{table}}%
\captionsetup{name=Table}%
\caption{Datasets used per instance class}
\label{tab:task-summary}
\end{figure}

We performed an experimental analysis of our proposed solution using a dedicated benchmark composed of real-life datasets and realistic scenarios, using the queries and constraints from prior work. We first examine \sysName's ability to produce refinements and the quality of those refinements, comparing results across different baselines. We measured the quality of the resulting refinement with respect to refinements produced by a task-specific dedicated solution for the instance class (i.e.,~\cite{campbell2024topk} for top-$k$ instances, \cite{moskovitch2023diversity} for diversity, and \cite{suraj2022range} for range). We conduct an ablation study to evaluate the impact of our two-step OPRO scheme and the history representation component used in our system implementation on the resulting refinements. We then present an assessment of the effect of the horizon ($T$) and the number of refinements per subspace ($K$) on the quality of the resulting refinement and the token costs of generating it. Finally, to demonstrate the usefulness of \sysName{}, we present three case studies of complex refining instances that are not supported by existing solutions.

\subsection{Experimental Setup}
\paragraph{Implementation} 
\sysName is implemented in Python 3 and uses DuckDB\cite{duckdb2025}. The structural prompt verification utilizes the Pydantic Python implementation~\cite{pydantic}. The SubspaceLM and AssignmentLM components are built into LangChain~\cite{langchain} and support various LLM provider APIs. Across all experiments, the decoding parameters were fixed to temperature $=0$ and top-$p=0$.
All experiments were run on a 24-core CPU server. The code and data are provided in~\cite{our_github_repository}. As default setting,  we set the horizon $T$ and the number of refinements per iteration $K$ to $5$, early stopping criterion $u=2$, and use GPT-5.6-Luna.

\paragraph{Benchmark.} 
To evaluate \sysName, we constructed a unified benchmark of \prob\ instances. Each instance comprises a dataset, a query, a constraint function along with a tolerance value, and a distance function. 
The benchmark encompasses $32$ instances utilizing $11$ real-world datasets of varying sizes and attribute dimensions, including adaptations of instances used in prior works~\cite{campbell2024topk,moskovitch2023diversity,suraj2022range}.
In particular, we consider $4$ instance classes: 
\textbf{(1) Top-$k$} queries as presented in~\cite{campbell2024topk} with constraint representation over the representation of different groups in the top-$k$ results, \textbf{(2) Range}~\cite{suraj2022range} queries with constraint over the ratio of data groups in the output, \textbf{(3) Diversity} in the output of SPJ queries as was studied in~\cite{moskovitch2023diversity}, and \textbf{(4) Complex} queries derived from each of the datasets using LLMs, and not supported by previous works.
The benchmark includes eight instances per class. For the instance classes (1)-(3), we used queries, constraints, and distance functions as defined in the corresponding prior work, with the addition of four new instances for the Range class and two additional instances to the Diversity class. Table~\ref{tab:task-summary} depicts the dataset used for each instance class, and their full description is available in~\cite{our_github_repository}.

\begin{figure*}[t]
\vspace{-3mm}
\centering

\includegraphics[width=0.92\linewidth]
{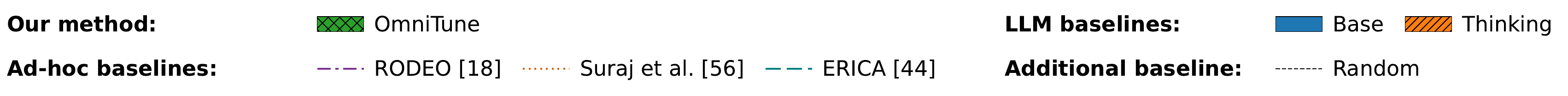}

\begin{subfigure}[b]{0.24\linewidth}
    \centering
    \includegraphics[width=\linewidth]
    {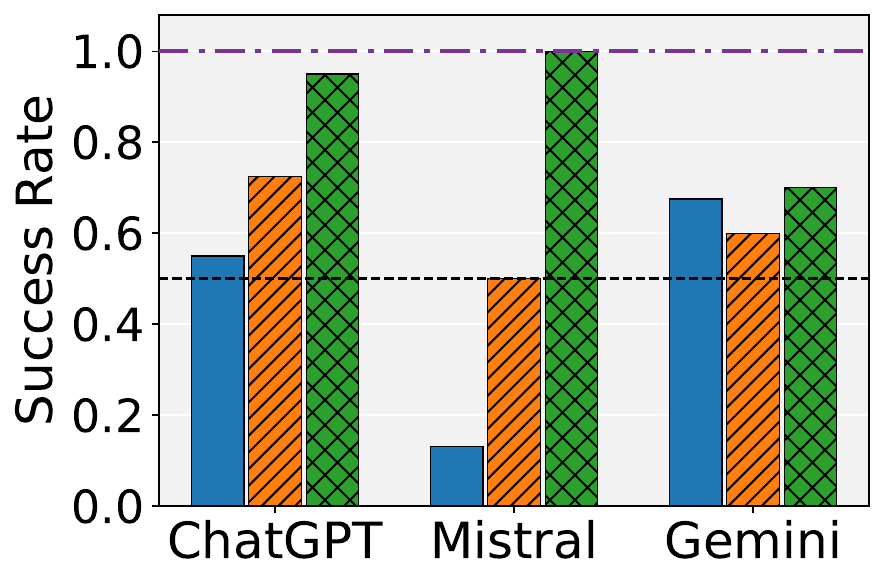}
    \caption{Top-$k$ Instances}
    \label{fig:baseline-comparison-success-top-k}
\end{subfigure}
\hfill
\begin{subfigure}[b]{0.24\linewidth}
    \centering
    \includegraphics[width=\linewidth]
    {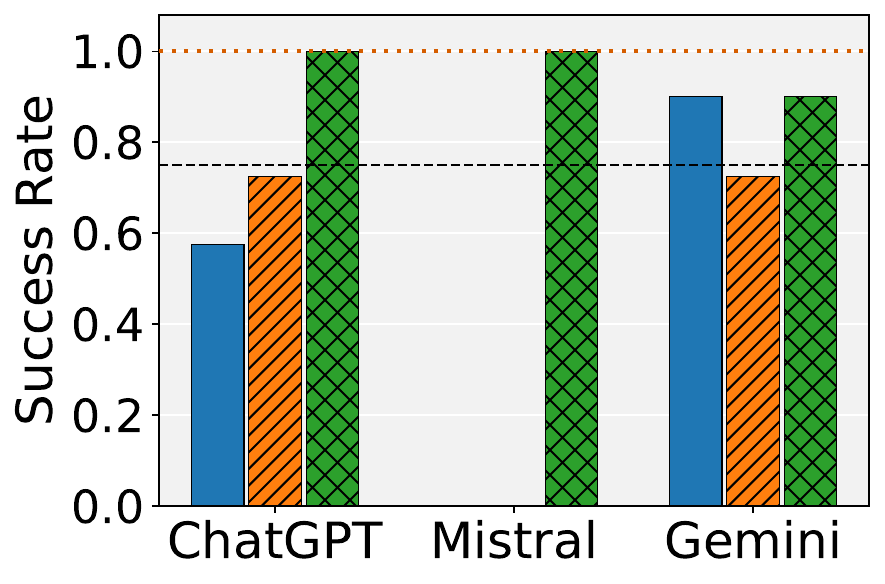}
    \caption{Range Instances}
    \label{fig:baseline-comparison-success-range}
\end{subfigure}
\hfill
\begin{subfigure}[b]{0.24\linewidth}
    \centering
    \includegraphics[width=\linewidth]
    {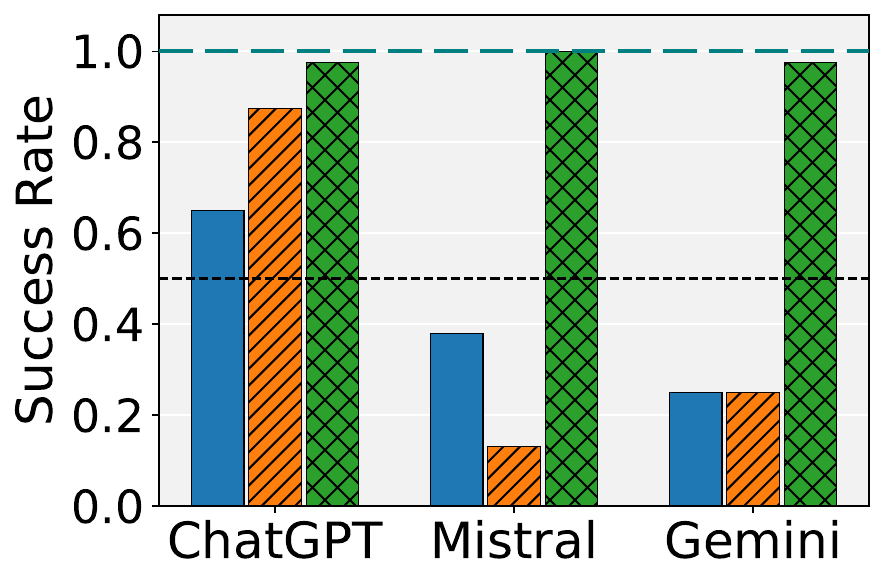}
    \caption{Diversity Instances}
    \label{fig:baseline-comparison-success-diversity}
\end{subfigure}
\hfill
\begin{subfigure}[b]{0.24\linewidth}
    \centering
    \includegraphics[width=\linewidth]
    {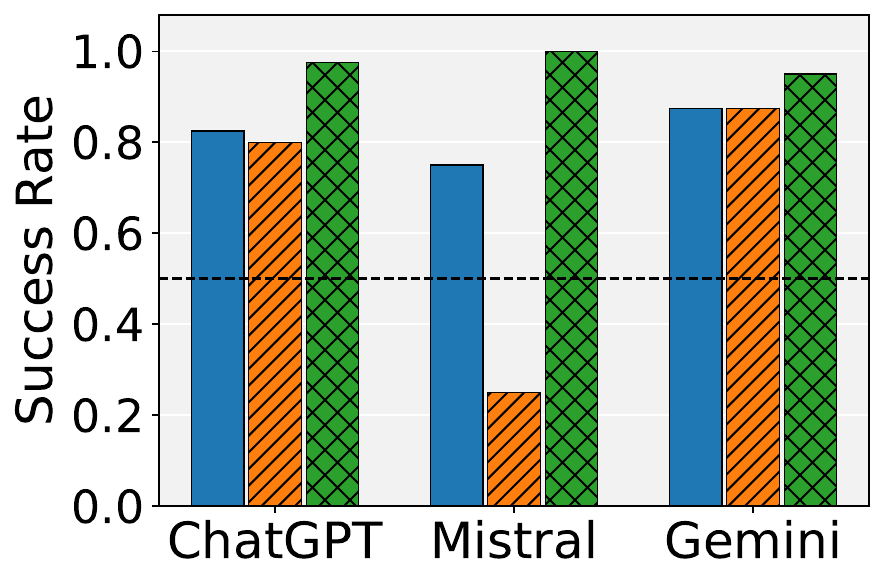}
    \caption{Complex Instances}
    \label{fig:baseline-comparison-success-complex}
\end{subfigure}

\vspace{2mm}

\begin{subfigure}[b]{0.24\linewidth}
    \centering
    \includegraphics[width=\linewidth]
    {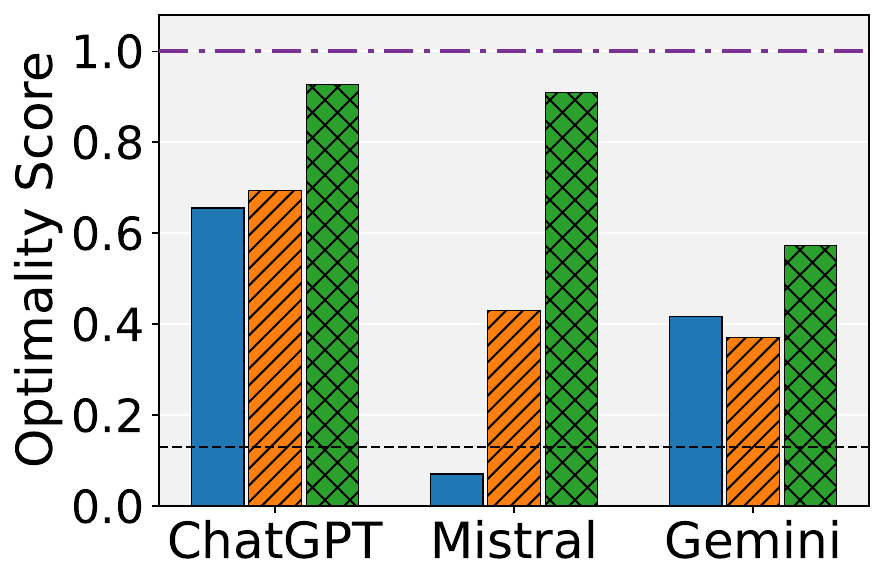}
    \caption{Top-$k$ Instances - Optimality}
    \label{fig:baseline-comparison-optimality-top-k}
\end{subfigure}
\hfill
\begin{subfigure}[b]{0.24\linewidth}
    \centering
    \includegraphics[width=\linewidth]
    {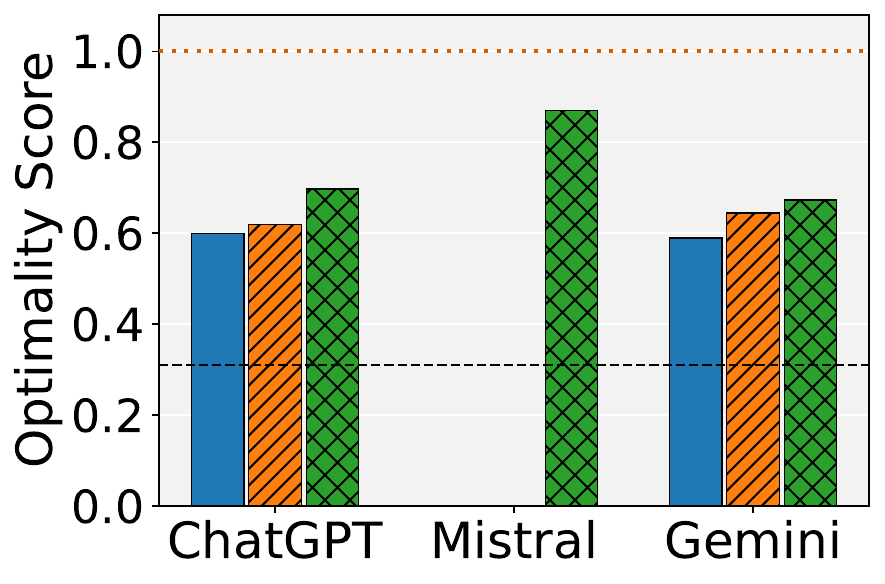}
    \caption{Range Instances}
    \label{fig:baseline-comparison-optimality-range}
\end{subfigure}
\hfill
\begin{subfigure}[b]{0.24\linewidth}
    \centering
    \includegraphics[width=\linewidth]
    {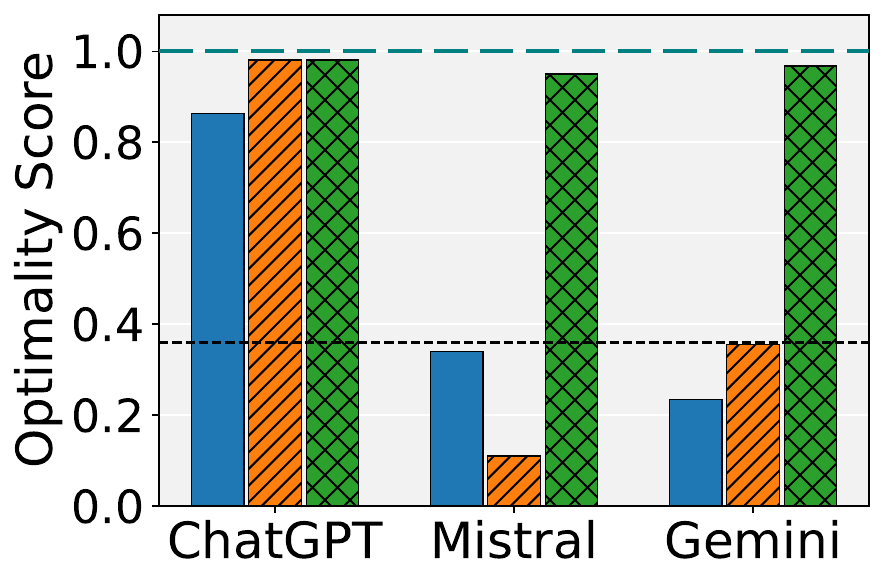}
    \caption{Diversity Instances}
    \label{fig:baseline-comparison-optimality-diversity}
\end{subfigure}
\hfill
\begin{subfigure}[b]{0.24\linewidth}
    \centering
    \includegraphics[width=\linewidth]
    {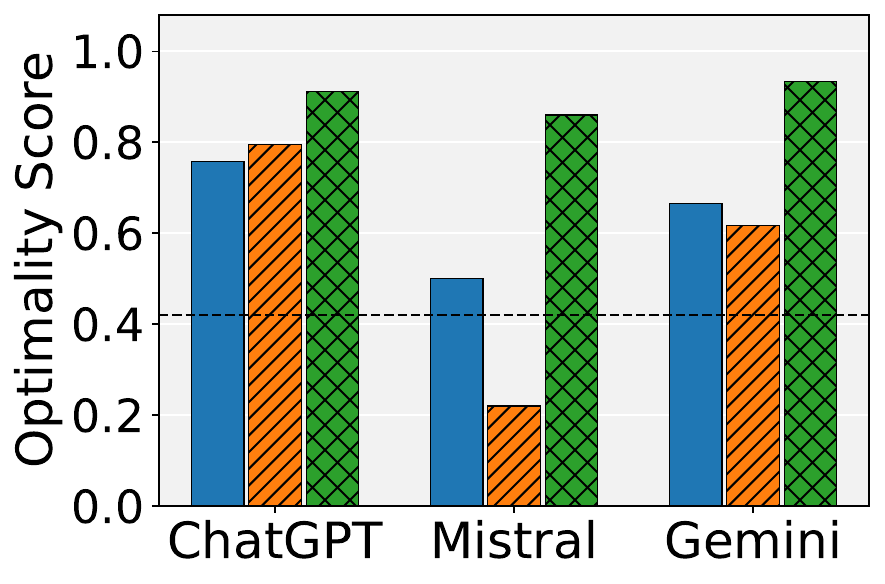}
    \caption{Complex Instances}
    \label{fig:baseline-comparison-optimality-complex}
\end{subfigure}

\caption{
\new{Comparison of \sysName{} and baselines across instance classes. Top: success rate; bottom: optimality. Base (blue) and Thinking (orange) denote the basic and reasoning LLM baselines; green denotes \sysName{} using the Base model. The dashed black line denotes random sampling, and colored lines denote task-specific methods~\cite{rodeo,suraj2022range,erica}, which define the optimality reference of $1.0$ for the top-k, range, and diversity classes (Complex instances are unsupported by prior work).}
}
\label{fig:baseline-comparison}

\end{figure*}

\paragraph{Evaluation Metrics.}
We evaluated the performance of the proposed solution using two criteria: \emph{success rate} and \emph{optimality}. The \emph{success rate} measures the number of tasks we were able to obtain a refinement that $\varepsilon$-satisfies the constraints function. For each successful execution, we measure the \emph{optimality} of the resulting refinement $\predVal$ with respect to the resulting refinement $\predVal_{opt}$ produced by a task-specific dedicated solution for the instance class (i.e.,~\cite{campbell2024topk} for top-$k$ instances, \cite{moskovitch2023diversity} for diversity and \cite{suraj2022range} for range). The optimality metric is defined as
$$\mathrm{Opt}(\predVal) = \frac{\dist_{\max} - \dist(\predVal)}{\dist_{\max} - \dist(\predVal_{opt})}$$
where $\Delta_{\max}$ is the max possible distance value for the given $\dist$ (e.g., $1$ for Jaccard-based distance function, or the  $|\pred|$ for predicate-based distances) and is used to normalize the result. Since there is no dedicated solution that can handle instances in the Complex class, we set $\dist(\predVal_{opt})$ to $0$ for these instances, which provides a lower bound on the possible distance value. Each experiment was repeated $5$ times, and we report the success rate and maximal $\mathrm{Opt}$ score across runs; average results show similar trends and are reported in~\new{\trcite{appendix:mean-opt}}.

 \paragraph*{Baselines}
 Our implementation of \sysName interacts with LLMs via a standard API, making it compatible with any general-purpose model. To evaluate the performance of \sysName, we used three lightweight, general-purpose LLMs:
 \begin{itemize}[leftmargin=*]
     \item \textbf{ChatGPT Base} OpenAI GPT-5.6-Luna low~\cite{openai2026gpt56systemcard}
     \item \textbf{Gemini Base} Google Gemini 3.0 Flash~\cite{google2025gemini3flash}
     \item \textbf{Mistral Base} Mistral 3.1 Small 24B-Instruct~\cite{mistralai2025magistral}
 \end{itemize}
To the best of our knowledge, there is no existing solution that can handle the \prob\ problem. As our framework utilizes LLMs, they serve as a natural baseline for performance comparison. To this end, we considered the above lightweight, general-purpose LLMs, which we used as part of \sysName, as standalone baseline solutions. Additionally, as our framework performs iterative multi-step computation, we further considered models that implement explicit chain-of-thought or multi-stage reasoning pipelines internally. To this end, we used 
 \begin{itemize} [leftmargin=*]
     \item \textbf{ChatGPT Thinking} OpenAI GPT-5.6-Luna high~\cite{openai2026gpt56systemcard}
     \item \textbf{Gemini Thinking} Google Gemini 3 Flash-Thinking~\cite{google2025gemini3flash}
     \item \textbf{Mistral Thinking} Mistral Magistral Small-Thinking~\cite{mistralai2025magistral}
 \end{itemize}
Both the Base and Thinking LLMs, were presented with $\probins$ (using the same representation as in \system{}), and asked to refine the query $Q$. At each iteration, we also provided the history of previous outputs. The models were stopped after $T\cdot K$ iterations (25), yet all terminated earlier, due to the limit of their context. See~\cite{our_github_repository} for the full implementation details of these baselines.  
In addition to the LLM-based baselines, we implemented a \textbf{sampling-based baseline}, which is given the search-space ranges for each of the refineable predicates and randomly samples $100$ assignments uniformly from the given ranges. The $\varepsilon$-satisfying assignment $\predVal$ (if such a refinement exists) with minimal $\dist$ is returned.

\subsection{Success rates and optimality}
The first set of experiments aims to evaluate the success rates and optimality scores of our solution compared with baseline approaches.
\new{Figure~\ref{fig:baseline-comparison} depicts the success rate and optimality, respectively, for all instances and baselines. }
The lightweight, general-purpose LLMs are denoted as Base (blue bars), multi-stage reasoning models are denoted as Thinking (orange bars), and the performance of \sysName is shown in green. Each category on the $x$-axis represents a model family, e.g., in the ChatGPT cluster group in Figure~\ref{fig:baseline-comparison-success-top-k}, the success rate of OpenAI GPT-5.6-Luna low is shown in blue, OpenAI GPT-5.6-Luna high in orange, and the success rate of \sysName using GPT-5.6-Luna low is shown in green. The success rate of the sampling-based baseline solution is shown as a dashed line.

\new{As shown in Figures~\ref{fig:baseline-comparison-success-top-k} --~\ref{fig:baseline-comparison-success-complex}, \sysName outperforms the baselines in nearly all settings (matching Gemini Base on Range). Averaged over the four instance classes, \sysName achieves a success rate of $97.5\%$ when using ChatGPT (the strongest model) and $100\%$ when using Mistral (the weakest model), compared to $65.0\%$ and $78.1\%$ for the ChatGPT Base and Thinking baselines, and $31.3\%$ and $21.9\%$ for the Mistral ones. With Gemini, \sysName reaches $88.1\%$, compared to $67.5\%$ and $61.3\%$. The gain is most striking for the weakest model: \sysName improves Mistral's success rate by more than $3\times$ over its Base baseline ($31.3\%$ to $100\%$) and $4.6\times$ over its Thinking baseline ($21.9\%$ to $100\%$), indicating that \sysName's structure can compensate for limited model capability. For reference, random sampling averages $56.3\%$, outperforming both Mistral baselines.}

\new {%
In terms of optimality, as with the success rate, \sysName outperforms the competitors in nearly all settings. The observed average optimality across the three LLM families generated by \sysName achieves $85.5\%$, compared with $47.4\%$ for the Base competitor and $48.6\%$ for the Thinking competitor. These correspond to relative improvements of $80.4\%$ and $75.8\%$, respectively. Per model, \sysName achieves $87.9\%$ with ChatGPT (compared to $71.9\%$ and $77.2\%$ for its Base and Thinking baselines), $78.7\%$ with Gemini ($47.6\%$ and $49.7\%$), and $89.9\%$ with Mistral ($22.7\%$ and $19.0\%$), i.e., an improvement of $4.0\times$ and $4.7\times$ for the weakest model. Random sampling averages $30.5\%$, above both Mistral baselines.}

\begin{table*}[t]
\vspace{-3mm}
\centering
\small
\setlength{\tabcolsep}{4pt}
\renewcommand{\arraystretch}{1.08}

\begin{tabular}{l l | cccc | cccc}
\toprule
 & & \multicolumn{4}{c|}{\textbf{Success Rate (\%)}} & \multicolumn{4}{c}{\textbf{Refinement Optimality (\%)}} \\
\textbf{Model} & \textbf{Setting} 
& Top-K & Range & Diversity & Complex
& Top-K & Range & Diversity & Complex \\
\midrule

\multirow{3}{*}{\shortstack{GPT-5.6-Luna\\(low effort)}}
 & \new{Single-Step OPRO (W/O SubspaceLM)}
   & 95.0 & 85.0 & 97.5 & 95.0
   & 79.09 & 68.18 & 96.84 & 90.24 \\

 & \new{SubspaceLM + Random (W/O AssignmentLM)}
   & 25.0 & 95.0 & 77.5 & 85.0
   & 22.94 & 65.78 & 93.18 & 80.50 \\

 & \textbf{\new{OmniTune (SubspaceLM+AssignmentLM)}}
   & \textbf{95.0} & \textbf{100.0} & \textbf{97.5} & \textbf{97.5}
   & \textbf{92.68} & \textbf{69.72} & \textbf{98.09} & \textbf{91.12} \\
\midrule

\multirow{3}{*}{Gemini-3.0-Flash}
 & \new{Single-Step OPRO (W/O SubspaceLM)}
   & 67.5 & 90.0 & 95.0 & 87.50
   & 55.55 & 66.81 & 94.40 & 89.45 \\

 & \new{SubspaceLM + Random (W/O AssignmentLM)}
   & 67.5 & 87.5 & 75.0 & 95.0
   & 57.09 & 46.63 & 67.16 & 84.65 \\

 & \textbf{\new{OmniTune (SubspaceLM+AssignmentLM)}}
   & \textbf{70.0} & \textbf{90.0} & \textbf{97.5} & \textbf{95.0}
   & \textbf{57.27} & \textbf{67.32} & \textbf{96.76} & \textbf{93.41} \\
\midrule

\multirow{3}{*}{Mistral-Small-3.1-24B}
 & \new{Single-Step OPRO (W/O SubspaceLM)}
   & 62.5 & 87.5 & 37.5 & 100.0
   & 48.63 & 63.07 & 34.59 & 85.55 \\

 & \new{SubspaceLM + Random (W/O AssignmentLM)}
   & 100.0 & 100.0 & 75.0 & 87.5
   & 77.50 & 63.10 & 67.74 & 63.29 \\

 & \textbf{\new{OmniTune (SubspaceLM+AssignmentLM)}}
   & \textbf{100.0} & \textbf{100.0} & \textbf{100.0} & \textbf{100.0}
   & \textbf{91.06} & \textbf{87.43} & \textbf{95.27} & \textbf{85.78} \\
\bottomrule
\end{tabular}

\caption{Two-step OPRO scheme ablation Study: Success Rate and Refinement Optimality. %
}
\label{tab:unified-success-optimality}
\end{table*}

\begin{table*}[t]
\centering
\small
\setlength{\tabcolsep}{4pt}
\renewcommand{\arraystretch}{1.08}

\begin{tabular}{l l | cccc | cccc}
\toprule
 & & \multicolumn{4}{c|}{\textbf{Success Rate (\%)}} & \multicolumn{4}{c}{\textbf{Refinement Optimality (\%)}} \\
\textbf{Model} & \textbf{Setting}
& Top-K & Range & Diversity & Complex
& Top-K & Range & Diversity & Complex \\
\midrule

\multirow{3}{*}{\shortstack{GPT-5.6-Luna}}
 & Complete History (W/O Skyline)
   & 50.0 & 85.0 & 45.0 & 90.0
   & 40.76 & 63.07 & 61.09 & 73.78 \\

 & Summarized History (W/O Skyline)
   & 90.0 & 87.5 & 92.5 & 97.5
   & 76.48 & 64.08 & 94.60 & 88.97 \\

 & \textbf{OmniTune (Summ. History+Skyline)}
   & \textbf{95.0} & \textbf{100.0} & \textbf{97.5} & \textbf{97.5}
   & \textbf{92.68} & \textbf{69.72} & \textbf{98.09} & \textbf{91.12} \\
\midrule

\multirow{3}{*}{Gemini-3.0-Flash}
 & Complete History (W/O Skyline)
   & 32.5 & 82.5 & 27.5 & 87.5
   & 28.95 & 57.21 & 36.65 & 54.01 \\

 & Summarized History (W/O Skyline)
   & 65.0 & 87.5 & 92.5 & 90.0
   & 42.94 & 59.11 & 94.97 & 63.77 \\

 & \textbf{OmniTune (Summ. History+Skyline)}
   & \textbf{70.0} & \textbf{90.0} & \textbf{97.5} & \textbf{95.0}
   & \textbf{57.27} & \textbf{67.32} & \textbf{96.76} & \textbf{93.41} \\
\midrule

\multirow{3}{*}{Mistral-Small-3.1-24B}
 & Complete History (W/O Skyline)
   & 62.5 & 87.5 & 75.0 & 75.0
   & 33.51 & 60.79 & 67.98 & 64.35 \\

 & Summarized History (W/O Skyline)
   & 100.0 & 75.0 & 62.5 & 87.5
   & 91.00 & 49.67 & 57.49 & 69.26 \\

 & \textbf{OmniTune (Summ. History+Skyline)}
   & \textbf{100.0} & \textbf{100.0} & \textbf{100.0} & \textbf{100.0}
   & \textbf{91.06} & \textbf{87.43} & \textbf{95.27} & \textbf{85.78} \\
\bottomrule
\end{tabular}

\vspace{0.4em}
\caption{History Structures Ablation Study: Success Rate and Refinement Optimality.%
}
\label{tab:history-skyline-ablation-unified}
\end{table*}

\subsection{Ablation Study}
The second set of experiments evaluates the usefulness of components in the OPRO scheme as well as the \sysName system. %

\paragraph{Ablation for the two-step OPRO scheme} To evaluate the effectiveness of the two-step OPRO scheme, we compared the performance of \sysName to two variants of the system: 
\begin{enumerate}[leftmargin=*]
    \item \emph{\new{Single-Step OPRO (W/O SubspaceLM)}}, which omits subspace selection, forcing the AssignmentLM to
explore the full refinement space $\allAss$. This corresponds to a naïve instantiation of the OPRO paradigm.

    \item \emph{\new{SubspaceLM + Random (W/O AssignmentLM)}}. In this variant of the system, the SubspaceLM agent generates subspaces, and the AssignmentLM is replaced with a simple random sampling process that selects five random refinements from the subspace.
\end{enumerate}
The success rate and the optimality values of the generated refinements are presented in Table~\ref{tab:unified-success-optimality}. The results highlight the need for both AssignmentLM and SubspaceLM. In particular, we observed a substantial drop in performance for \new{the Single-Step OPRO variant}, with up to a $62.5\%$ decrease in success rate and a $60.7\%$ decrease in optimality values for the Diversity instances when using the Mistral model, reflecting the difficulty of unguided exploration in large refinement spaces.
\new{Similarly, the SubspaceLM + Random variant achieves lower success scores than the full system in all but one case.}

\paragraph{Ablation for \sysName history component implementation}
The second ablation study focused on the effect of the history representation component presented in Section ~\ref{sec:history} on the performance of \sysName. In particular, we aim to study the advantage of using the skyline component of \sysName. We therefore considered the following alternative representation of the history:
\begin{enumerate}[leftmargin=*]
    \item \emph{Complete History, W/O Skyline}, feeds the full raw conversation history of the LLM at every iteration and disables the skyline component concatenated to the AssignmentLM's prompt. 
    \item \emph{Summarized History, W/O Skyline}, utilizes history-summarization instead of the full conversation log, with skyline tracking by the AssignmentLM disabled.
\end{enumerate}

\new{Table~\ref{tab:history-skyline-ablation-unified} presents the success rates and optimality values. Across all models and refinement classes, the complete \sysName (with both history summarization and the skyline), achieves the highest scores, demonstrating the importance of these components. We observed an average improvement of $42.6\%$ in success rate when using \sysName compared to the complete history and without the skyline component, with a most prominent improvement of $3.5\times$ (from $27.5\%$ to $97.5\%$) for Gemini, in the Diversity class. 
}

 \begin{figure*}[t]
\vspace{-2mm}
  \centering

  \begin{minipage}{\linewidth}
    \hspace{0.05\linewidth}\includegraphics[width=0.26\linewidth]{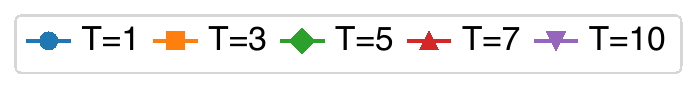}
  \end{minipage}
  \vspace{-0.5em}

  \begin{subfigure}[b]{0.25\linewidth}
    \includegraphics[width=\linewidth]{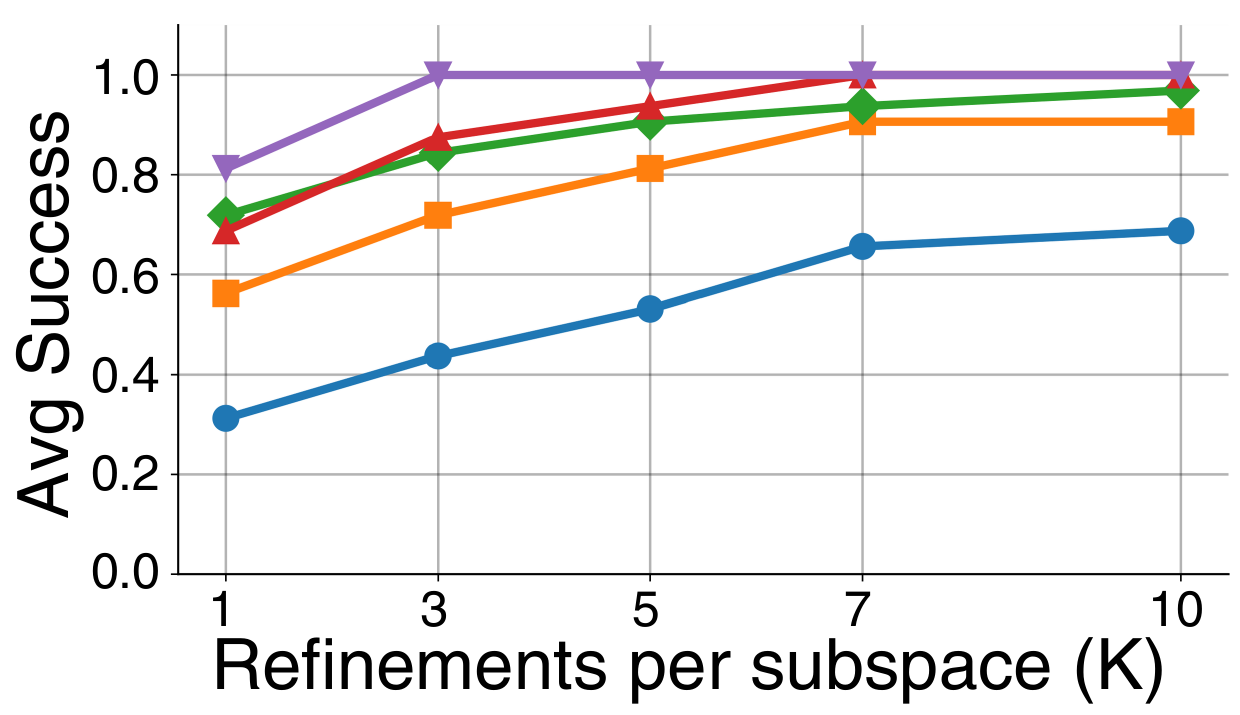}
    \caption{Success rate}
    \label{fig:suc-ref-a}
  \end{subfigure}
  \hspace{0.06\linewidth}
  \begin{subfigure}[b]{0.25\linewidth}
    \includegraphics[width=\linewidth]{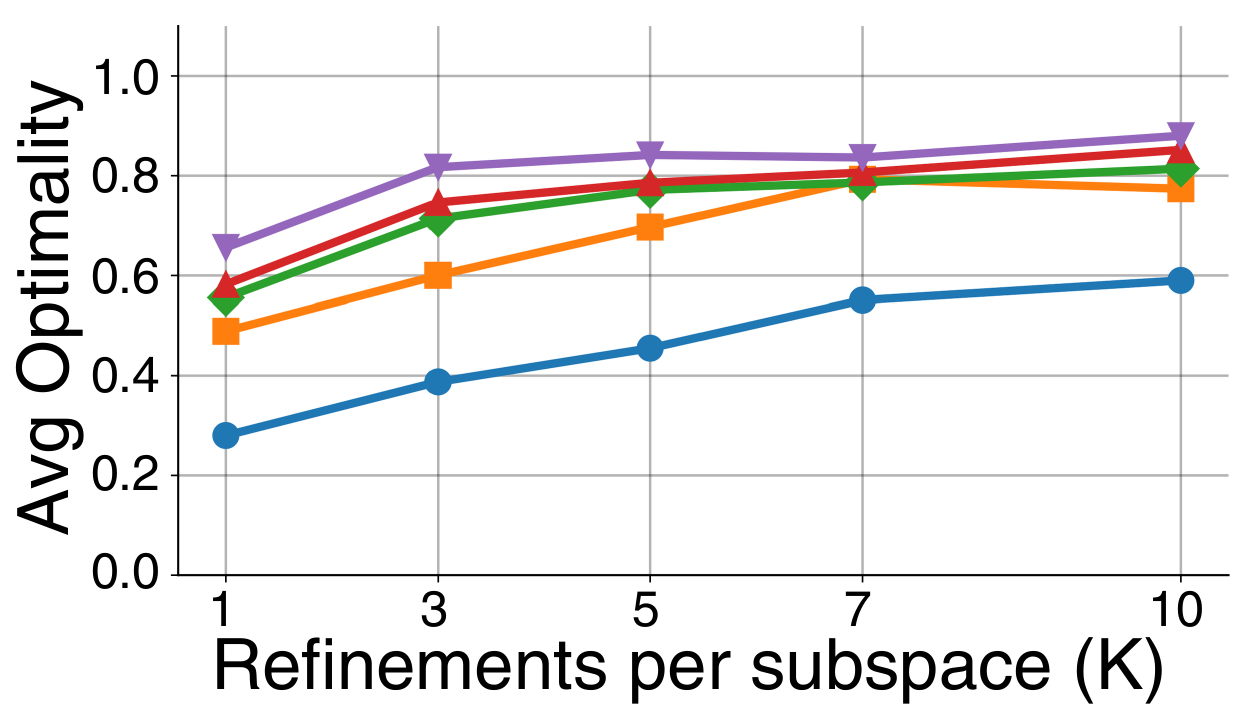}
    \caption{Optimality Score}
    \label{fig:opt-ref-b}
  \end{subfigure}
  \hspace{0.06\linewidth}
  \begin{subfigure}[b]{0.25\linewidth}
    \includegraphics[width=\linewidth]{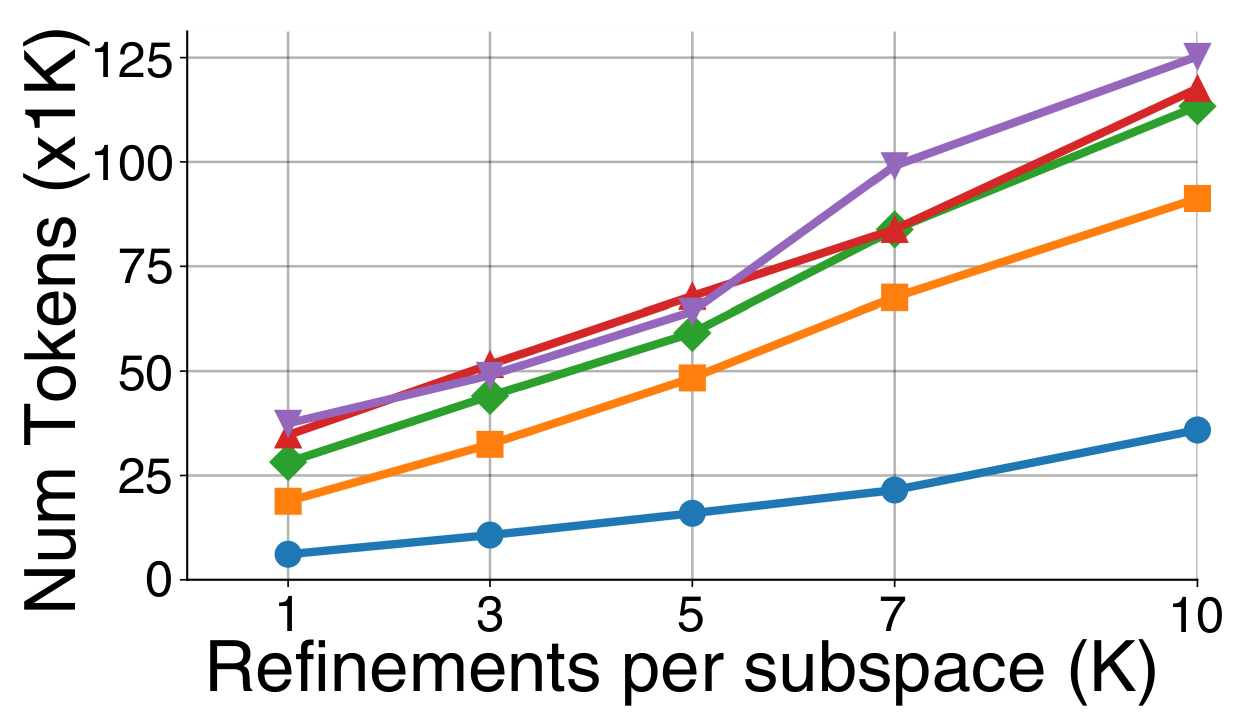}
    \caption{Token Cost}
    \label{fig:tokens-ref-c}
  \end{subfigure}  
  \caption{
  Success rate, Optimality score and Token-cost as a function of refinement per subspace $K$}
  \label{fig:refinements-num-effect}
\end{figure*}
\begin{figure*}[t]
\vspace{-2mm}
  \centering
  \begin{minipage}{\linewidth}
    \hspace{0.05\linewidth}\includegraphics[width=0.23\linewidth]{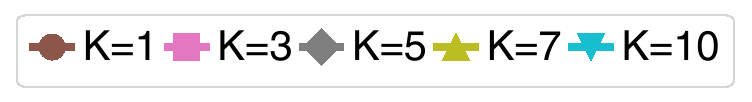}
  \end{minipage}
  \begin{subfigure}[b]{0.25\linewidth}
    \includegraphics[width=\linewidth]{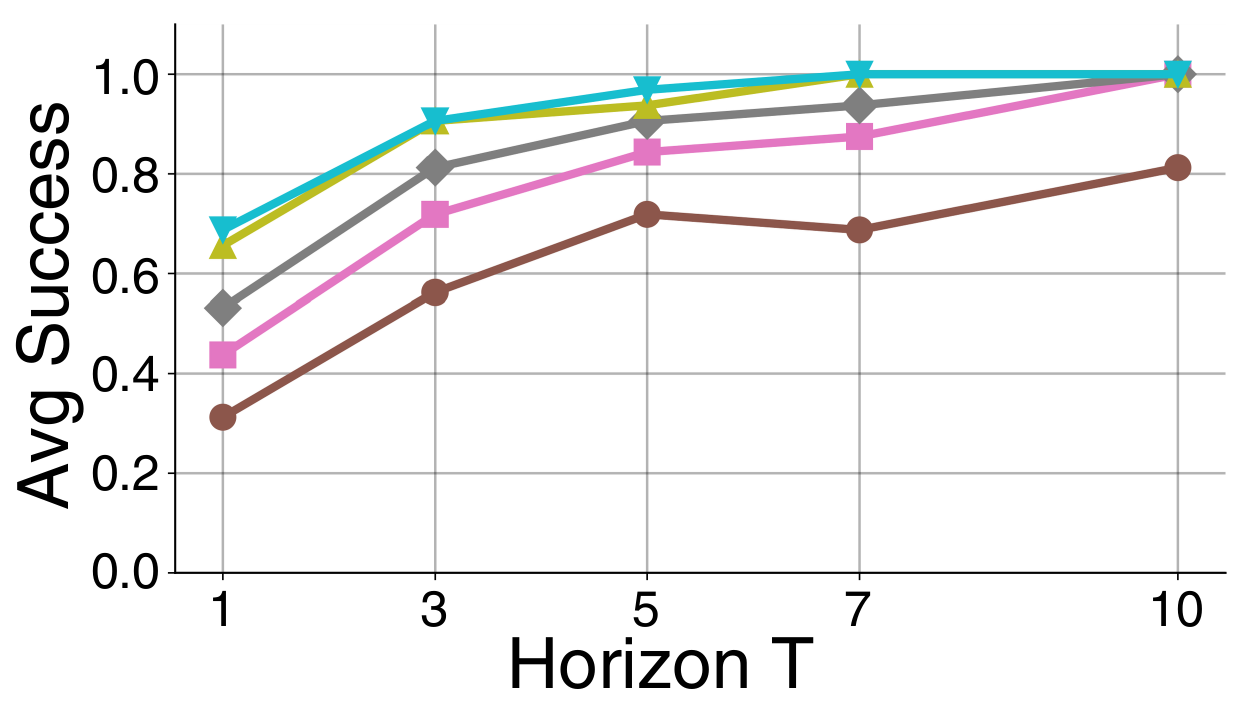}
    \caption{ Success Rate}
    \label{fig:suc-sub-a}
  \end{subfigure}
  \hspace{0.06\linewidth}
  \begin{subfigure}[b]{0.25\linewidth}
    \includegraphics[width=\linewidth]{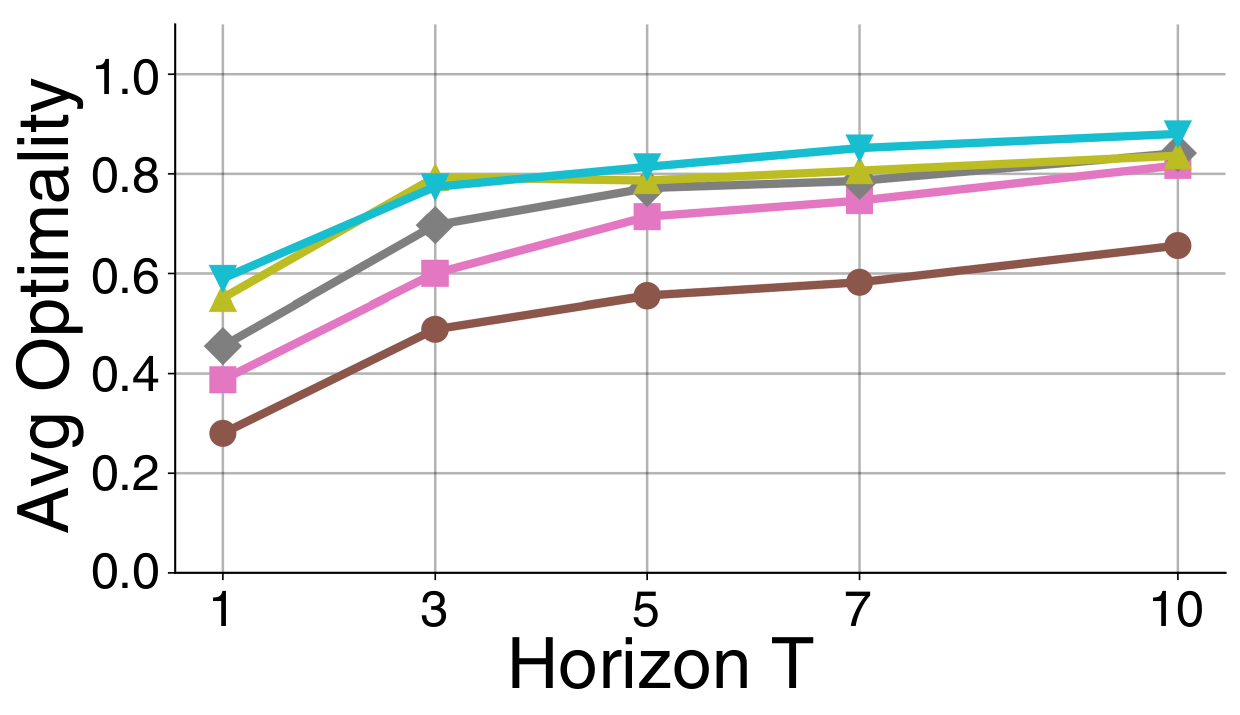}
      \caption{Optimality Score}
    \label{fig:opt-sub-b}
  \end{subfigure}
  \hspace{0.06\linewidth}
  \begin{subfigure}[b]{0.25\linewidth}
    \includegraphics[width=\linewidth]{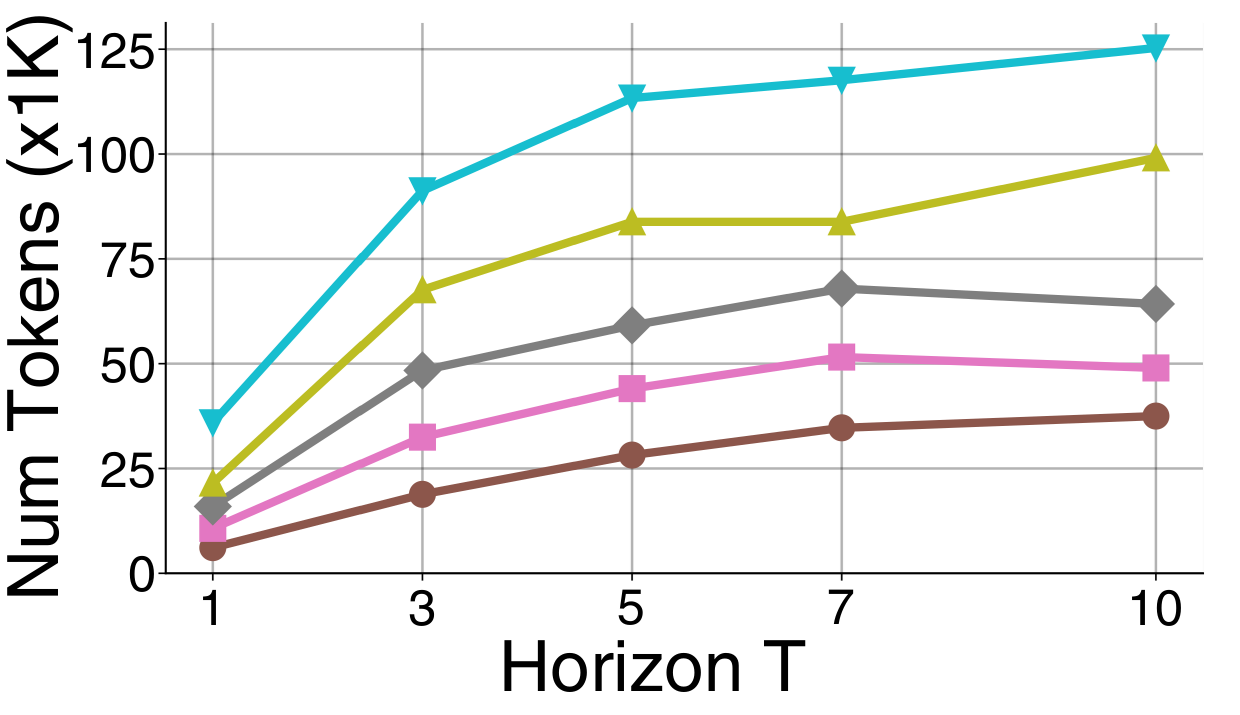}
      \caption{Token Cost}
    \label{fig:tok-sub-c}
  \end{subfigure}

  \caption{
  Success rate, Optimality score and Token-cost as a function of the number of subspaces $T$.}
  \label{fig:horizon-t-effect}
\end{figure*}

\subsection{Effect of Parameters}%
Next, we analyze the impact of the horizon $T$ and per-subspace refinements $K$ on \new{success rate, optimality, and token cost}.

\paragraph{Refinements per Iteration ($K$)} 
Figure~\ref{fig:refinements-num-effect}  shows the success rates, optimality values, and token cost, respectively, when ranging $K$ from $1$ to $10$ using different (fixed) numbers of subspaces: ($T \in \{1,3,5,7,10\}$). We report the average over all considered instance classes\new{; for per-instance results, refer to \trref{appendix:parameters}}.
As expected, we observed an overall improvement in the success rates and optimality values with the cost of a growing amount of tokens sent to the LLM model as $K$ increases. 
In Figure~\ref{fig:suc-ref-a}, we see a notable convergence of success rate for $K>5$, a saturation that is achieved earlier as horizon $T$ grows larger, starting from $T=3$. 
Moreover, we note that for any horizon $T\geq5$, the increase in success rate for any value of $K>5$ is negligible. A similar pattern can be seen for the measures of average optimality in Figure~\ref{fig:opt-ref-b}. This pattern suggests that, when the horizon $T > 1$, the resulting optimality and success diminish when the number of refinements crosses the value of $K=5$.
The observed token cost (Figure~\ref{fig:tokens-ref-c}) showed, as expected, consistent, linear growth in tokens used as $K$ increases, for all different values of horizon $T$.

\paragraph{Horizon ($T$)}
Figure~\ref{fig:horizon-t-effect} shows the effect of varying the horizon $T$ for different (fixed) numbers of refinements per iteration \new($K \in \{1, 3, 5, 7, 10 \}$). Consistent with the impact of increasing $K$ as described above, 
when increasing $T$, the overall trend is growth in both the quality of the solution (success rate and optimality values) as well as the token cost. 
As with $K$, we observed a diminishing improvement in the success rate (Figure~\ref{fig:suc-sub-a}) and optimality (Figure~\ref{fig:opt-sub-b}) beyond the point of $T=5$ when using $5$ or more refinements per iteration. In contrast to the linear growth observed in the token cost when increasing $K$, we observed only a moderate growth for increasing values of $T$ (Figure~\ref{fig:tok-sub-c}). This is due to the effect of a higher number of refinements per subspace, leading to faster convergence in finding the minimal refinement. Thus, coupled with the early stopping mechanism, the system requires fewer subspace iterations to yield a successful refinement and terminate. \new{For a per-instance analysis see~\trcite{appendix:parameters}.}

  \begin{graybox}{Results Summary}
\new{\begin{itemize}[leftmargin=*]
\setlength{\itemsep}{3pt}
    \item \sysName handles a wide range of refinement problems, including ones not supported by existing solutions, and outperforms the baselines in nearly all settings.
    \item With its best-performing LLM (Mistral, the weakest model), \sysName finds an $\varepsilon$-satisfying refinement (success rate) in \good{$100\%$} of the tasks with an average optimality of \good{$89.8\%$}.
    \item The two-step OPRO scheme and the history representation of \sysName consistently improve results across all LLMs, over Single-Step OPRO and complete history, respectively. The two-step scheme improves Mistral's success rate and optimality by \good{$39\%$} and \good{$55\%$}, and the history representation improves Gemini's by \good{$53\%$} and \good{$78\%$}.
    \item Increasing the horizon $T$ and the number of refinements per iteration $K$ improves quality at a higher token cost, with $T=K=5$ yielding the best tradeoff.
\end{itemize}}
\end{graybox}

\subsection{Use cases}
Prior query-refinement methods \cite{campbell2024topk,moskovitch2023diversity,suraj2022range} are restricted in the queries and constraints they can support, leaving many practically important analytical scenarios unaddressed.
To demonstrate the usefulness of \sysName, we present three real-life use cases of previously unsupported settings, including \texttt{HAVING}-clause predicates, nested subqueries, window-function filters, and output-level constraints defined over cross-attribute statistics. A summary of the use cases is presented in Table \ref{tab:case-studies-short-queries}. \new{For a detailed description of each setting, refer to \trref{appendix:case-studies}; all additional material is available in our GitHub repository~\cite{our_github_repository}.} In all cases, we use the predicate-based refinement distance described in Example~\ref{ex:distance}.

\begin{table*}[t]
\small
\setlength{\tabcolsep}{3pt}
\centering
\begin{tabular}{>{\raggedright\arraybackslash}p{0.4cm}>{\raggedright\arraybackslash}p{3.0cm} >{\raggedright\arraybackslash}p{2.4cm} >{\raggedright\arraybackslash}p{6.8cm} >{\raggedright\arraybackslash}p{3.6cm}}
\toprule
\textbf{\#} & \textbf{Task} & \textbf{Database} & \textbf{Query} & \textbf{Constraints} \\
\midrule
1) & Identify multiple customer segments with frequent purchases and high average order value. &
Olist e-commerce~\cite{olistdataset} (\texttt{customers}, \texttt{orders}, \texttt{order\_items}, \texttt{products}) &
\begin{tabular}[t]{@{}l@{}}
\texttt{SELECT ... FROM customers c JOIN ...} \\
\texttt{GROUP BY product\_category, customer\_id} \\
\texttt{HAVING COUNT(*) >= \origv{6} \refv{3}} \\
\texttt{AND AVG(price) >= \origv{300} \refv{400}} \\
\texttt{AND AVG(price) <= \origv{800} \refv{900}}
\end{tabular} &
At least 4 product segments; cross-segment average order value $\geq$ \$500. \\
\midrule
2) & Construct a geographically diversified home portfolio under a cost-per-sqft budget. &
U.S. residential listings~\cite{realestatedataset} (\texttt{HomesForSale}) &
\begin{tabular}[t]{@{}l@{}}
\texttt{SELECT city, price, sqft, type FROM HomesForSale} \\
\texttt{WHERE sqft >= \origv{1500} \refv{1400}} \\
\texttt{AND type IN ('sfr','condo',\refv{'townhouse'})} \\
\texttt{AND city IN (SELECT city FROM ... GROUP BY city} \\
\texttt{HAVING COUNT(*) >= \origv{20} \refv{15}} \\
\texttt{AND AVG(price) <= \origv{750000} \refv{820000})}
\end{tabular} &
At least 6 cities; average cost per square foot $\leq$ \$520. \\
\midrule
3) & Compare compensation across
ethnic groups while excluding noisy salary anomalies. &
Texas Tribune state employee salaries ~\cite{texastribune} (\texttt{TexasTribune})  &
\begin{tabular}[t]{@{}l@{}}
\texttt{WITH ranked AS (SELECT ...,} \\
\texttt{PERCENT\_RANK() OVER (...) AS pr} \\
\texttt{FROM TexasTribune} \\
\texttt{WHERE HRSWKD > \origv{40} \refv{25} AND ANNUAL > \origv{50000} \refv{48000})} \\
\texttt{SELECT ETHNICITY, AVG(ANNUAL), SUM(MONTHLY)} \\
\texttt{FROM ranked} \\
\texttt{WHERE pr >=  \origv{0.40} \refv{0.30} AND pr <= \origv{0.90} \refv{0.92}} \\
\texttt{GROUP BY ETHNICITY}

\end{tabular} &
For each ethnicity: total monthly payroll $\geq$ \$3{,}500{,}000;
$\tfrac{\max(\texttt{AVG(ANNUAL)})}{\min(\texttt{AVG(ANNUAL)})} \leq 1.15$. \\
\bottomrule
\end{tabular}
\caption{Summary of real-world use cases. The queries are represented in a compact way; the original query values are presented in \origv{red}, and the refined query values are in \refv{green}.} %
\label{tab:case-studies-short-queries}
\end{table*}

\paragraph{Case Study 1: Value-Based Customer Segmentation (Olist~\cite{olistdataset}).}
The Olist~\cite{olistdataset} dataset is a publicly available Brazilian e-commerce dataset of orders placed at Olist Store\footnote{\url{https://www.olist.com/}}.
An e-commerce analytics team wants to identify customer segments with frequent purchases and high average order value for targeted campaigns. For this purpose, they formulate a query that groups transactions by product category and customer, requiring at least $6$ purchases and an average order value between $\$300$ and $\$800$ (Query 1 in Table~\ref{tab:case-studies-short-queries}). 
Senior management requires the result to comply with the following company-level constraints: (1) having at least $4$ product segments and a (2) cross-segment average order value of at least $\$500$. 
The results of the query on the \textit{Olist} dataset do not meet these constraints: it returns only two segments, with a cross-segment average order value of $\$385$. \sysName{} suggests refining the threshold for required purchases to 3 and adjusting the allowed range for average order value to between $\$400$ and $\$900$. The resulting refined query 
satisfies both constraints, as it contains $4$ qualifying segments with a cross-segment average order value of $\$540$.

\paragraph{Case Study 2: Diversified City-Level Housing Investment.}
A real estate investment firm plans to construct an acquisition portfolio under a cost-per-square-foot budget. To this end, they use the \textit{HomesForSale} dataset~\cite{realestatedataset}, an open U.S. real-estate listings dataset and formulate a query that
selects single-family rental (sfr) and condo candidate properties with area at or above $1500$ square-foot, while from cities with at least $20$ overall properties for sale, and average price of at most $750$k (Query 2 in Table~\ref{tab:case-studies-short-queries}). To ensure geographic diversification while maintaining cost efficiency, they require the result to comply with the following constraints: (1) cover at least six different cities and (2) maintain an average cost per square foot of at most $\$520$. The results of the original query fails to satisfy the required constraints, yielding properties from only three cities, with average price per square foot of $\$548$. \sysName{} refines the query to include townhouse properties, sets the minimal area to $1400$ and refines the retained city criteria by reducing required property count to $15$ and maximal average price to $820$k. The output of the resulting query satisfies the constraints, yielding candidates from six cities and reducing the average price per square foot to $\$509$.

\paragraph{Case Study 3: Pay-Equity Analysis Across Ethnic Groups.}
A public-sector analytics team seeks to compare compensation across ethnic groups while excluding noisy salary anomalies. To this end, they formulate a query on the \textit{Texas Tribune} salary dataset~\cite{texastribune} that filters employees by minimum weekly work-hours and annual salary thresholds, computes \texttt{PERCENT\_RANK()} within each ethnicity, and retains only employees whose salaries fall between the 40th and 90th percentiles, before aggregating by ethnicity (Query 3 in Table~\ref{tab:case-studies-short-queries}).
Senior management requires the result to comply with the following pay-equity constraints: (1) each ethnic group must have a total monthly payroll of at least $3.5$M, and (2) the ratio between the maximum and minimum group-average annual salaries must be at most $1.15$.
After executing the query on the \textit{Texas Tribune} dataset, the team finds the result does not meet these constraints: the minimum group payroll is only $2.0$M, and the max/min salary ratio is $1.65$.
The team therefore provides the query and constraints to \sysName{}, which minimally refines the percentile bounds from $[0.40, 0.90]$ to $[0.30, 0.92]$, while refining the work-hours threshold from $40$ to $25$ and the annual salary threshold from $50$k to $48$k.
The refined query satisfies both constraints, with minimum group payroll of $3.55M$ and max/min average-salary ratio of $1.11$

In all three cases, \sysName{} finds a constraint-satisfying refinement after \textit{1-2} iterations. These examples show that \sysName{} can successfully refine realistic analytical queries involving \texttt{HAVING} predicates, nested subqueries, and window-based filters. \new{For full refinement traces and additional details, refer to \trref{appendix:case-studies}.}

%% file: related.tex
\section{Related Work}\label{sec:related}
We first survey the space of query refinement research, and then discuss related work on LLMs for SQL query generation and black-box optimization.

\vspace{2mm}
\noindent\textbf{Query refinement.}
The problem of query refinement, i.e., modifying queries to satisfy some output constraints, was studied in a line of work~\cite{mishra2009refinement,KoudasLTV06,MusleaL05,ChuC94,TranC10,TranCP09,moskovitch2023diversity,campbell2024topk,suraj2022range,erica,rodeo,MoskovitchLJ22}. For example, a subset of works \cite{MusleaL05,KoudasLTV06} aim to relax queries with an empty result set to produce some answers. Other works~\cite{mishra2009refinement,ChuC94} address the issues of too many or too few answers by refining queries to meet specific cardinality constraints on the result's size. A recent line of work has studied the use of refinement to satisfy diversity constraints~\cite{moskovitch2023diversity,campbell2024topk,suraj2022range,erica,rodeo} and primarily focused on SPJ queries, where a refinement of a query is a set of modifications to the constants in the WHERE clause. The authors of Shetiya et al.~\cite{suraj2022range} consider range queries, namely, selection queries with conjunction of range conditions over predicates of the data, and Campbell et al.~\cite{campbell2024topk,rodeo} proposed an automated approach using linear optimization techniques to refine SPJ queries with an ORDER BY clause. Different notions of similarity measures used to define the minimality of refinements were defined in these works. E.g., in Shetiya et al.~\cite{suraj2022range}, the similarity between queries is measured by the Jaccard similarity of their outputs, whereas Li at al.~\cite{moskovitch2023diversity,erica} defines the similarity between queries with respect to the values of the constants in their WHERE clause. Our proposed solution generalizes the concept of query refinements to support a broader class of queries, constraints, and similarity measures. Imposing constraints over answer sets and performing query relaxation has also been studied in the context of package queries \cite{BrucatoBAM16,brucato2014improving}. However, because package queries are not standard SQL, supporting them was outside the scope of this paper.
\vspace{3mm}
\noindent\textbf{LLMs for SQL Generation.}
Recent work has demonstrated that large language models (LLMs) achieve state-of-the-art performance on the text-to-SQL task~\cite{bench2sql,nl2sqldb2spider}, in which users express their information needs in natural language and the system generates a SQL query whose execution answers those needs~\cite{wherearewetodatnl2sql,evoschema}. 
LLM-based solutions surpass earlier dedicated architectures~\cite{yu-etal-2018-syntaxsqlnet,text2sqlnonllm}, primarily due to their strong semantic reasoning capabilities, which enable them to jointly interpret underspecified user intent and database schemas.
Accordingly, most text-to-SQL systems are designed around single-pass query synthesis pipelines, commonly emphasizing schema linking~\cite{redsql,genlink2025,wang2025linkalign}, clause-level SQL generation~\cite{lee-2019-clause}, and execution-based validation mechanisms~\cite{yang2025auto} that are primarily used to detect and correct syntactic errors or mismatches with the database schema.

In contrast, our setting considers a substantially different problem.
Rather than generating a SQL query from a natural language description, the model is given an existing query and is tasked with refining it so as to satisfy a set of constraints while minimizing deviation from the original query.
This refinement process inherently requires iterative reasoning over query modifications, explicit trade-offs between competing objectives, and feedback from previous attempts, which goes beyond execution-based correctness checks and fundamentally differs from the one-shot generation paradigm underlying most text-to-SQL approaches.

\vspace{3mm}
\noindent\textbf{Black-Box and LLM-Based Optimization.} In contrast to \emph{white-box} optimization, where the objective function and constraints
are explicitly specified and structurally accessible to the solver, \emph{black-box} optimization treats the objective function as an oracle that can only be evaluated at sampled points, without access to its internal structure.
Representative approaches include Monte-Carlo sampling, multi-armed bandits, evolutionary algorithms, and related population-based methods~\cite{Kirkpatrick1983,Kennedy1995,Hansen2001,montecarlo1,montecarlo2}.
While broadly applicable, these methods rely on the ability to meaningfully sample the search space and learn solely from observed objective values, often requiring extensive
exploration and struggling in large, discrete, or highly discontinuous domains.

Recently, large language models have been proposed as optimization engines in settings
where both the search space and the objective can be expressed in natural language.
Prior work shows that LLMs can optimize prompts, heuristics, and symbolic programs—through iterative prompting~\cite{Meyerson2023,Chen2022,Chen2023a,Lehman2022,yang2024opro}.
However, existing OPRO-style approaches degrade in large, discrete, and combinatorial
domains, where naïve candidate generation and unstructured interaction histories provide
weak guidance and limit scalability~\cite{yang2024opro,zhang2024revisiting}.

To address these limitations, we build on the OPRO framework and introduce a dedicated
two-step scheme for query refinement, explicitly structuring both the refinement search
space and the optimization feedback to guide LLMs more effectively.

%% file: conclusion.tex
\section{Conclusion}
\label{sec:conclusion}

We presented \sysName, a general framework for SQL query refinement that seeks refinements
minimizing distance to an original query while approximately satisfying user-defined
constraints.
Our approach adapts the OPRO paradigm to query refinement via a dedicated two-step scheme
that combines structured subspace exploration with focused refinement sampling, guided by
compact history summarization and skyline-based feedback.
An extensive empirical evaluation demonstrates that \sysName{} is effective and robust
across a broad range of existing and novel query refinement scenarios beyond the scope of
prior solutions.

In future work, we plan to further enhance the OPRO scheme and the \system{} design, for example by incorporating skyline-aware exploration–exploitation strategies and extending the LLM modules into agentic components capable of issuing and analyzing auxiliary
analytical queries.

%% file: case_studies_appendix.tex
\section{Case Studies}
\label{appendix:case-studies}

\subsection{Case Study 1: Refining \texttt{HAVING} Predicates in Value-Based Customer Segmentation (Olist)}
\label{sec:case-study-1}

\paragraph{Setup.}
\textbf{Database.} We use the Olist Brazilian E-Commerce public dataset, consisting of multiple relations including \texttt{customers}, \texttt{orders}, \texttt{order\_items}, and \texttt{products}. The \texttt{products} relation contains \texttt{product\_category\_name}, and \texttt{order\_items} provides transaction-level \texttt{price} values. \\

\textbf{User intent.} A marketplace growth team analyzes purchasing behavior across product categories to identify customer segments worthy of targeted marketing and retention investment. Mirroring real e-commerce practice, where customer segments are often defined using aggregate behavioral signals to prioritize higher-value customers and allocate resources more effectively, the team groups Olist transactions by product category and customer, seeking customer-category pairs that show repeated purchase activity and high average transaction value. 

To support meaningful cross-category planning, the business requires the query result to contain \textbf{at least 4 product segments}. At the same time, to ensure the expected revenue from the selected segments justifies the marketing expenses, \textbf{the average order value across returned segments must be at least 500\$}. 

After applying the initial thresholds, however, the team notices that only \textbf{2 product segments} are present. Additionally, \textbf{the average order value across returned segments is 385\$}. To this end, the team must seek a minimal refinement of the aggregate thresholds that expands category coverage while preserving value and stability.

\paragraph{Refinement objective.}
In this scenario, only \texttt{HAVING} thresholds can be refined, while the join structure and grouping keys remain fixed.

In this case, to define the subspace domain $\allSS$, \sysName{} first generates auxiliary SQL queries to identify the max and min values for the refinable derived attributes, \texttt{avg\_price} and \texttt{COUNT(*)}. Then, \sysName{} uses these values as boundaries to the subspace domain it then allows SubspaceLM to sample subspaces from.
Columns (d)--(e) in the Case 1 row of \new{Table}~\ref{fig:case-studies-overview} capture both the distance metric and the required output constraints.

Prior refinement methods are not applicable in these case studies because they typically assume refinable predicates are restricted to top-level SPJ \texttt{WHERE} conditions, whereas our scenarios require refining richer SQL constructs (e.g., \texttt{HAVING} aggregates, nested-subquery aggregate predicates, and window-derived filters such as \texttt{PERCENT\_RANK()}) together with output-level constraints over grouped and cross-attribute statistics. In contrast, \sysName{} supports all these settings through the same two-level OPRO process: SubspaceLM proposes bounded refinement regions, AssignmentLM samples concrete predicate assignments within each region, each candidate query is executed on $D$, required output statistics are computed, and the skyline of non-dominated (deviation, distance) trade-offs is updated (See row 1 in \new{Table}~\ref{fig:case-studies-overview}).

\paragraph{Running summary.}
In the round shown in column (c) of case study 1 in \new{Table}~\ref{fig:case-studies-overview}, SubspaceLM first bounded the \texttt{HAVING} search region to \texttt{COUNT(*)} in $[2,6]$ and average-price bounds in $[200,500]$/$[700,900]$. AssignmentLM then tested $k_0$--$k_3$: $k_0=\{3,400,900\}$ is the first assignment with csat $=0.00$ (and $d=0.32$), while the nearby alternatives keep nonzero csat violations. Thus, $k_0$ is selected as the minimal refinement for this case.
\input{casestudyfigs}
\paragraph{Interpretation.}
Given the initial shortfall in both segment coverage and average order value, SubspaceLM focused on the two \texttt{HAVING} aggregates that most directly control those outcomes: purchase frequency and price band. It therefore proposed a bounded region where \texttt{COUNT(*)} stays in $[2,6]$, the lower price threshold in $[200,500]$, and the upper threshold in $[700,900]$, reflecting the hypothesis that moderate relaxation of support with a value-oriented price window would improve constraint satisfaction. Within this subspace, AssignmentLM first tested lower-distance alternatives that still produced nonzero csat, then shifted toward \{3,400,900\}, which is the first fully feasible point; this explains why the selected refinement is slightly farther than some probes yet operationally better.

\paragraph{Outcome.}
The refinement process converges to a solution satisfying all constraints while minimizing distance, as shown by the in-place updates in the query above.

This refinement result yields \textbf{4 different product segments}, while preserving a \textbf{cross-segment average order-value of 540\$}, satisfying both of the abovementioned constraints, while remaining relatively close to the original query with a refinement distance of $d=0.32$.

\subsection{Case Study 2: Diversified City-Level Investment in Residential Housing}
\label{sec:case-study-2}

\paragraph{Setup.}
\textbf{Database.} We use an open U.S. real-estate listings dataset, modeled as a single relation \texttt{HomesForSale} with attributes including \texttt{city}, \texttt{price}, \texttt{sqft}, and \texttt{type}. \\

\textbf{User intent.} A residential real-estate investment firm screens home listings to construct a \textbf{geographically diversified} acquisition pipeline. To this end, they first want to identify cities that exhibit sufficient market depth and acceptable average prices, and then select individual homes within those cities that satisfy its portfolio criteria: mid-market homes with sufficient living space and perceived resale stability. 

To satisfy internal diversification policy, the final investment set must span \textbf{at least 6 cities}. At the same time, to ensure profitability the firm does not want diversification to come at the expense of cost efficiency, meaning that the selected properties should maintain an \textbf{average cost per square foot of at most 520\$}. \\

After applying initial filters, the firm's analysts find out that the candidate portfolio remains concentrated in only \textbf{3 different cities}, while the \textbf{average cost per square foot is as high as 548\$}.

\paragraph{Refinement objective.}
To satisfy diversification requirements while preserving the original acquisition logic, the team applies \emph{minimal} adjustments to the numeric predicates (\texttt{sqft}, nested \texttt{AVG(price)}), the categorical predicate (\texttt{type}), and the nested support threshold \texttt{COUNT(*)}. 
To define the subspace domain $\allSS$, \sysName{} first generates auxiliary SQL queries to identify the max and min values for the nested \texttt{HAVING}-clause predicates, \texttt{avg\_price} and \texttt{COUNT(*)}. 
The Case 2 row of \new{Table}~\ref{fig:case-studies-overview} (columns (d)--(e)) summarizes the objective and output requirements.

Prior refinement methods are not applicable in these case studies because they typically assume refinable predicates are restricted to top-level SPJ \texttt{WHERE} conditions, whereas our scenarios require refining richer SQL constructs (e.g., \texttt{HAVING} aggregates, nested-subquery aggregate predicates, and window-derived filters such as \texttt{PERCENT\_RANK()}) together with output-level constraints over grouped and cross-attribute statistics. In contrast, \sysName{} supports all these settings through the same two-level OPRO process: SubspaceLM proposes bounded refinement regions, AssignmentLM samples concrete predicate assignments within each region, each candidate query is executed on $D$, required output statistics are computed, and the skyline of non-dominated (deviation, distance) trade-offs is updated (See row 2 of \new{Table}~\ref{fig:case-studies-overview}).

\paragraph{Running summary.}
In the step shown in column (c) of case study 2 in \new{Table}~\ref{fig:case-studies-overview}, SubspaceLM defines a bounded region over \texttt{sqft}, nested \texttt{COUNT(*)}/\texttt{AVG(price)}, and the \texttt{type} set to directly target diversification and cost constraints. AssignmentLM evaluates three candidates in that region and finds that $k_1=\{1400,15,820\text{k},(sfr,c,t)\}$ is the first fully feasible point (csat $=0.00$, $d=0.11$). Although $k_2$ is slightly closer ($d=0.10$), its positive csat indicates unresolved constraints, so $k_1$ is correctly identified as a minimal constraint satisfying refinement.

\paragraph{Interpretation.}
Given that the baseline failed on city diversity and cost efficiency, SubspaceLM concentrated on predicates that govern market breadth: \texttt{sqft}, nested city support \texttt{COUNT(*)}, nested \texttt{AVG(price)}, and the \texttt{type} set. It bounded these controls to \texttt{sqft} in $[1400,1600]$, \texttt{COUNT(*)} in $[15,25]$, \texttt{AVG(price)} in $[75\text{k},85\text{k}]$, and type sets from \{\texttt{sfr},\texttt{condo}\} to \{\texttt{sfr},\texttt{condo},\texttt{townhouse}\}, consistent with the idea that broader inventory and slightly looser city viability filters correlate with better geographic coverage. Within this subspace, AssignmentLM first evaluated near-original assignments that remained infeasible, then moved to \{1400,15,820\text{k},(sfr,c,t)\}, which is the first assignment satisfying all constraints while remaining close to the original query.

\paragraph{Outcome.}
After a couple of \textit{two-level refinement} iterations, \sysName{} converges to the in-place refined thresholds shown in the query above.

This refinement yields properties from \textbf{6 different cities}, while maintaining an \textbf{average price per sqft of 509\$}, successfully satisfying both constraints, keeping the threshold close to the original query's values with a refinement distance of $d=0.11$.

\subsection{Case Study 3: Fairness-Aware Salary Analysis with Window-Based Filtering (Texas Tribune Salaries)}
\label{sec:case-study-3}

\paragraph{Setup.}
\textbf{Database.} We use the Texas Tribune state employee salary dataset, modeled as a single relation with attributes including \texttt{ETHNICITY}, \texttt{ANNUAL}, \texttt{MONTHLY}, and \texttt{HRSWKD}.

\textbf{User intent.} A public-sector pay-equity task force analyzes the Texas Tribune state employee salary dataset to screen for possible ethnicity-based pay disparities. To prevent possible bias, the team seeks to compare compensation across ethnic groups among employees who are more directly comparable, rather than across the entire workforce.
To avoid unwanted anomalous individuals, the analysts want to exclude both very low-paid tails and extreme top-end outliers, thus they choose to focus on employees whose salaries fall within a \textbf{target percentile range}.

To ensure that the resulting cohort is suitable for a preliminary pay-equity review, the analysts come up with the following output-level constraints on the query's result: First, to ensure sufficient monetary volume, \textbf{the minimum total monthly support for each selected group must be at least $3{,}500{,}000$}. Second, to enforce disparity between group-average salaries, they require the ratio between the maximum and minimum average annual salaries, $\frac{\max(\texttt{avg\_annual})}{\min(\texttt{avg\_annual})}$, to be at most $1.15$.

However, the initial query's grouped output still suffers from weak payroll support for some ethnicities: the \textbf{lowest monthly payroll across ethnicity groups is only $2{,}000{,}000\$$}, and the \textbf{ratio between maximum and minimum average annual salaries is $1.65$}.

\paragraph{Refinement objective.}
To satisfy fairness and stability requirements while preserving the analysts' intent, the task force applies \emph{minimal} adjustments to the numerical predicates (\texttt{HRSWKD} and \texttt{ANNUAL}) as well as to the percentile window bounds used to filter the CTE output. Columns (d)--(e) in the Case 3 row of \new{Table}~\ref{fig:case-studies-overview} summarize the objective and required grouped-output properties.

Prior refinement methods are not applicable in these case studies because they typically assume refinable predicates are restricted to top-level SPJ \texttt{WHERE} conditions, whereas our scenarios require refining richer SQL constructs (e.g., \texttt{HAVING} aggregates, nested-subquery aggregate predicates, and window-derived filters such as \texttt{PERCENT\_RANK()}) together with output-level constraints over grouped and cross-attribute statistics. In contrast, \sysName{} supports all these settings through the same two-level OPRO process: SubspaceLM proposes bounded refinement regions, AssignmentLM samples concrete predicate assignments within each region, each candidate query is executed on $D$, required output statistics are computed, and the skyline of non-dominated (deviation, distance) trade-offs is updated (See row 3 in \new{Table}~\ref{fig:case-studies-overview}).

\paragraph{Running summary.}
As seen in column (c) of case study 3 in \new{Table}~\ref{fig:case-studies-overview}, given the low payroll support and large cross-group salary ratio in the initial result, SubspaceLM narrows the search to a cohort-shaping region over \texttt{HRSWKD}, \texttt{ANNUAL}, and percentile-window bounds, with \texttt{HRSWKD} in $[25,40]$, \texttt{ANNUAL} in $[45\text{K},55\text{K}]$, and percentile limits in $[0.30,0.85]$ and $[0.45,0.95]$, under the hypothesis that controlled expansion of the comparable-employee pool would improve both support mass and salary-ratio stability. AssignmentLM then probes $k_0$--$k_2$ within that subspace and first reaches full feasibility at $k_1=\{25,48000,0.30,0.92\}$ with csat $=0.00$ and $d=0.18$; the closer alternatives ($k_0$, $k_2$) are retained as diagnostics, but because their csat values remain positive, they are not selected.

\paragraph{Outcome.}
\sysName{} converges to the in-place refined thresholds shown in the query above.

Under this refined query, the estimated grouped-output statistics become
\[
\begin{aligned}
\min(\texttt{total\_monthly}) &\approx 3{,}550{,}000, \\
\frac{\max(\texttt{avg\_annual})}{\min(\texttt{avg\_annual})} &\approx 1.11.
\end{aligned}
\]
effectively satisfying both constraints, while the refinement distance remains low as $d=0.18$, remaining practically close to the original query.

%% file: casestudyfigs.tex
\begin{table*}[t]
\centering
\footnotesize
\setlength{\tabcolsep}{2pt}
\renewcommand{\arraystretch}{1.0}
\begin{tabular}{@{}
  p{0.07\textwidth}
  p{0.22\textwidth}
  p{0.32\textwidth}
  p{0.19\textwidth}
  p{0.20\textwidth}@{}}
\toprule
\textbf{(a) Case \#}
  & \textbf{(b) Query $Q \rightarrow Q'$}
  & \textbf{(c) Excerpt (csat, $d$)}
  & \textbf{(d) Refinement distance}
  & \textbf{(e) Formalized constraints} \\
\midrule

\begin{minipage}[t]{\linewidth}\vspace{0pt}
\textbf{1}
\end{minipage}

&

\begin{minipage}[t]{\linewidth}\vspace{0pt}
\begin{alltt}
\scriptsize
SELECT segment, customer,
  COUNT(*) AS cnt,
  AVG(price) AS avg_price
FROM Olist joins
GROUP BY segment, customer
HAVING COUNT(*) >= \origv{6} \refv{3}
  AND avg_price >= \origv{300} \refv{400}
  AND avg_price <= \origv{800} \refv{900}
\end{alltt}
\end{minipage}

&

\begin{minipage}[t]{\linewidth}\vspace{0pt}
  \textbf{Subspace $t=1$}\\
  $\texttt{COUNT(*)}\in[2,6]$,\\
  $\texttt{avg\_price}_\ell\in[200,500]$,
  $\texttt{avg\_price}_u\in[700,900]$\\[3pt]
  $k_0:\{3,400,900\}$ \hfill(0.00, 0.32)\\
  $k_1:\{2,300,700\}$ \hfill(0.29, 0.38)\\
  $k_2:\{2,300,800\}$ \hfill(0.33, 0.33)\\
  $k_3:\{3,400,700\}$ \hfill(0.24, 0.54)
\end{minipage}

&

\begin{minipage}[t]{\linewidth}\vspace{0pt}
\[
\begin{aligned}
&\;\;\;\;\;\;\;\Delta=\left|\tfrac{\Delta\texttt{COUNT(*)}}{\texttt{COUNT(*)}}\right|+\\&\left|\tfrac{\Delta\texttt{avg\_price}_\ell}{\texttt{avg\_price}_\ell}\right|+\left|\tfrac{\Delta\texttt{avg\_price}_u}
                      {\texttt{avg\_price}_u}\right|
\end{aligned}
\]
\end{minipage}

&

\begin{minipage}[t]{\linewidth}\vspace{0pt}
\[|\texttt{segs}(Q'(D))|\ge 4\]
\[\texttt{AVG(ord\_val)}\ge 500\]
\end{minipage}

\\[6pt]\midrule

\begin{minipage}[t]{\linewidth}\vspace{0pt}
\textbf{2}
\end{minipage}

&

\begin{minipage}[t]{\linewidth}\vspace{0pt}
\begin{alltt}
\scriptsize
SELECT city, price,
  sqft, type
FROM HomesForSale
WHERE sqft >= \origv{1500} \refv{1400}
  AND type IN (\origv{'sfr','condo'},
    \refv{'sfr','condo', 'townhouse'})
  AND city IN (
    SELECT city
    FROM HomesForSale
    GROUP BY city
    HAVING COUNT(*) >= \origv{20} \refv{15}
      AND AVG(price) <= \origv{750000} \refv{820000})
\end{alltt}
\end{minipage}

&

\begin{minipage}[t]{\linewidth}\vspace{0pt}
  \textbf{Subspace $t=2$}\\
  $\texttt{sqft}\in[1400,1600]$,\\
  $\texttt{COUNT(*)}\in[15,25]$,
  $\texttt{AVG(price)}\in[75\text{k},85\text{k}]$,\\
  $\texttt{types}\in[\{sfr,condo\},\{sfr,condo,townhouse\}]$\\[3pt]
  $k_0:\{1450,18,800\text{k},(sfr,condo)\}$
    \hfill(0.24, 0.13)\\
  $k_1:\{1400,15,820\text{k},(sfr,c,t)\}$
    \hfill(0.00, 0.11)\\
  $k_2:\{1500,17,780\text{k},(sfr,condo)\}$
    \hfill(0.19, 0.10)
\end{minipage}

&

\begin{minipage}[t]{\linewidth}\vspace{0pt}
\[
\begin{aligned}
\Delta={}&\left|\tfrac{\Delta\texttt{sqft}}{\texttt{sqft}}\right|+\delta_{Jac}(\texttt{types},\texttt{types'})\\
        &+\left|\tfrac{\Delta\texttt{COUNT(*)}}{\texttt{COUNT(*)}}\right|+\left|\tfrac{\Delta\texttt{AVG}(p)}{\texttt{AVG}(p)}\right|
\end{aligned}
\]
\end{minipage}

&

\begin{minipage}[t]{\linewidth}\vspace{0pt}
\[|\texttt{cities}(Q'(D))|\ge 6\]
\[\overline{\texttt{price}/\texttt{sqft}}\le 520\]
\end{minipage}

\\[6pt]\midrule

\begin{minipage}[t]{\linewidth}\vspace{0pt}
\textbf{3}
\end{minipage}

&

\begin{minipage}[t]{\linewidth}\vspace{0pt}
\begin{alltt}
\scriptsize
WITH ranked AS (
  SELECT *,
    PERCENT_RANK() OVER (...) AS pr
  FROM TexasTribune
  WHERE HRSWKD > \origv{40} \refv{25}
    AND ANNUAL > \origv{50000} \refv{48000}
)
SELECT ETHNICITY,
  AVG(ANNUAL) AS avg_annual,
  SUM(MONTHLY) AS total_monthly
FROM ranked
WHERE pr >= \origv{0.40} \refv{0.30} AND pr <= \origv{0.90} \refv{0.92}
GROUP BY ETHNICITY
\end{alltt}
\end{minipage}

&

\begin{minipage}[t]{\linewidth}\vspace{0pt}
  \textbf{Subspace $t = 2$}\\
  $\texttt{HRSWKD}\in[25,40]$,
  $\texttt{ANNUAL}\in[45\text{K},55\text{K}]$,\\
  $\texttt{pr}_l\in[0.30,0.85]$,
  $\texttt{pr}_u\in[0.45,0.95]$\\[3pt]
  $k_0:\{30,50000,0.35,0.90\}$ \hfill(0.29, 0.16)\\
  $k_1:\{25,48000,0.30,0.92\}$ \hfill(0.00, 0.18)\\
  $k_2:\{35,52000,0.40,0.85\}$ \hfill(0.34, 0.14)
\end{minipage}

&

\begin{minipage}[t]{\linewidth}\vspace{0pt}
\[
\begin{aligned}
\Delta={}&\left|\tfrac{\Delta\texttt{HRSWKD}}{\texttt{HRSWKD}}\right|+\left|\tfrac{\Delta\texttt{ANNUAL}}{\texttt{ANNUAL}}\right|\\
        &+\left|\tfrac{\Delta\texttt{pr}_\ell}{\texttt{pr}_\ell}\right|+\left|\tfrac{\Delta\texttt{pr}_u}{\texttt{pr}_u}\right|
\end{aligned}
\]
\end{minipage}

&

\begin{minipage}[t]{\linewidth}\vspace{0pt}
\[\min(\texttt{tot\_m})\ge 3.5\text{M}\]
\[\tfrac{\max(\texttt{avg\_a})}{\min(\texttt{avg\_a})}\le 1.15\]
\end{minipage}

\\
\bottomrule
\end{tabular}
\caption{Overview of all three case studies. Each row corresponds to one case study.
Column~(a) shows the case number;
column~(b) shows the in-place query refinement (\origv{red crossed-out} $\to$ \refv{green bold});
column~(c) shows an excerpt of one OPRO subspace with sampled assignments and $(csat,d)$ pairs;
column~(d) formalizes the refinement distance objective;
column~(e) lists the formalized output constraints.}
\label{fig:case-studies-overview}
\Description{A unified five-column table presenting all three case studies side by side,
with case numbers, SQL query refinements, candidate subspace excerpts, distance objectives, and output constraints.}
\end{table*}

%% file: mean-optimality-appendix.tex
\section{Mean-Trial Optimality}
\label{appendix:mean-opt}

\begin{figure*}[t]
\vspace{-3mm}
\centering

\includegraphics[width=0.92\linewidth]
{images/baselines/success/legend_only.pdf}

\begin{subfigure}[b]{0.24\linewidth}
    \centering
    \includegraphics[width=\linewidth]
    {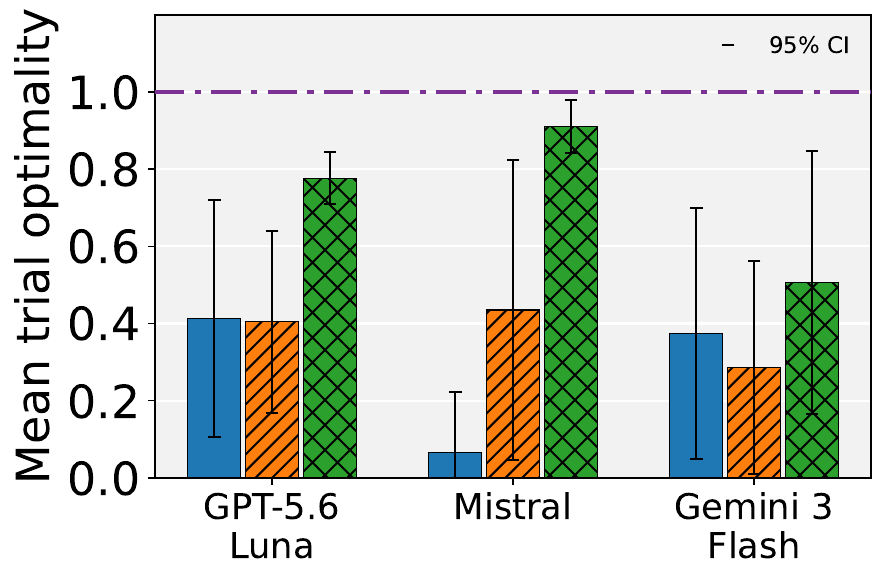}
    \caption{Top-$k$ Instances - Optimality}
    \label{fig:baseline-comparison-mean-optimality-top-k}
\end{subfigure}
\hfill
\begin{subfigure}[b]{0.24\linewidth}
    \centering
    \includegraphics[width=\linewidth]
    {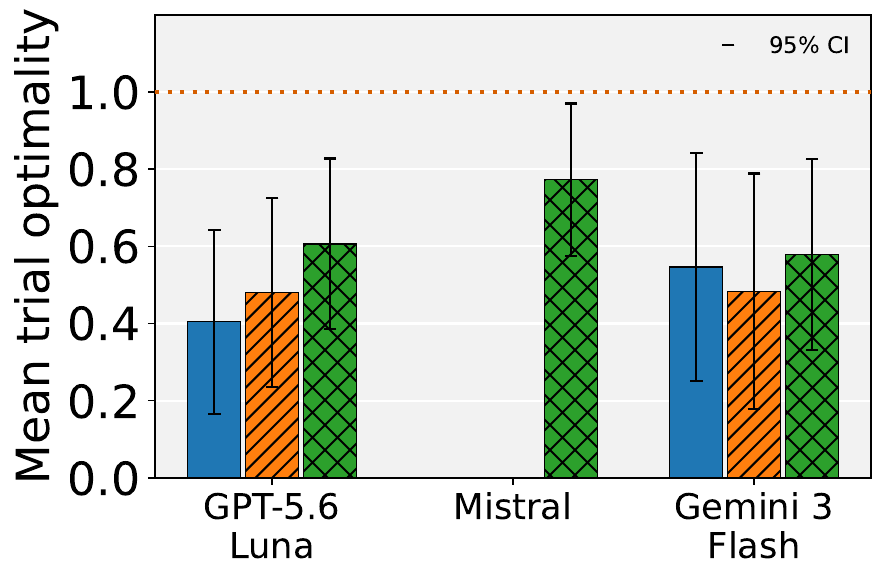}
    \caption{Range Instances}
    \label{fig:baseline-comparison-mean-optimality-range}
\end{subfigure}
\hfill
\begin{subfigure}[b]{0.24\linewidth}
    \centering
    \includegraphics[width=\linewidth]
    {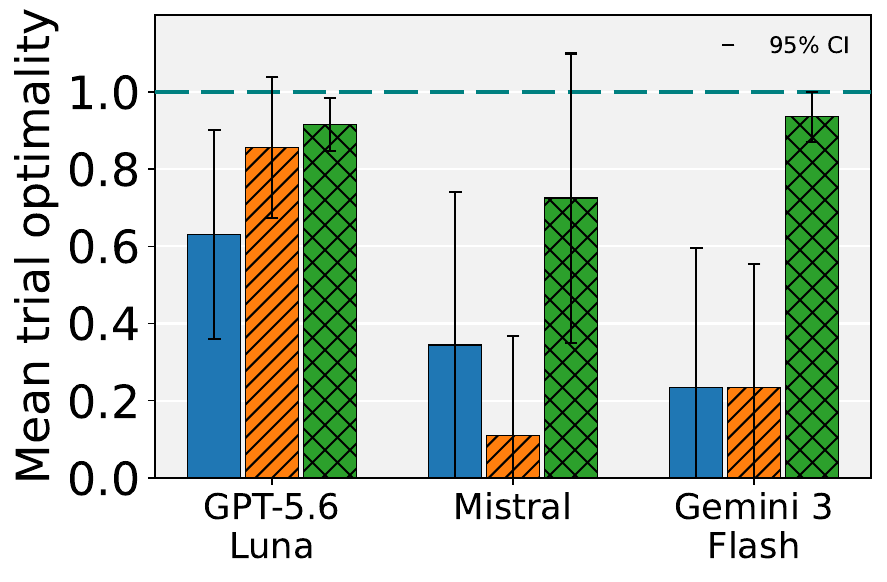}
    \caption{Diversity Instances}
    \label{fig:baseline-comparison-mean-optimality-diversity}
\end{subfigure}
\hfill
\begin{subfigure}[b]{0.24\linewidth}
    \centering
    \includegraphics[width=\linewidth]
    {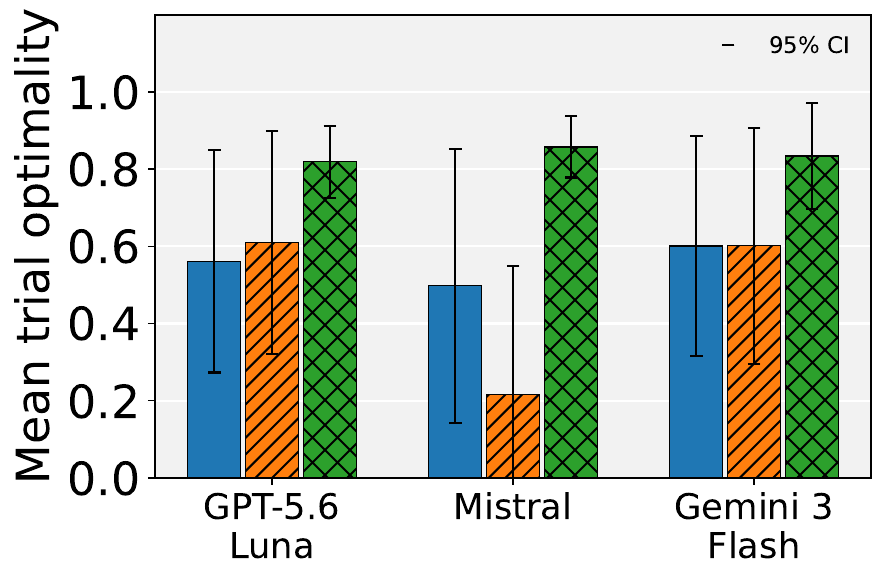}
    \caption{Complex Instances}
    \label{fig:baseline-comparison-mean-optimality-complex}
\end{subfigure}

\caption{
Mean trial optimality of \sysName{} and the Base and Thinking LLM baselines across instance classes. Blue and orange bars represent Base and Thinking, respectively; green bars represent \sysName{} using the Base model. Unsuccessful trials contribute zero optimality. Error bars show two-sided $95\%$ Student-$t$ confidence intervals across eight task means per class, with five trials per task for \new{GPT-5.6-Luna and Gemini-3-Flash} and one recovered historical trial per task for Mistral. Displayed lower confidence bounds are clipped at zero. Colored horizontal lines denote task-specific reference methods: RODEO for Top-$k$~\cite{rodeo}, the \new{fairness-aware} range-query method for Range~\cite{suraj2022range}, and ERICA for Diversity~\cite{erica}. These methods define the optimality reference of $1.0$ for their respective classes; no corresponding reference is shown for Complex instances\new{.}}
\label{fig:baseline-comparison-mean-optimality}
\end{figure*}

We also report mean-trial optimality, averaging optimality scores over recorded trials and assigning zero to unsuccessful trials. In \new{Figure}~\ref{fig:baseline-comparison-mean-optimality}, \new{\sysName{}} exceeds both the Base and Thinking LLM baselines for every model and instance class. Its mean scores range from 50.71\% to 93.63\%. The task-specific reference methods reach 100\% on their respective classes, so \new{\sysName{}} does not match those reference scores.

Table~\ref{tab:ablation-mean-trial-optimality} shows that the full method has the highest mean in 9 of the 12 model--class combinations. It exceeds both history ablations in all 12. The exceptions to its overall lead are GPT-5.6-Luna on Diversity, where Single-Step OPRO scores 92.64\% versus 91.61\%; and Gemini-3-Flash on Top-$k$ and Range, where SubspaceLM + Random and Single-Step OPRO score 57.09\% and 61.74\%, respectively, versus OmniTune's 50.71\% and 57.82\%.

\begin{table*}[t]
\centering
\small
\begin{tabular}{llrrrr}
\toprule
Model & Setting & Top-$k$ & Range & Diversity & Complex \\
\midrule
GPT-5.6-Luna
 & Single-Step OPRO (W/O SubspaceLM) & 59.71 & 57.97 & \textbf{92.64} & 75.88 \\
 & SubspaceLM + Random (W/O AssignmentLM) & 21.89 & 55.61 & 65.40 & 73.62 \\
 & Complete History (W/O Skyline) & 24.99 & 47.83 & 43.34 & 56.22 \\
 & Summarized History (W/O Skyline) & 55.84 & 55.26 & 84.60 & 72.21 \\
 & OmniTune (Full) & \textbf{77.62} & \textbf{60.64} & 91.61 & \textbf{81.91} \\
\midrule
Gemini-3-Flash
 & Single-Step OPRO (W/O SubspaceLM) & 50.69 & \textbf{61.74} & 90.27 & 71.33 \\
 & SubspaceLM + Random (W/O AssignmentLM) & \textbf{57.09} & 46.39 & 46.86 & 81.16 \\
 & Complete History (W/O Skyline) & 22.26 & 44.94 & 26.27 & 53.68 \\
 & Summarized History (W/O Skyline) & 42.94 & 55.23 & 91.73 & 63.77 \\
 & OmniTune (Full) & 50.71 & 57.82 & \textbf{93.63} & \textbf{83.42} \\
\midrule
Mistral
 & Single-Step OPRO (W/O SubspaceLM) & 48.63 & 59.92 & 34.59 & 85.55 \\
 & SubspaceLM + Random (W/O AssignmentLM) & 77.50 & 59.91 & 67.74 & 63.29 \\
 & Complete History (W/O Skyline) & 33.51 & 60.79 & 67.98 & 64.35 \\
 & Summarized History (W/O Skyline) & 91.00 & 49.67 & 57.49 & 69.26 \\ 
 & OmniTune (Full) & \textbf{91.06} & \textbf{77.24} & \textbf{72.52} & \textbf{85.78} \\
\bottomrule
\end{tabular}
\caption{Mean trial optimality (\%) across instance classes, including unsuccessful trials as zero. Optimality is averaged across the recorded trials for each model, setting, and instance class, assigning zero to unsuccessful trials and excluding unfinished trials.
}
\label{tab:ablation-mean-trial-optimality}
\end{table*}

\paragraph{Comparison with best-of-five optimality.}
\new{The best-of-five results in Figure~\ref{fig:baseline-comparison} (Section~\ref{sec:eval})} report the strongest trial, whereas mean-trial optimality includes unsuccessful trials as zeros. Accordingly, \new{\sysName{}'s} mean is lower than or equal to its best-of-five score in every model--class combination. The largest difference is for Mistral on Diversity: 72.52\% mean versus 95.27\% best-of-five. GPT-5.6-Luna on Top-$k$ likewise falls from 92.68\% to 77.62\%. These differences show how much the strongest attempt can exceed typical recorded performance.

The ablation ranking is also more nuanced under the mean-trial measure. \new{\sysName{}} leads both two-step ablations in every combination in the best-of-five results (Table~\ref{tab:unified-success-optimality}), but leads all settings in only 9 of 12 combinations by mean-trial optimality. Its advantage over both history ablations persists in all 12 combinations under both measures (Table~\ref{tab:history-skyline-ablation-unified}).

%% file: parameters_appendix.tex
\section{Per-Instance Class Breakdown of the Parameters Effect}
\label{appendix:parameters}

In Section~\ref{sec:eval} we reported the effect of the horizon $T$ and per-subspace refinements $K$ on \sysName{} using averages over all instance classes. Here we present the corresponding per-class breakdown (Top-$k$, Range, Diversity, and Complex), which reveals class-specific trends that the averaged view does not capture.

\begin{figure*}[t]
\vspace{-2mm}
  \centering

  \begin{subfigure}[b]{\linewidth}
    \includegraphics[width=0.25\linewidth]{images/tokens/refinement/legend_only.pdf}
  \end{subfigure}
  \vspace{-0.8em}

  \begin{subfigure}[b]{0.24\linewidth}
    \includegraphics[width=\linewidth]{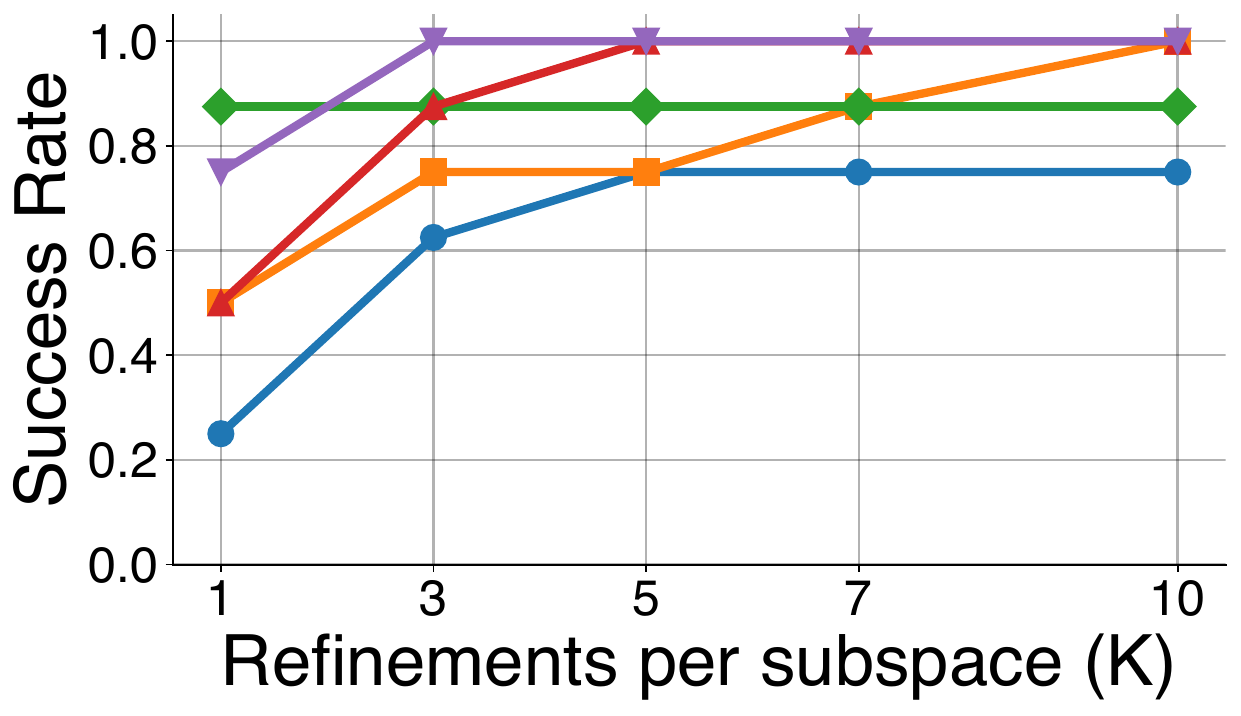}
    \caption{Success: Top-$k$}
    \label{fig:app-suc-ref-a}
  \end{subfigure}
  \hfill
  \begin{subfigure}[b]{0.24\linewidth}
    \includegraphics[width=\linewidth]{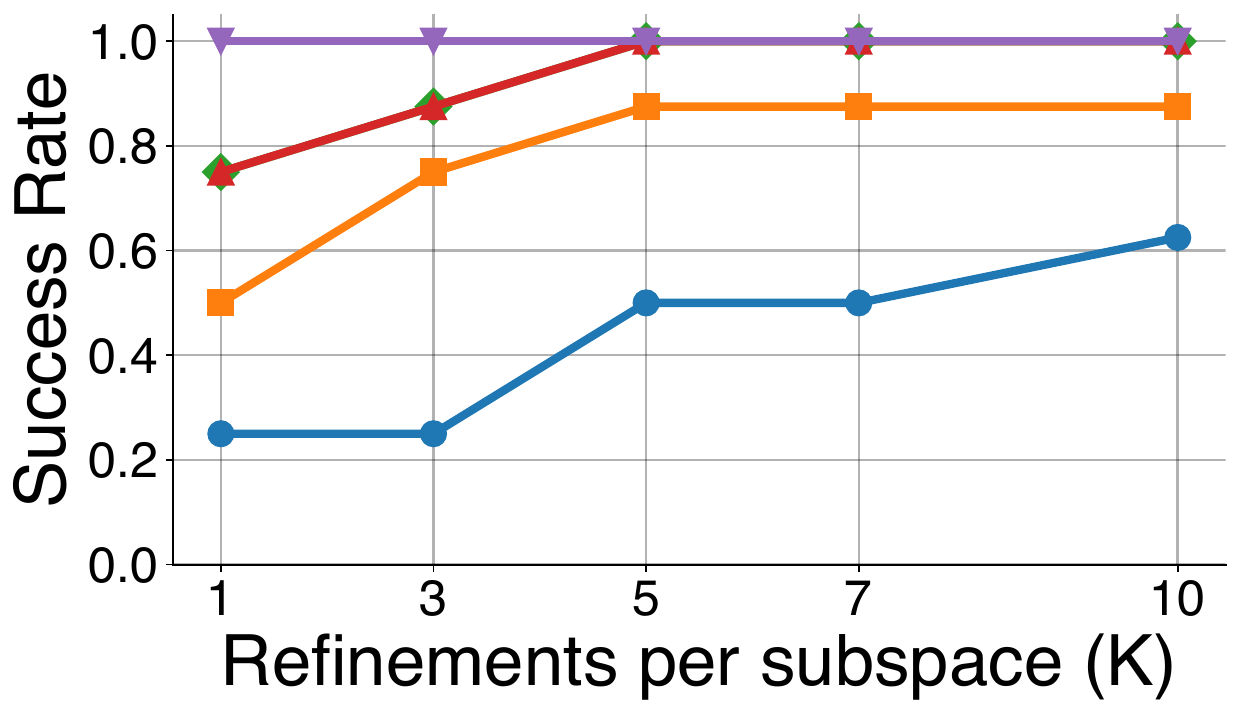}
    \caption{Success: Range}
    \label{fig:app-suc-ref-b}
  \end{subfigure}
  \hfill
  \begin{subfigure}[b]{0.24\linewidth}
    \includegraphics[width=\linewidth]{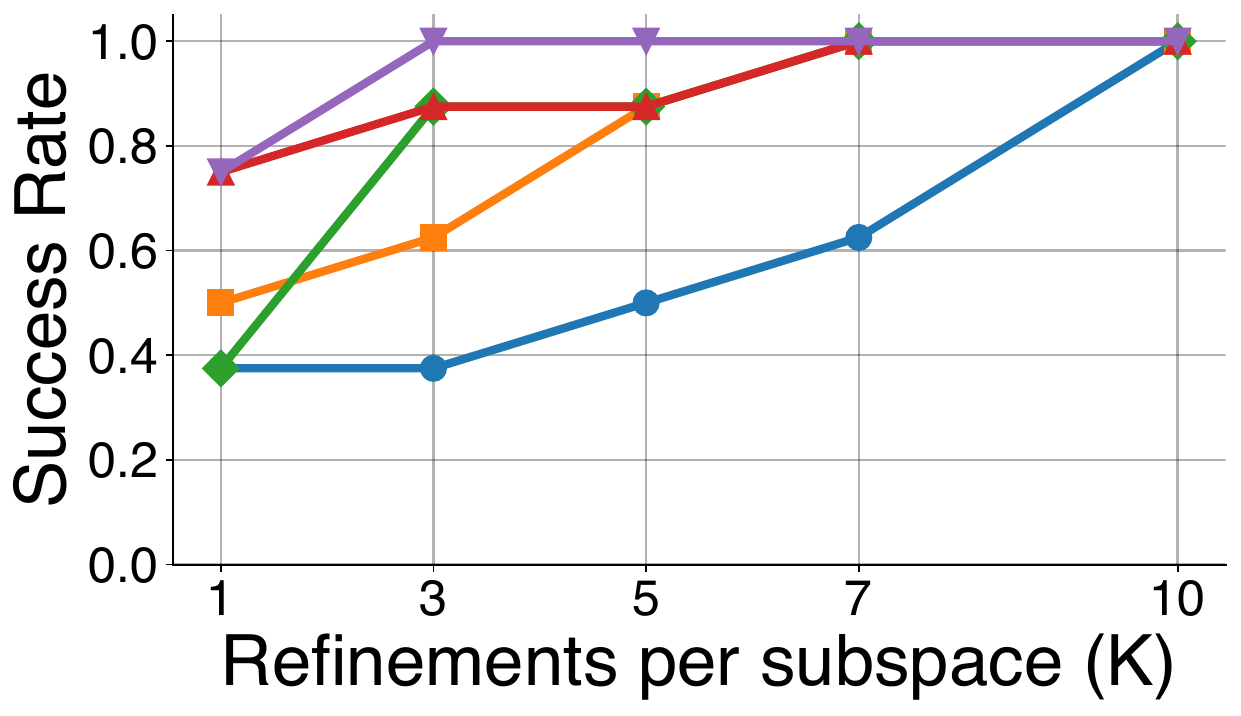}
    \caption{Success: Diversity}
    \label{fig:app-suc-ref-c}
  \end{subfigure}
  \hfill
  \begin{subfigure}[b]{0.24\linewidth}
    \includegraphics[width=\linewidth]{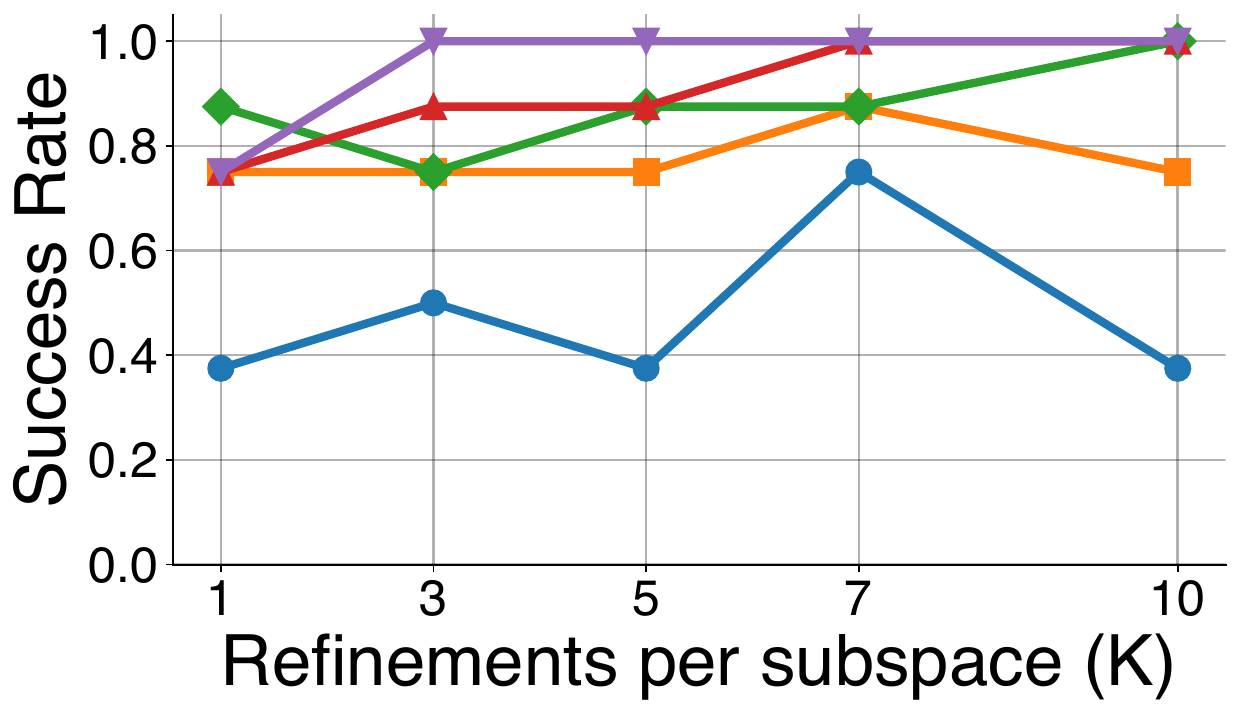}
    \caption{Success: Complex}
    \label{fig:app-suc-ref-d}
  \end{subfigure}

  \vspace{0.5em}

  \begin{subfigure}[b]{0.24\linewidth}
    \includegraphics[width=\linewidth]{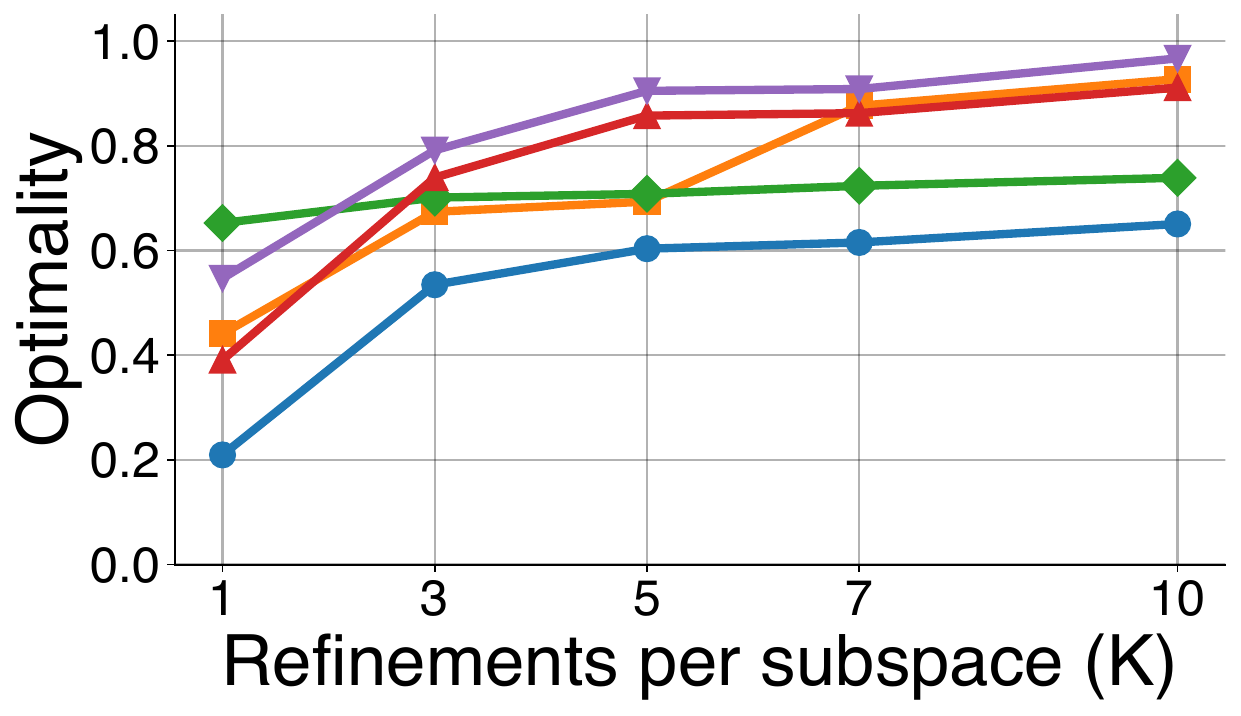}
    \caption{Optimality: Top-$k$}
    \label{fig:app-opt-ref-a}
  \end{subfigure}
  \hfill
  \begin{subfigure}[b]{0.24\linewidth}
    \includegraphics[width=\linewidth]{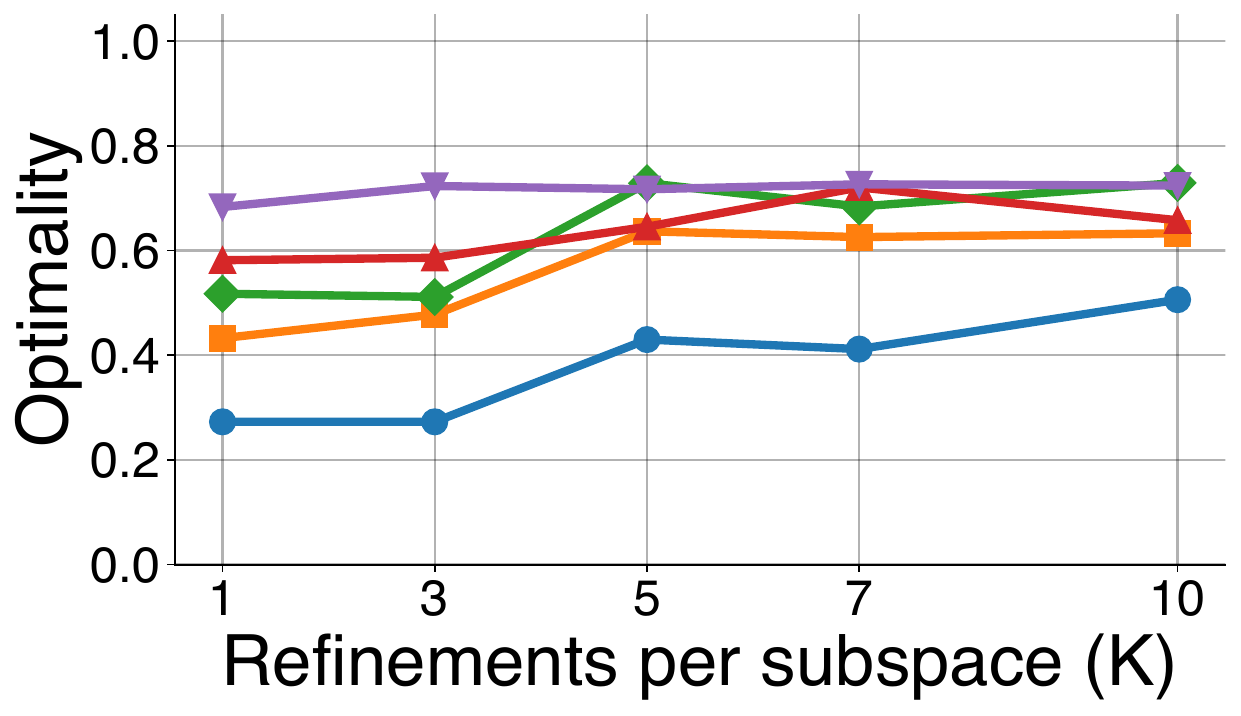}
    \caption{Optimality: Range}
    \label{fig:app-opt-ref-b}
  \end{subfigure}
  \hfill
  \begin{subfigure}[b]{0.24\linewidth}
    \includegraphics[width=\linewidth]{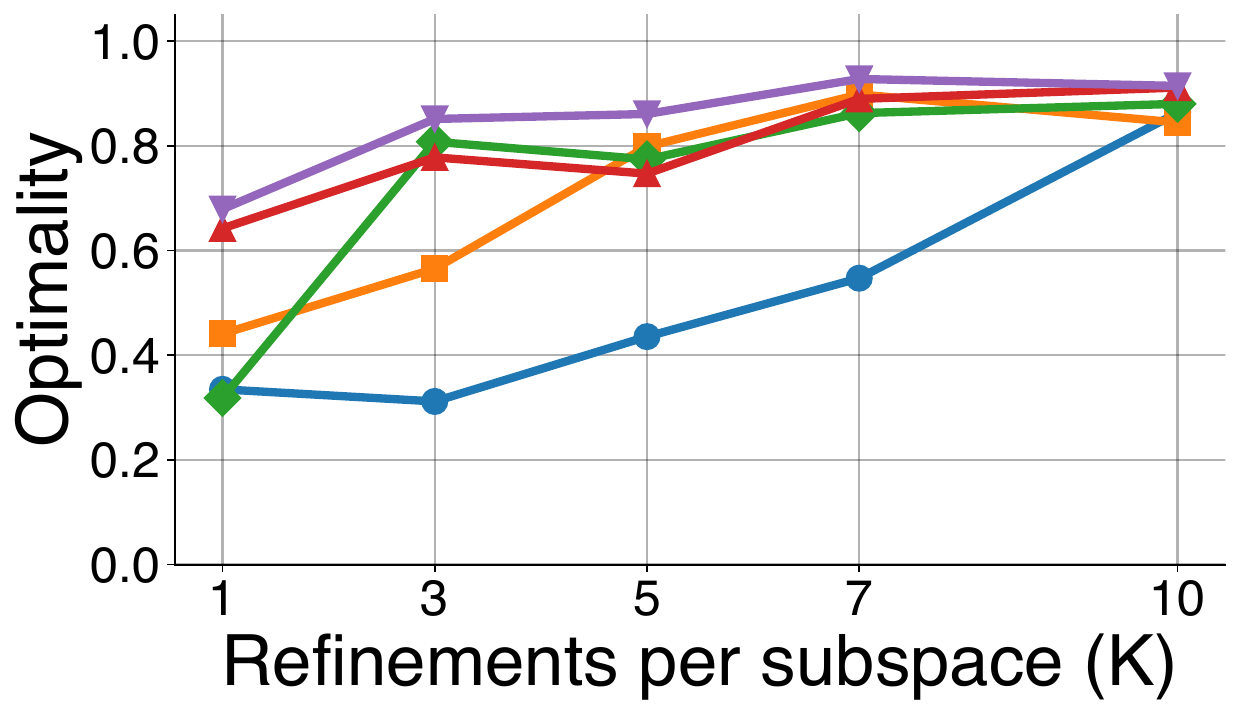}
    \caption{Optimality: Diversity}
    \label{fig:app-opt-ref-c}
  \end{subfigure}
  \hfill
  \begin{subfigure}[b]{0.24\linewidth}
    \includegraphics[width=\linewidth]{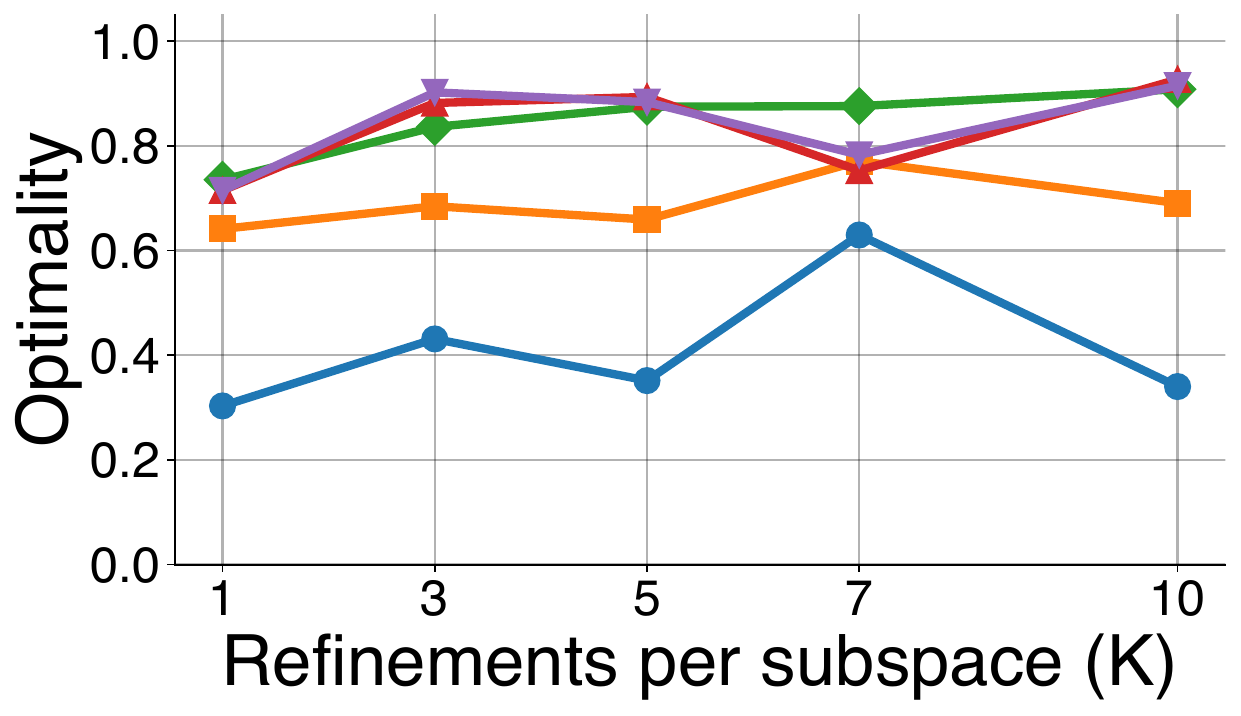}
    \caption{Optimality: Complex}
    \label{fig:app-opt-ref-d}
  \end{subfigure}

  \vspace{0.5em}

  \begin{subfigure}[b]{0.24\linewidth}
    \includegraphics[width=\linewidth]{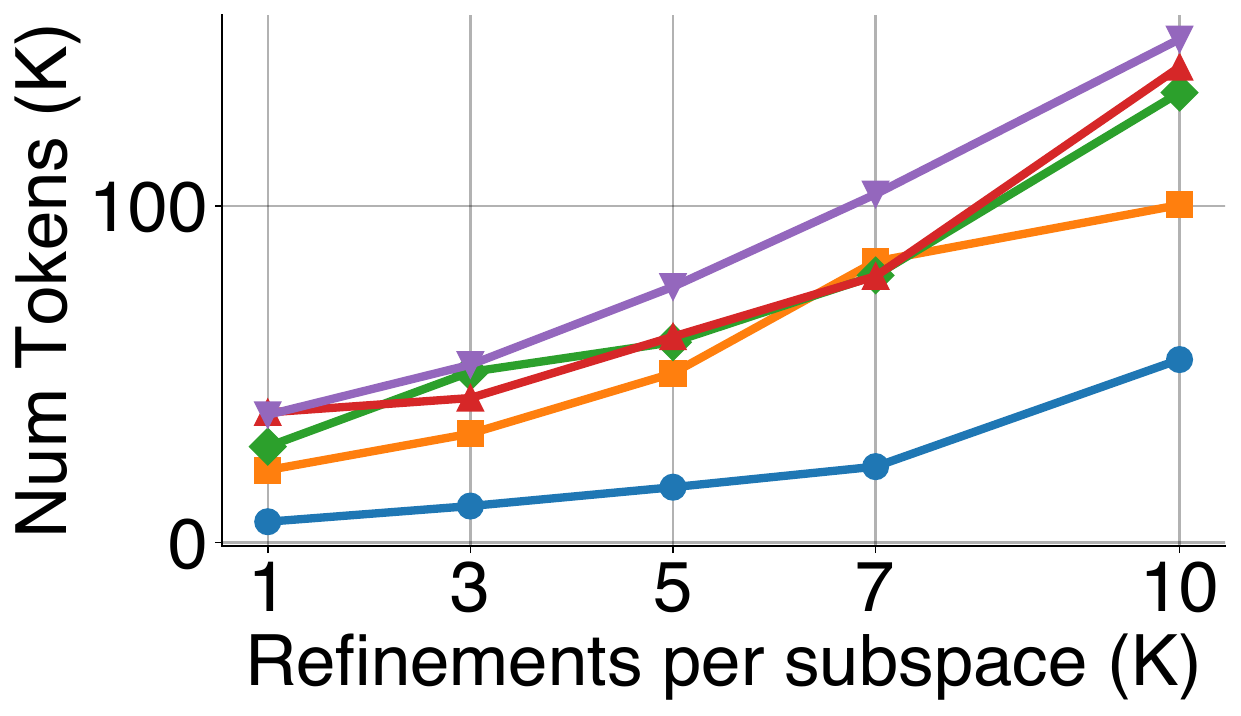}
    \caption{Tokens: Top-$k$}
    \label{fig:app-tok-ref-a}
  \end{subfigure}
  \hfill
  \begin{subfigure}[b]{0.24\linewidth}
    \includegraphics[width=\linewidth]{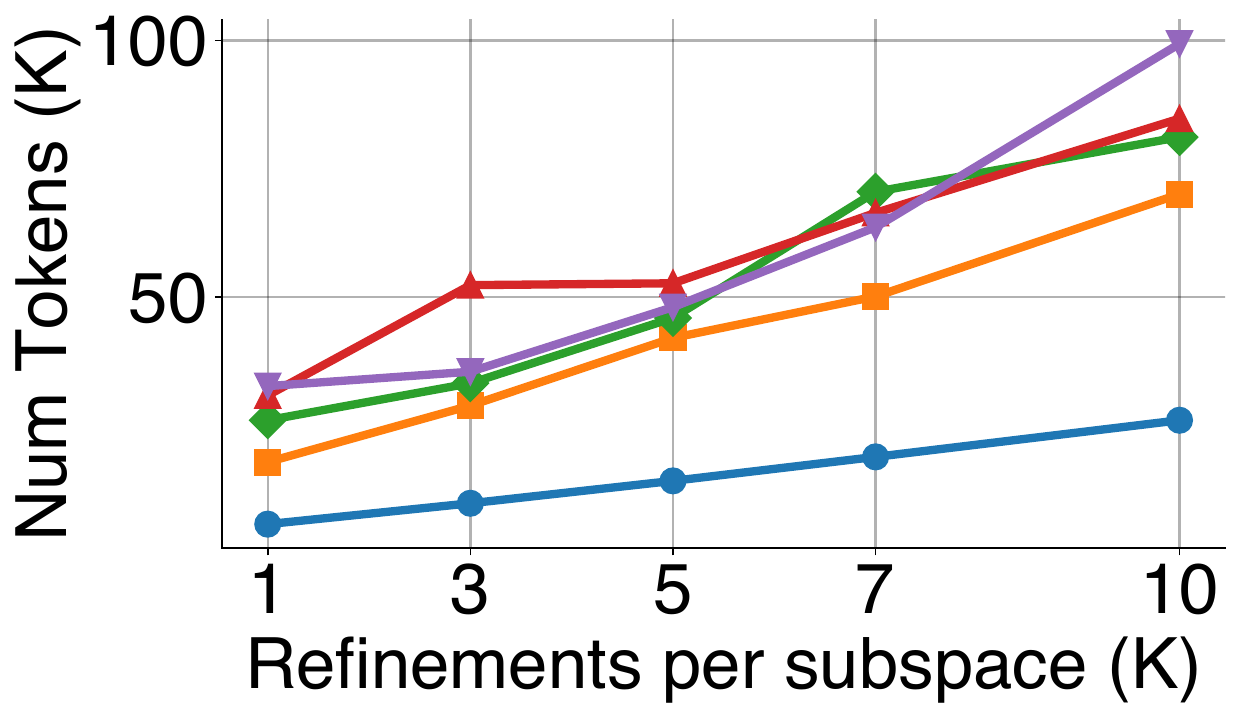}
    \caption{Tokens: Range}
    \label{fig:app-tok-ref-b}
  \end{subfigure}
  \hfill
  \begin{subfigure}[b]{0.24\linewidth}
    \includegraphics[width=\linewidth]{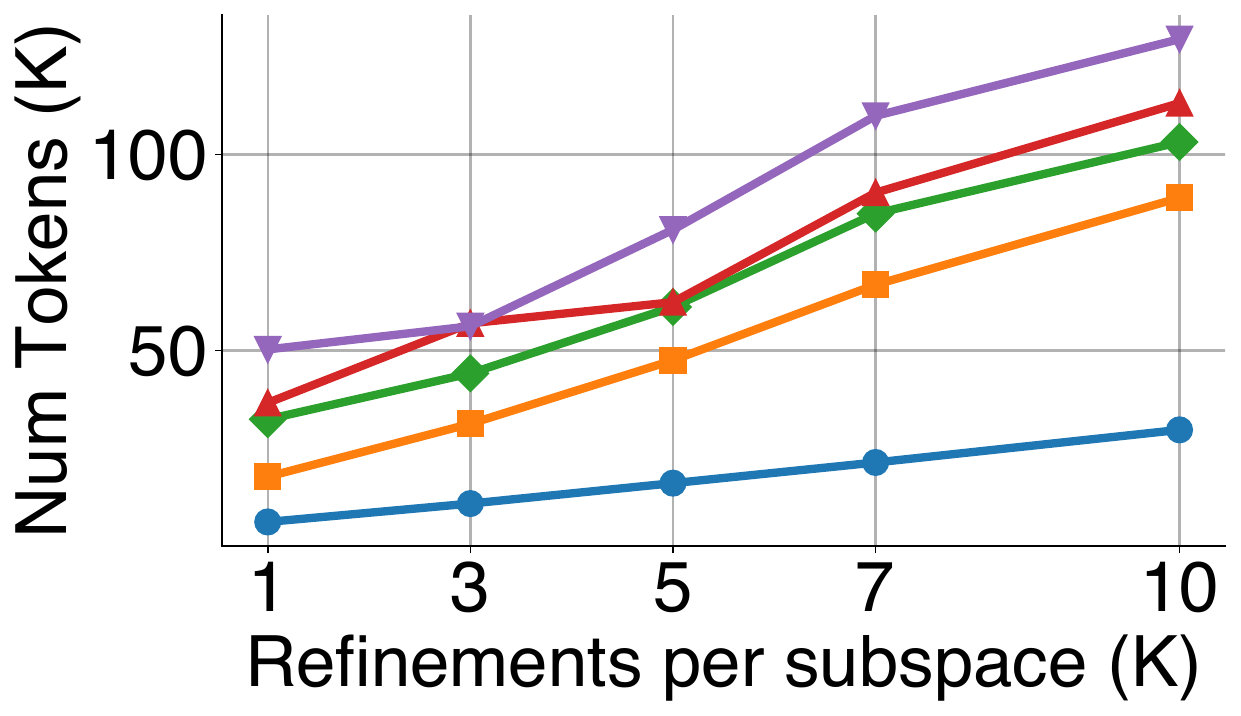}
    \caption{Tokens: Diversity}
    \label{fig:app-tok-ref-c}
  \end{subfigure}
  \hfill
  \begin{subfigure}[b]{0.24\linewidth}
    \includegraphics[width=\linewidth]{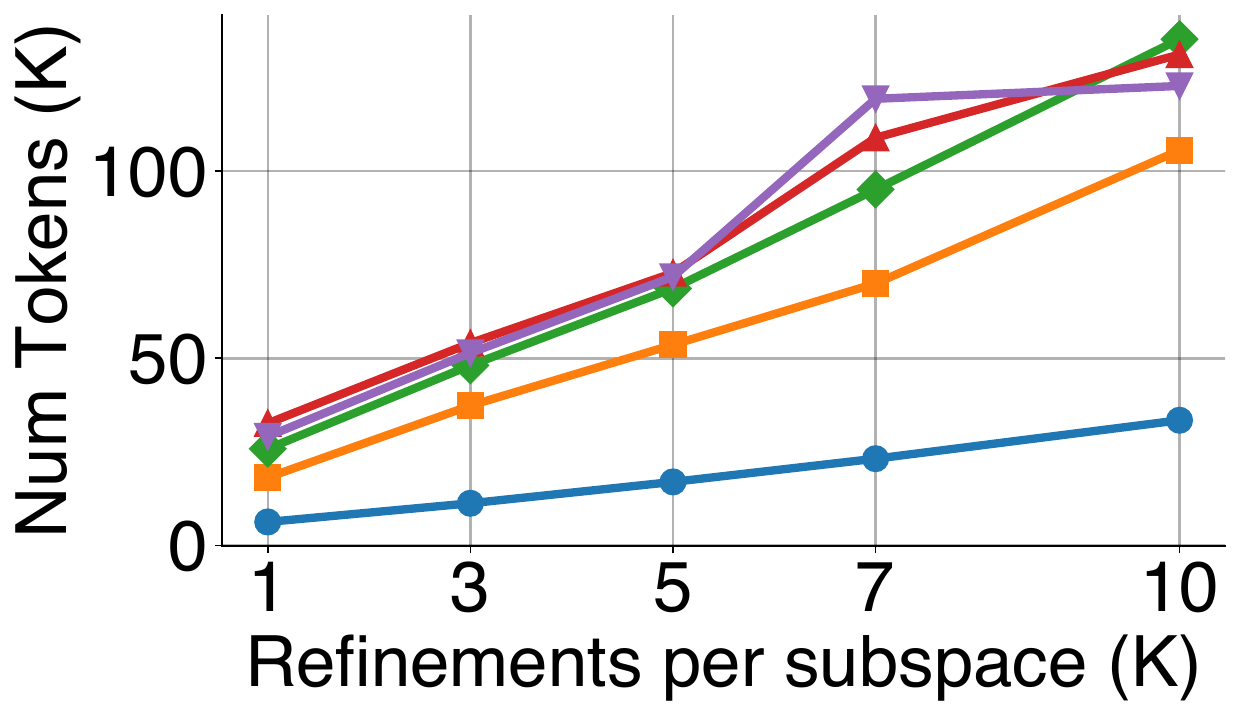}
    \caption{Tokens: Complex}
    \label{fig:app-tok-ref-d}
  \end{subfigure}

  \caption{Per-class success rate (top row), optimality score (middle row), and token cost (bottom row) as a function of refinements per subspace $K$.}
  \label{fig:app-refinements-all}
\end{figure*}

\paragraph{Refinements per Iteration ($K$)}
Figure~\ref{fig:app-refinements-all} presents the per-class success rates, optimality values, and token cost when ranging $K$ from $1$ to $10$ for different fixed horizons ($T$).
The convergence pattern observed in the averaged results (Figure~\ref{fig:refinements-num-effect}) holds consistently for the Top-$k$, Range and Diversity classes (Figures~\ref{fig:app-suc-ref-a}--\ref{fig:app-suc-ref-c} and~\ref{fig:app-opt-ref-a}--\ref{fig:app-opt-ref-c}): when the horizon is set to $5$ or more, incremental gains in both success rate and optimality become negligible beyond $K=5$.
For Complex instances (Figures~\ref{fig:app-suc-ref-d} and~\ref{fig:app-opt-ref-d}), however, the results did not show a clear convergence trend when using a low number of subspaces (i.e., $T \leq 3$). This is due to the more involved structure of the queries in this class, e.g., including \texttt{HAVING}-clause predicates, which require a larger number of iterations to improve the quality of the output.
The per-class token cost (Figures~\ref{fig:app-tok-ref-a}--\ref{fig:app-tok-ref-d}) is consistent with the averaged view, showing linear growth in tokens as $K$ increases across all classes.

\begin{figure*}[t]
\vspace{-2mm}
  \centering

  \begin{subfigure}[b]{\linewidth}
    \includegraphics[width=0.25\linewidth]{images/tokens/subspace/legend_only.pdf}
  \end{subfigure}
  \vspace{-0.8em}

  \begin{subfigure}[b]{0.24\linewidth}
    \includegraphics[width=\linewidth]{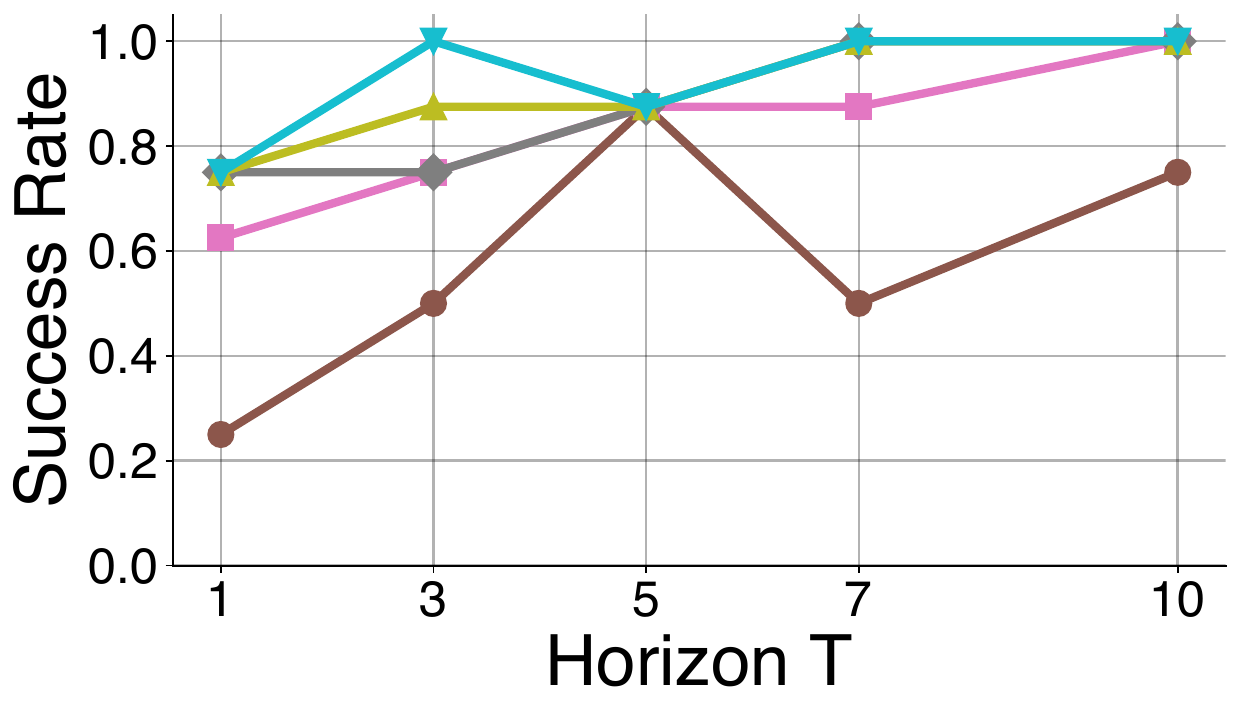}
    \caption{Success: Top-$k$}
    \label{fig:app-suc-sub-a}
  \end{subfigure}
  \hfill
  \begin{subfigure}[b]{0.24\linewidth}
    \includegraphics[width=\linewidth]{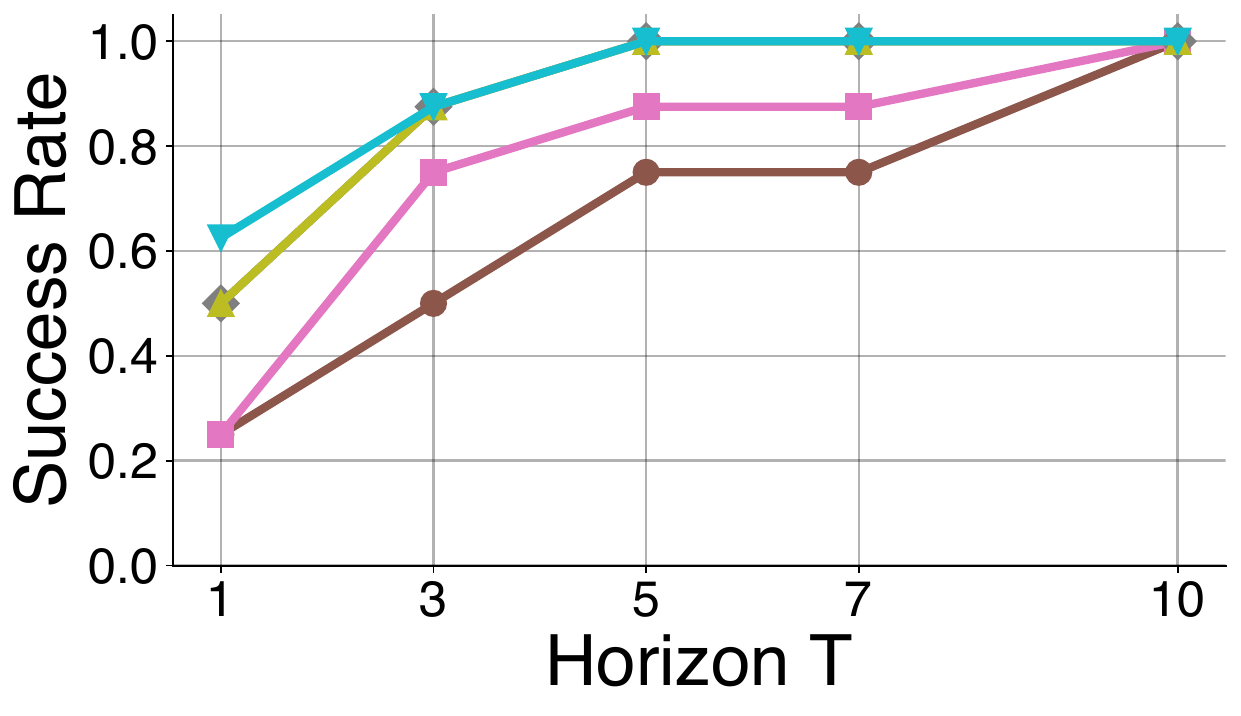}
    \caption{Success: Range}
    \label{fig:app-suc-sub-b}
  \end{subfigure}
  \hfill
  \begin{subfigure}[b]{0.24\linewidth}
    \includegraphics[width=\linewidth]{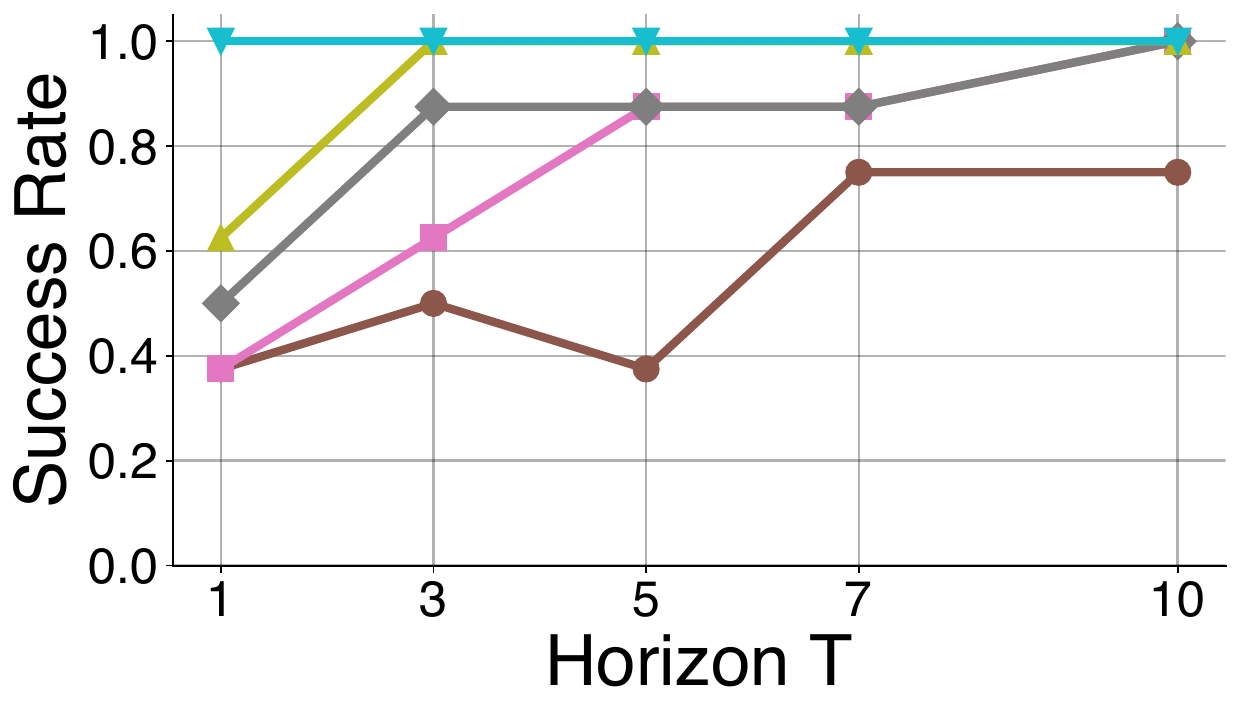}
    \caption{Success: Diversity}
    \label{fig:app-suc-sub-c}
  \end{subfigure}
  \hfill
  \begin{subfigure}[b]{0.24\linewidth}
    \includegraphics[width=\linewidth]{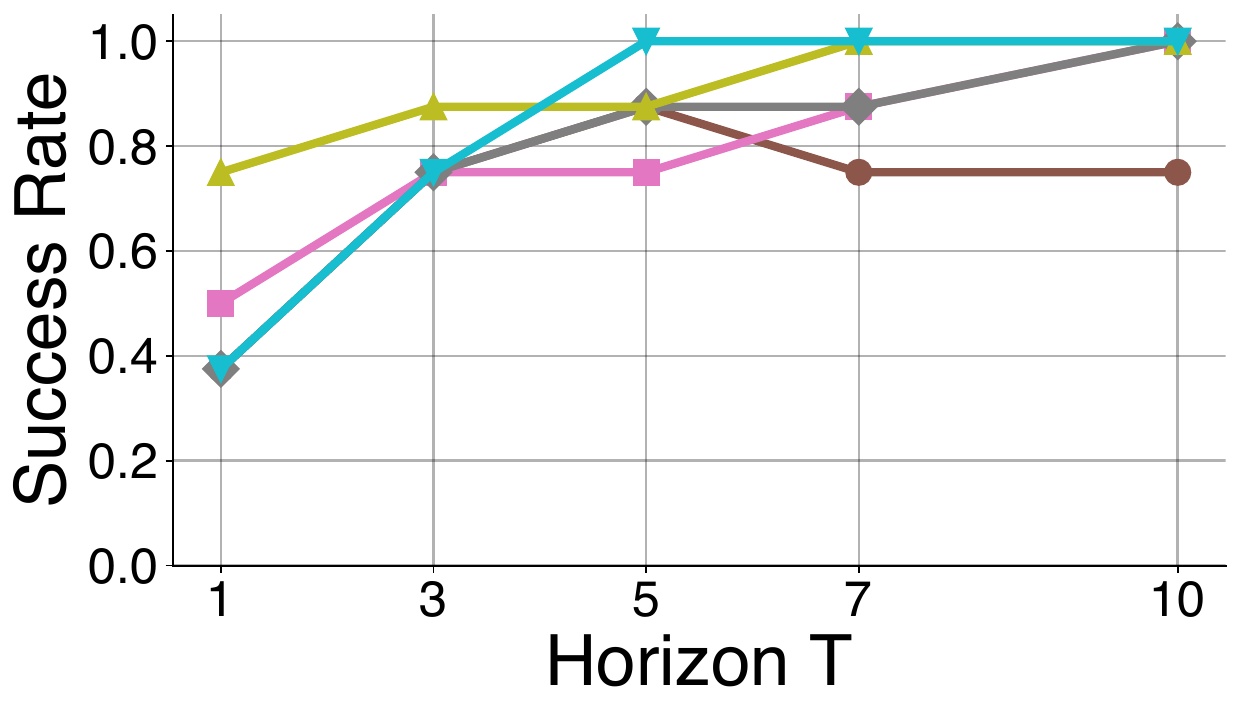}
    \caption{Success: Complex}
    \label{fig:app-suc-sub-d}
  \end{subfigure}

  \vspace{0.5em}

  \begin{subfigure}[b]{0.24\linewidth}
    \includegraphics[width=\linewidth]{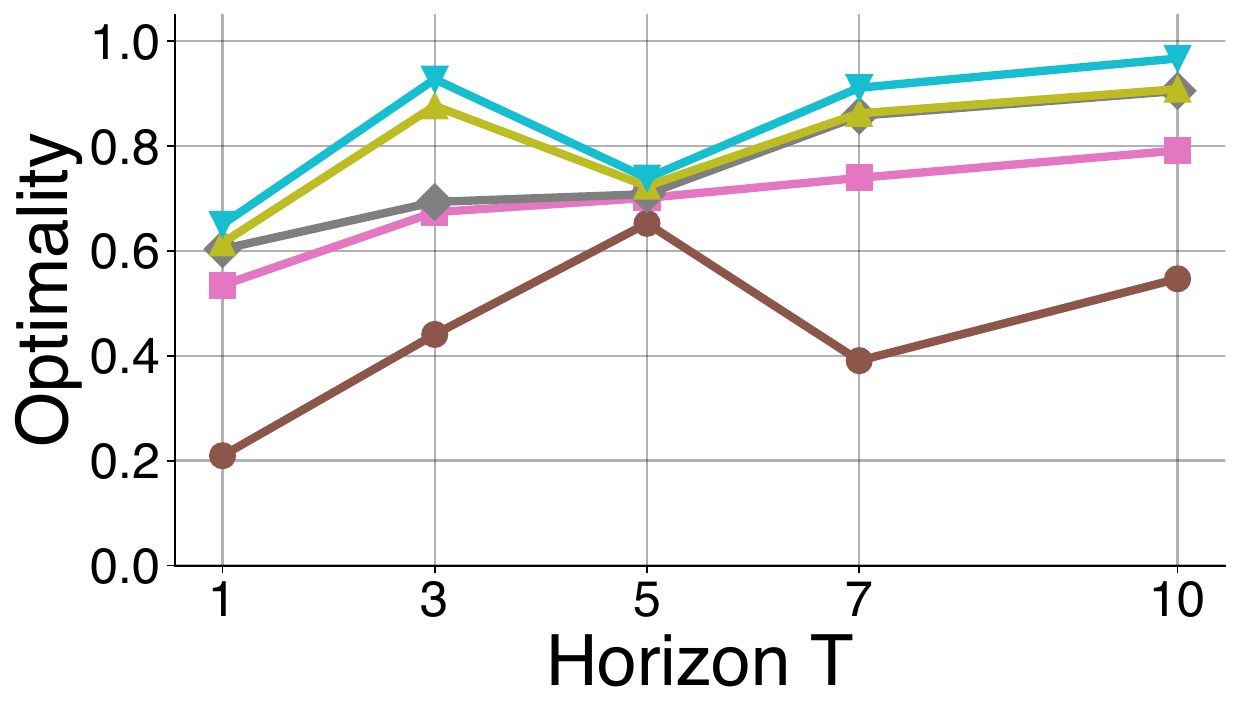}
    \caption{Optimality: Top-$k$}
    \label{fig:app-opt-sub-a}
  \end{subfigure}
  \hfill
  \begin{subfigure}[b]{0.24\linewidth}
    \includegraphics[width=\linewidth]{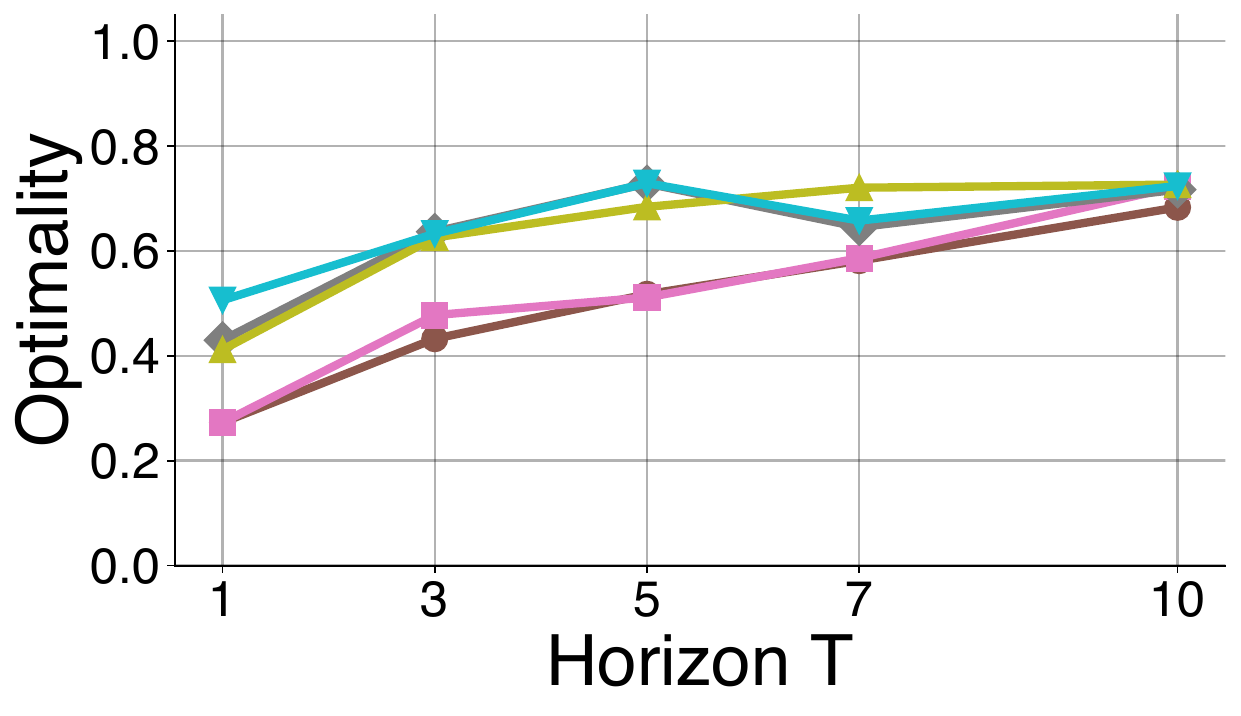}
    \caption{Optimality: Range}
    \label{fig:app-opt-sub-b}
  \end{subfigure}
  \hfill
  \begin{subfigure}[b]{0.24\linewidth}
    \includegraphics[width=\linewidth]{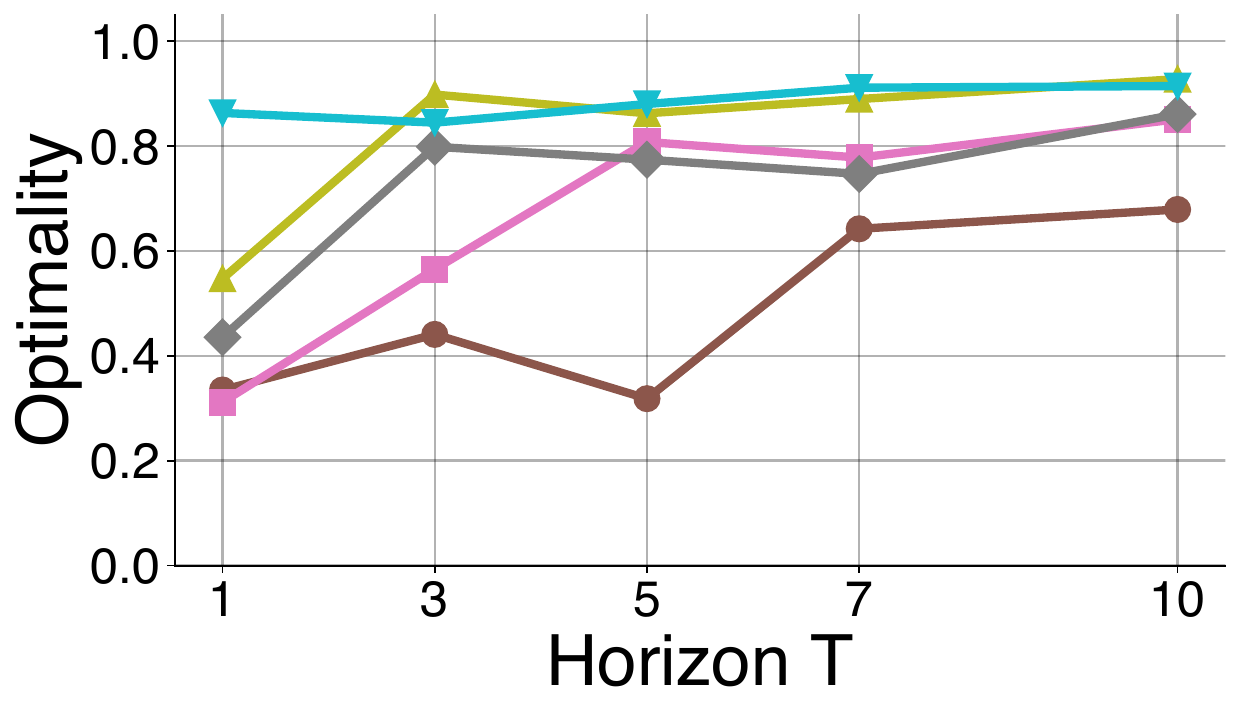}
    \caption{Optimality: Diversity}
    \label{fig:app-opt-sub-c}
  \end{subfigure}
  \hfill
  \begin{subfigure}[b]{0.24\linewidth}
    \includegraphics[width=\linewidth]{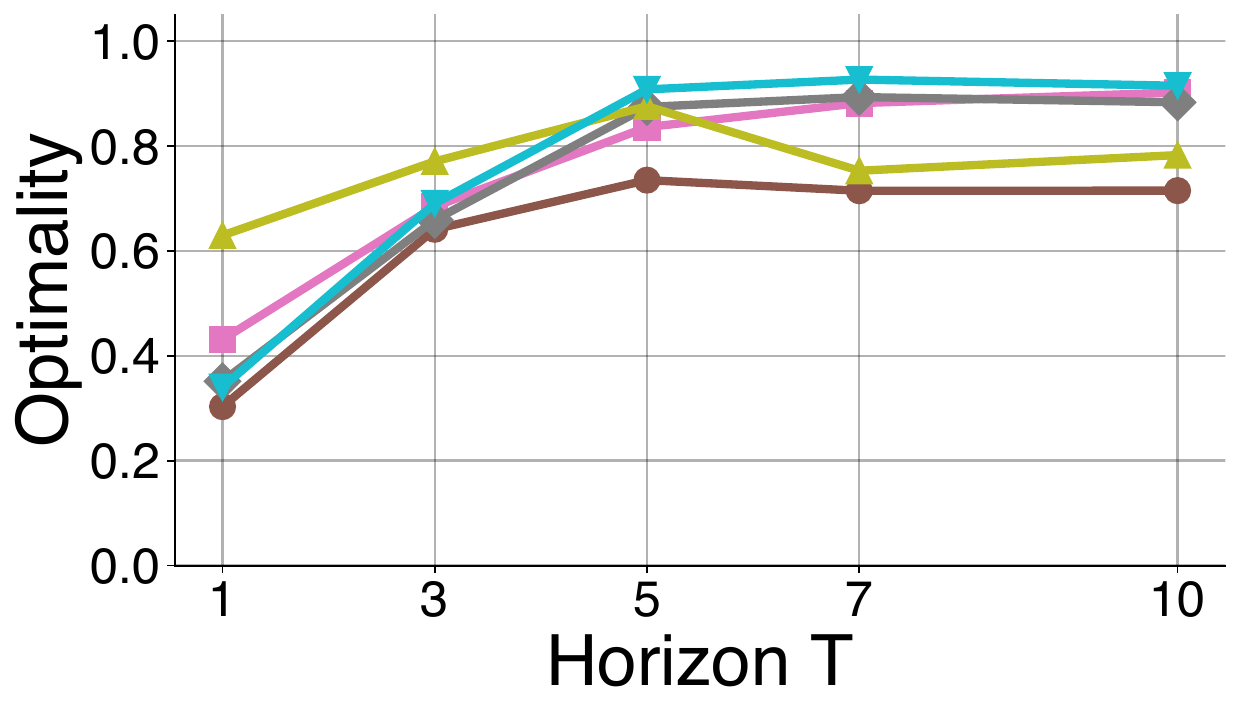}
    \caption{Optimality: Complex}
    \label{fig:app-opt-sub-d}
  \end{subfigure}

  \vspace{0.5em}

  \begin{subfigure}[b]{0.24\linewidth}
    \includegraphics[width=\linewidth]{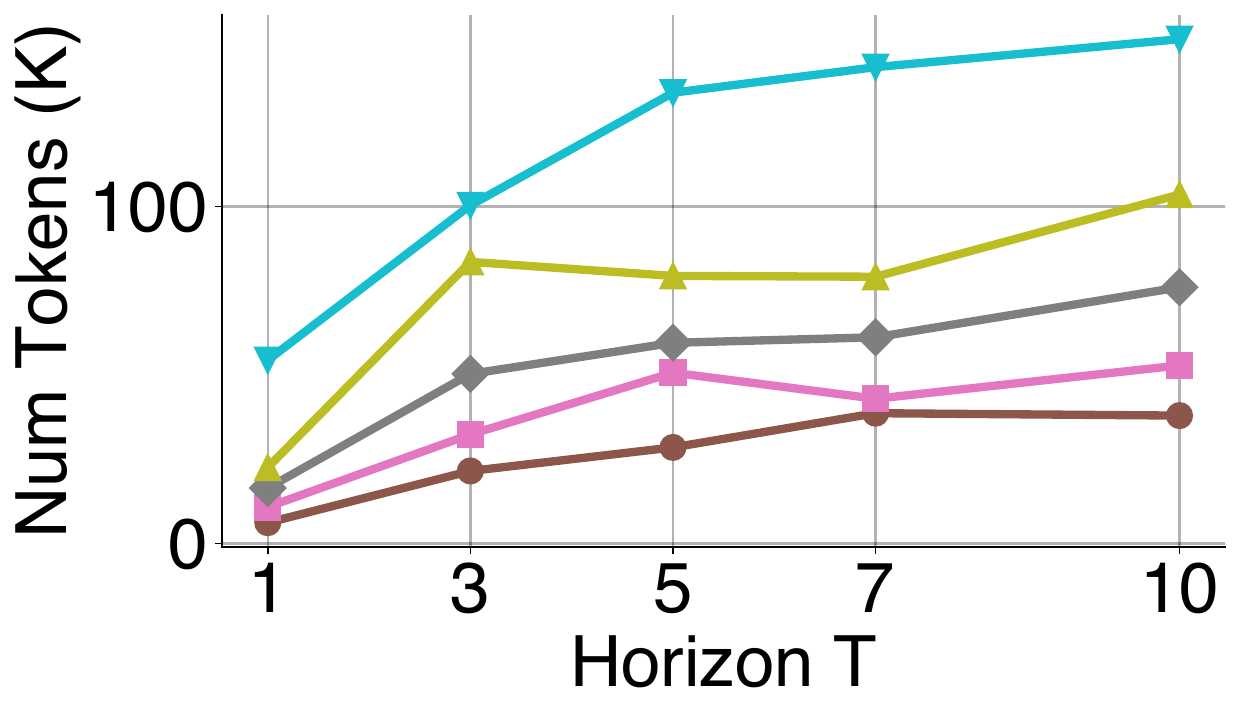}
    \caption{Tokens: Top-$k$}
    \label{fig:app-tok-sub-a}
  \end{subfigure}
  \hfill
  \begin{subfigure}[b]{0.24\linewidth}
    \includegraphics[width=\linewidth]{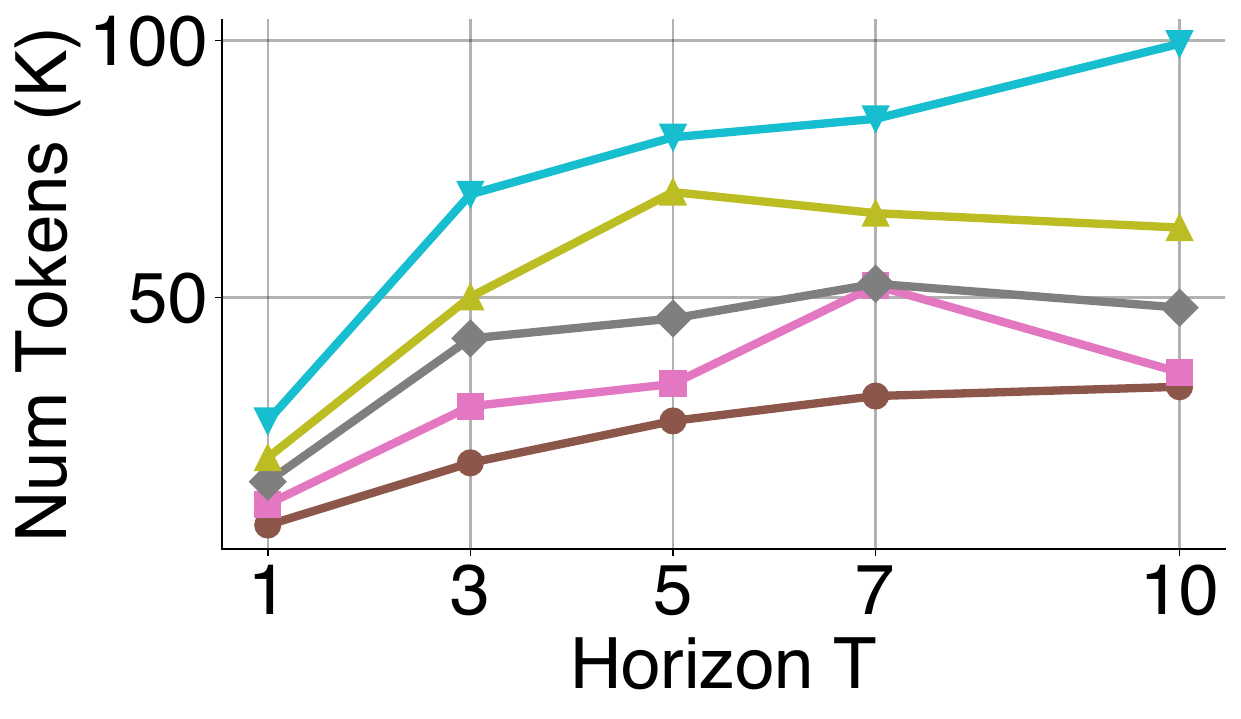}
    \caption{Tokens: Range}
    \label{fig:app-tok-sub-b}
  \end{subfigure}
  \hfill
  \begin{subfigure}[b]{0.24\linewidth}
    \includegraphics[width=\linewidth]{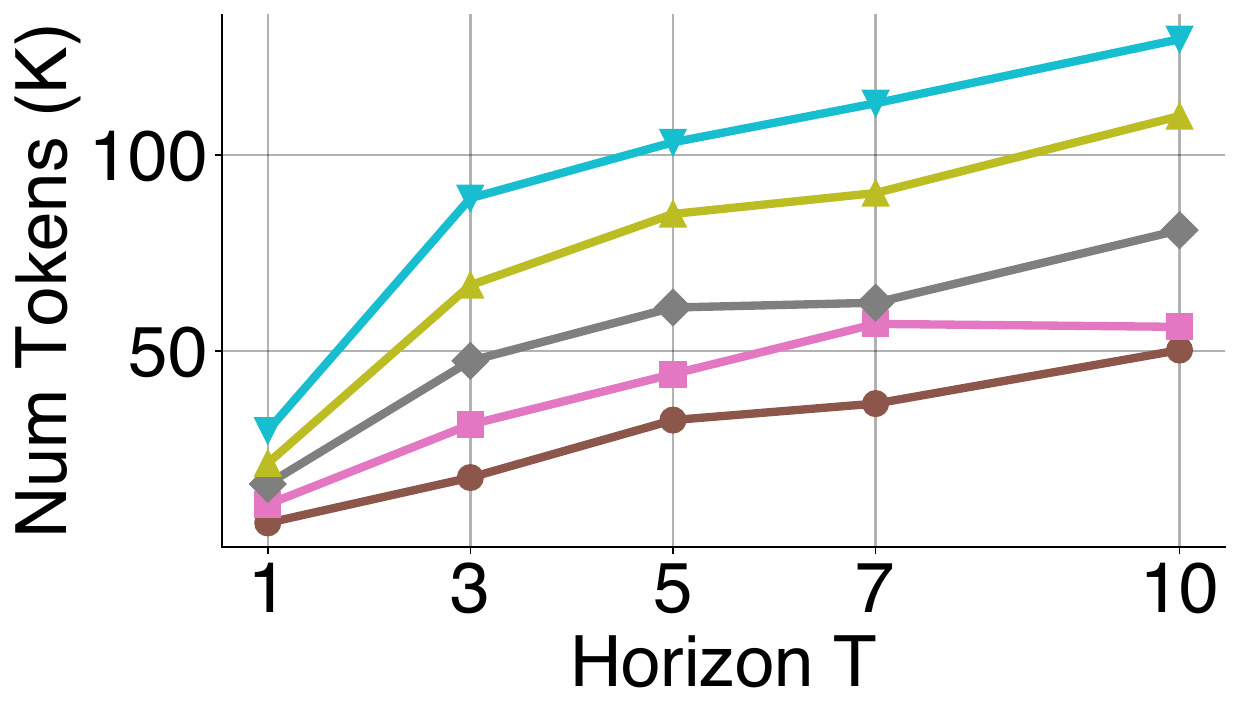}
    \caption{Tokens: Diversity}
    \label{fig:app-tok-sub-c}
  \end{subfigure}
  \hfill
  \begin{subfigure}[b]{0.24\linewidth}
    \includegraphics[width=\linewidth]{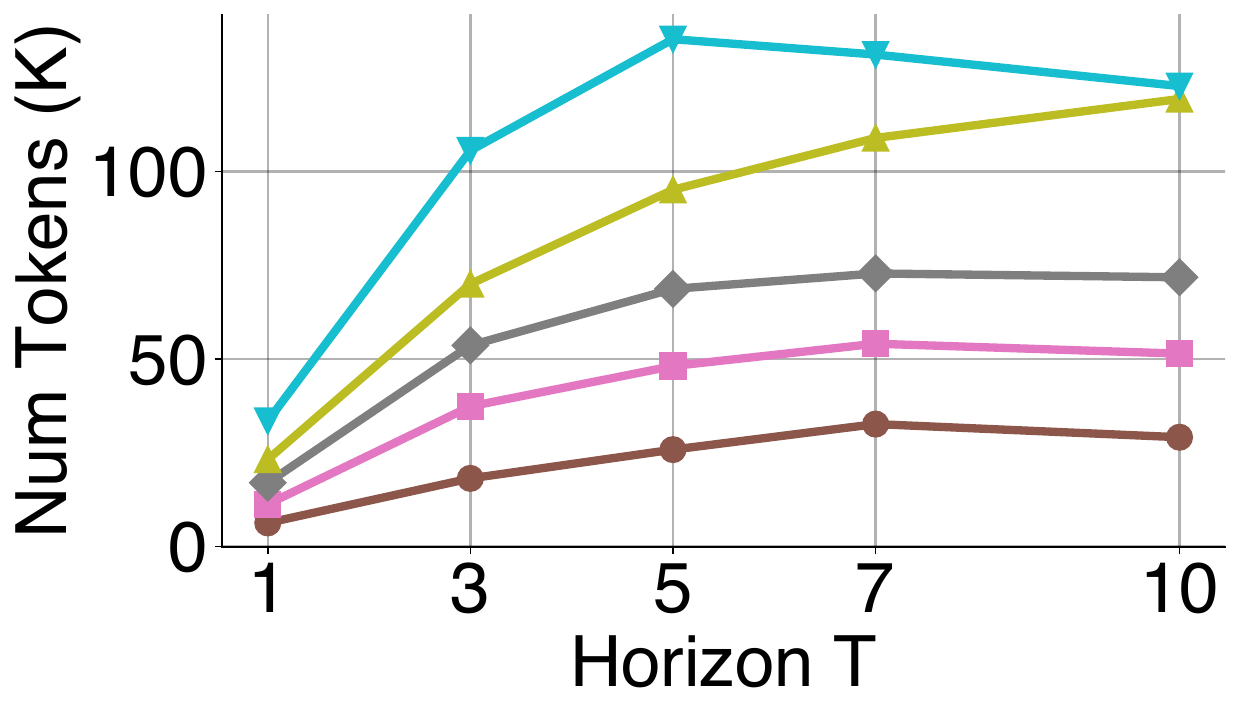}
    \caption{Tokens: Complex}
    \label{fig:app-tok-sub-d}
  \end{subfigure}

  \caption{Per-class success rate (top row), optimality score (middle row), and token cost (bottom row) as a function of the number of subspaces (horizon $T$).}
  \label{fig:app-subspaces-all}
\end{figure*}

\paragraph{Horizon ($T$)}
Figure~\ref{fig:app-subspaces-all} provides the per-class view of the effect of varying the horizon $T$, complementing the averaged results in Figure~\ref{fig:horizon-t-effect}.
The general trend of improving quality with growing $T$ holds across all classes, yet we observed some deviations for the Top-$k$ class, likely due to the inherent complexity of the constraint objectives in this class, which can involve internal trade-offs in certain scenarios. 
As with $K$, the per-class results confirm a diminishing improvement in success rate and optimality beyond $T=5$ when using $5$ or more refinements per iteration. The moderate growth in token cost for increasing $T$, as opposed to the linear growth observed for $K$, is also consistent across all four classes (Figures~\ref{fig:app-tok-sub-a}--\ref{fig:app-tok-sub-d}), further supporting the explanation that higher refinements per subspace lead to faster convergence, reducing the number of subspace iterations needed before early stopping.

